\documentclass[11pt]{article} 

\makeatletter\renewcommand{\section}{\@startsection
	{section}{1}{\z@}{-3.5ex plus -1ex minus
		-.2ex}{2.3ex plus .2ex}{\bf }}

\makeatletter\renewcommand{\subsection}{\@startsection{subsection}{2}{\z@}{-3.25ex
    plus -1ex minus
    -.2ex}{1.5ex plus .2ex}{\it }}
\makeatletter\renewcommand{\subsubsection}{\@startsection{subsubsection}{3}{-2.45ex}{-3.25ex
    plus -1ex minus -.2ex}{1.5ex plus .2ex}{\it }}
\renewcommand{\thesection}{\arabic{section}}
\renewcommand{\thesubsection}{\arabic{section}.\arabic{subsection}.}

\renewcommand{\theequation}{\thesection.\arabic{equation}}
\makeatletter \@addtoreset{equation}{section}

\usepackage{graphicx}
\usepackage{float} 
\usepackage{epsfig}
\usepackage{verbatim}
\usepackage{easyfig}
\usepackage{bbm}
\usepackage{amsmath}
\usepackage{amsfonts}
\usepackage[compact]{titlesec}
\usepackage{subdepth}
\usepackage{dsfont}
\usepackage{tikz}
\usepackage[utf8]{inputenc}
\usepackage[T1]{fontenc}
\usepackage{amsmath}
\usepackage[toc]{glossaries}
\usepackage[countmax]{subfloat}
\usepackage{caption}
\usepackage{subcaption}
\usepackage{setspace}
\usepackage{physics}
\usepackage{amsfonts}
\usepackage{amssymb}
\usepackage{mathrsfs}
\usepackage{amsmath,amssymb}
\usepackage{bm}
\usepackage{caption}
\usepackage{amsmath}
\usepackage{graphicx}
\usepackage{cite}
\usepackage{pdfpages}
\usepackage{caption}
\usepackage{subcaption}
\usepackage{bbm}
\usepackage{braket}
\usepackage{slashed}
\usepackage{bm}
\usepackage{cite}

\usepackage[left=3cm, right=3cm, top=4cm, bottom=4cm]{geometry}
\usepackage[utf8]{inputenc}
\usepackage{color}
\usepackage{caption}
\usepackage{subcaption}
\usepackage{minted} 
\usepackage{xcolor} 
\usepackage{array}
\usepackage{graphicx}
\usepackage{braket, amsfonts, amsmath}
\usepackage{forest}
\usepackage{subcaption}
\usepackage{diagbox}
\usepackage{booktabs}

\usepackage{xcolor}

\usepackage[splitrule]{footmisc}

\AddToHook{env/*/before}{\unskip}
\AddToHook{env/*/after}{\ignorespaces}

\newcommand{\sixj}[6]{\left\{ \begin{matrix}
    #1 & #2 & #3 \\
    #4 & #5 & #6
  \end{matrix} \right\}}

\graphicspath{{figures_new/}}

\renewenvironment{thebibliography}[1]
{\baselineskip=16pt plus 2pt minus 1pt
	\section*{\large\refname
		\@mkboth{\MakeUppercase\refname}{\MakeUppercase\refname}}%
	\list{\@biblabel{\@arabic\c@enumiv}}%
	{\settowidth\labelwidth{\@biblabel{#1}}%
		\leftmargin\labelwidth
		\advance\leftmargin\labelsep
		\@openbib@code
		\usecounter{enumiv}%
		\let\p@enumiv\@empty
		\renewcommand\theenumiv{\@arabic\c@enumiv}}%
	\sloppy
	\clubpenalty4000
	\@clubpenalty \clubpenalty
	\widowpenalty4000%
	\sfcode`\.\@m}

\let\fn\footnote
\renewcommand{\footnote}[1]{\linespread{1.1}\fn{#1}\linespread{1.29}}

 \newcommand{\appendices}{\section*{Appendices}\setcounter{section}{0} \setcounter{equation}{0}
 \renewcommand{\thesection}{\Alph{section}.}
\renewcommand{\thesubsection}{\Alph{section}.\arabic{subsection}.}
\renewcommand{\theequation}{\thesection\arabic{equation}}}

\def\tyng(#1){\hbox{\tiny$\yng(#1)$}}

\newcommand{\be}{\begin{equation}}
	\newcommand{\ee}{\end{equation}}
\newcommand{\bea}{\begin{array}}
	\newcommand{\ea}{\end{array}}
\newcommand{\beqa}{\begin{eqnarray}}
	\newcommand{\eeqa}{\end{eqnarray}}
\newcommand{\nn}{\nonumber}

\DeclareCaptionSubType*{figure}

\DeclareCaptionLabelFormat{subalph}{(\alph{subfigure})}
\numberwithin{equation}{section}

\begin{document}
	\fontfamily{bch}\fontsize{11pt}{15pt}\selectfont
	\begin{titlepage}
		\begin{flushright}
		\end{flushright}
		\vspace{-2cm} 
		\begin{center}
			{\Large Real-Time Quantum Dynamics on the Fuzzy Sphere: Chaos and Entanglement}\\
			~\\
\centerline{\bf{S. Kürkcüoğlu}, \bf{B. Özcan}}			
\vskip 1em
\centerline{\sl Middle East Technical University, Department of Physics,}
\centerline{\sl Dumlupınar Boulevard, 06800, Ankara, Turkey}
\centerline{\sl}
\centerline{\sl }
\begin{tabular}{r l}
	E-mails:
	&\!\!\!{\fontfamily{cmtt}\fontsize{11pt}{15pt}\selectfont kseckin@metu.edu.tr  }\\
	&\!\!\!{\fontfamily{cmtt}\fontsize{11pt}{15pt}\selectfont berk@metu.edu.tr} 
\end{tabular}
		\end{center}
\begin{quote}
  \begin{center}
     {\bf Abstract}
  \end{center}
We study the real-time quantum dynamics of a matrix model consisting two bosonic fields on the fuzzy sphere $S_F^2$ using the Gaussian state approximation. Starting from the Hamiltonian formulation and using Wick's theorem, we derive a closed set of coupled nonlinear differential equations governing the time evolution of the one- and two-point correlation functions. The thermal equation of state is found by maximizing the von Neumann entropy over Gaussian states and solving the algebraic self-consistency equation(s) leading to a complete determination of the symplectic spectrum of the covariance matrix. We identify near-thermal initial conditions and use them to solve the equations of motion and employ our findings to probe chaos by calculating the largest Lyapunov exponent at various temperatures. Our results demonstrate that the latter tends to zero at a finite temperature indicating that the quantum dynamics respect the Maldacena–Shenker–Stanford bound across all temperatures, while approaching toward the classically chaotic regime at high temperatures. Finally, we examine the entanglement dynamics of the model in real-time by considering a sequence of bipartitions of the Hilbert space and computing the entanglement entropy and clearly exhibit the fast scrambling features that emerge in due detail.
\vskip 1em
\end{quote}

\vskip 1em

\begin{center}
    \itshape Dedicated to A. P. Balachandran, mentor, friend and an inspiring seeker of knowledge.
\end{center}

\end{titlepage}
\setcounter{footnote}{0}
\pagestyle{plain} \setcounter{page}{2}
\newpage

\section{Introduction and Summary of Results}
\label{sec:1}

Recently, there has been ample interest in exploring and quantifying chaos emerging from various matrix models \cite{Sekino:2008he,  Asplund:2011qj,  Shenker:2013pqa, Gur-Ari:2015rcq, Berenstein:2016zgj, Maldacena:2015waa, Maldacena:2016hyu, Aoki:2015uha, Asano:2015eha, Asplund:2015osa, Berkowitz:2016jlq, Berkowitz:2016znt, Buividovich:2017kfk, Buividovich:2018scl, Coskun:2018wmz, Baskan:2019qsb, Baskan:2022dys }.  Interest in this area is especially propelled by a result due to Maldacena-Shenker-Stanford (MSS) \cite{Maldacena:2015waa}, which briefly states that the largest Lyapunov exponent for quantum chaotic dynamics is controlled by a temperature dependent bound and is given by $\lambda_L \leq 2\pi T$. It is demonstrated that the 
the Sachdev-Ye-Kitaev (SYK) \cite{Sachdev:2010um, Maldacena:2016hyu}  model saturates this bound, while it is also expected to be so for the Banks-Fischler-Shenker-Susskind (BFSS) model \cite{Banks:1996vh}, it appears to be rather a formidable task to prove it. Classical chaotic dynamics of the BFSS model is studied in \cite{Gur-Ari:2015rcq} as an approximation to the quantum dynamics of the system in the high temperature regime where it is found that the largest Lyapunov exponent is given as $\lambda_L = 0.2924(3) (\lambda_{'t \, Hooft} T)^{1/4}$, meaning that the MSS bound is not obeyed only below a critical temperature, namely $T_c \approx 0.015$, while it remains parametrically smaller than $2 \pi T$ for $T > T_c$. Some other recent work on matrix models exploring aspects of chaotic dynamics and related issues within the classical regime are given in \cite{Iizuka:2013yla, Aoki:2015uha,Asano:2015eha,Asplund:2015osa,Coskun:2018wmz,Baskan:2019qsb,Baskan:2021aap,Amore:2024ihm}. Investigations of the BFSS  and related matrix models in the Euclidean signature dates back to early and mid 2000's \cite{Kawahara:2007fn,Delgadillo-Blando:2007mqd,Delgadillo-Blando:2008cuz} and more recently to \cite{Filev:2015hia, OConnor:2016gbq, Asano:2018nol}. These works have shown that the BFSS model and similar variants exhibit two distinct phases. For the BFSS model, one of these is described by essentially a large number of coincident $D0$-branes in the ’t Hooft limit, for which the system is holographically dual to a black-brane configuration whose low-temperature geometry can be obtained using gauge–gravity duality. The thermodynamic properties and phase structure of these models have been investigated both analytically and numerically using Euclidean time compactification and Monte-Carlo techniques and comprehensive expositions to these results can be found in \cite{Ydri:2017ncg,Ydri:2016dmy}. It is also important to mention the treatment of fuzzy scalar field as a multitrace matrix model \cite{OConnor:2007ibg} and generalized hermitian matrix model \cite{Nair:2011ux} in the spirit of \cite{Brezin:1977sv}. Nevertheless, it does not seem possible in any immediate way to quantitatively connect them to the chaotic dynamics these models are expected to exhibit. 

Thus a key challenge in more rigorously testing the MSS bound in BFSS and similar matrix models is the need for proper real-time techniques that may be used to explore quantum dynamics and chaos within such models. Recently, a promising new alternative approach based on an approximation using Gaussian states has been proposed and applied to the BFSS model \cite{Buividovich:2017kfk, Buividovich:2018scl} and related matrix models. This method allows for studying real-time quantum dynamics in a non-perturbative framework and is related to the self-consistent 2PI techniques in quantum field theory \cite{Berges:2004yj} as we will remark in section 2.4. and also used in quantum chemistry as well as for analyzing many-body systems \cite{Broeckhove, Heller}. Within this approximation the quantum state is represented by a Gaussian density matrix (or equivalently a Gaussian Wigner distribution) which allows for the entire dynamics to be expressed in terms of one- and two-point correlation functions, which, as is well-known, suffice to completely characterize Gaussian states \cite{Mukunda:1988cb, Simon:1994hgu, Bertlmann:2023wol, Weedbrook:2011wxo}. Higher-order correlation functions are expressed in terms of the latter via Wick contraction, which essentially amounts to reducing the infinite hierarchy of Schwinger–Dyson like equations stemming from the Heisenberg equations of motion for such models to a finite set of nonlinear differential equations that can be solved numerically. Applications of this approach to the BFSS model and its bosonic part have shown that the largest Lyapunov exponent satisfies the MSS bound $\lambda_L\le 2\pi T$ for all temperatures and vanishes at a low but a finite temperature. It does not, however by any means, allow to demonstrate that the MSS bound is saturated for the BFSS model since it does not capture the full quantum dynamics, nor does it preserve the supersymmetry of the latter. Nevertheless, it provides a powerful framework for taking the quantum effects into account, albeit within the restricted domain of Gaussian states for studying chaos in matrix models beyond the purely classical treatment. 
 
 In this work, we apply the Gaussian state approximation (GSA) to investigate the real-time quantum dynamics on the fuzzy-sphere $S_F^2$ \cite{Madore:1991bw} (see \cite{Balachandran:2005ew} for a review of the literature on $S_F^2$). In particular, we focus our attention to a bosonic two-matrix model on $S_F^2 \times \mathbb{R}$ where $\mathbb{R}$ stands for time. Starting from the Hamiltonian, we compute the Heisenberg equations of motion and using  Wick's theorem we obtain a truncated set of ordinary nonlinear coupled differential equations for the one- and two-point correlation functions in the most general Gaussian state. These equations govern the quantum dynamics of the model, while a key ingredient in extracting physically relevant solutions out of them in order to probe chaotic dynamics is the suitable selection of the initial conditions. This is also necessary to be able to link our results to what may be expected in the classical limit, i.e. at high temperatures.  Thus, firstly, a detailed statistical mechanics analysis is called for to obtain the thermalization conditions and the equation of state to describe the thermodynamic properties of the system. This is achieved by maximizing the von Neumann entropy for the most general Gaussian density matrix at fixed energy, which allows us to find and characterize the thermal equilibrium in terms of the symplectic eigenvalues of the covariance matrix of the Gaussian (i.e. the connected part of the 2-point functions). These are expressed in terms of characteristic frequencies $\omega_l$, whose form depends whether the interaction potential is $- \frac{\lambda}{2} \sum_{i,j} [\hat{X}_i,\hat{X}_j]^2$ which is $O(2)$-symmetic and involve both planar and non-planar contributions via the Wick contraction or of the form $\frac{\lambda}{2} Tr ({\hat X}_1  {\hat X}_1 {\hat X}_2 {\hat X}_2)$, which breaks the $O(2)$ symmetry to the dihedral group $D_4$ and lead to solely planar terms. In the latter case, we find that  $\omega_l := \sqrt{ \left (\mu^2 + \frac{l(l+1)}{R^2} + \frac{\lambda C}{N}\right)}$ involving the quantity $C$ defined through the equation \eqref{eq:C}.  We determine that $C$ satisfies a self-consistency equation, as derived in the text in \eqref{eq:consistency-equation} and solve it numerically  and obtain its value at  any given temperature $T$ and the other parameters, namely the mass $\mu$ of the scalar fields, $R$ the radius of $S_F^2$ and the matrix level, $N$, of the system. This avails us to unambiguously determine the symplectic eigenvalues $f_l$ of the covariance matrix  and hence the von Neumann entropy as well as other physically relevant quantities, such as the energy $E$ at a given temperature $T$ and the total coordinate dispersion $\frac{1}{N} \braket{ Tr \, {\hat X}_i \, {\hat X}_i}$ which takes the value $C$ at thermal equilibrium. For the $O(2)$-symmetric potential the same analysis leads to system of $N$ coupled algebraic equations on the covariances of the modes, which we have also successfully solved numerically. Nevertheless, as may be expected, thermal equilibrium conditions lead to a static solution of the truncated equations of motion, which is easily verified both analytically and numerically. To proceed, we determine configurations which are drawn randomly from classical Gaussian ensembles of suitable variances but are close enough to the thermal equation of state and hence dynamical so that the time evolution generated by the equations of motion with these initial data allows us to study the emerging chaotic behavior of the model. Energy as well as the total coordinate dispersion versus temperature plots of the thermal equilibrium and the latter {\it dynamical thermal equilibrium states} are graphically represented by the plots given in Fig.\ref{fig:E_CDvT} which clearly depict the thermal similarity of these configurations. Averaging over several (typically between seven and nine) such initial configurations we have determined the largest Lyapunov exponent $\lambda_L$ of the model at a given value of the parameters, $\mu, R$ in the 't Hooft limit, at the matrix level $N=5$ as function of the temperature. Our results, presented in section \ref{sec:DandC} clearly demonstrate the dependence of $\lambda_L$ over a wide range of temperatures and its decrease toward zero at a finite $T$ value, which clearly demonstrate the effect of quantum dynamics  becoming more attenuated at lower temperatures and driving the system toward a non-chaotic regime. Thus, the quantum dynamics of the model within GSA respects the MSS bound at all temperatures. At higher $T$ values, we find that both the classical and the quantum dynamics are significantly chaotic and similar in this respect. We also seize the opportunity to model how the $\lambda_L$ changes with curvature, by plotting its response to the radius $R$ at several different values of $T$. 
 
Finally we thoroughly examine the entanglement dynamics of our model by decomposing the system into two subsystems in a sequential manner. Although the GSA does not lead to unitary time evolution, it ensures that pure states evolve into pure states — a property that makes computing entanglement entropy feasible \cite{Buividovich:2018scl}. In the most general setting the latter is obtained via the von Neumann entropy of the reduced density matrix $\hat{\rho}_A = \mathrm{Tr}_B\,\hat{\rho}$, where the full Hilbert space admits a bipartite decomposition $\mathcal{H} = \mathcal{H}_A \otimes \mathcal{H}_B$. Taking into consideration that our system consists of $2N^2$ canonical modes, we introduce a sequence of bipartitions $\mathcal{H} = \mathcal{H}_{\leq L} \otimes \mathcal{H}_{>L}$, where $\mathcal{H}_{\leq L}$ contains the modes with $l \leq L$ and $\mathcal{H}_{>L}$ those with $l > L$. The entanglement entropy $S_{\leq L}$ is then computed from the symplectic eigenvalues of the reduced covariance matrix obtained by restricting that of the full system to the first $(L+1)^2$ modes. In section \ref{sec:ent} we  present our results for $N=5$, $\lambda = \frac{1}{5}$, across bipartite levels $L = 0, 1, 2, 3, 4$ for the planar and $L = 1, 2, 3, 4$ for the full interaction both of which demonstrate the fast scrambling features comprising of an initial sharp increase followed by a saturation toward a constant value expected from a chaotic quantum system as well as other related aspects in due detail.

\section{Basics}\label{sec:model}

In this section we describe the model on which we focus our attention in this paper. We start with the Hamiltonian description, obtain the Heisenberg equations of motion, introduce the Gaussian state approximation which will be used to obtain a truncated set of equations of motion approximating the quantum dynamics.

\subsection{The Model and the Equations of Motion}\label{ssec:hamiltonian}

We consider a two-matrix model on $S_F^2 \times \mathbb{R}$ where $S_F^2$ stands for the fuzzy $2$-sphere and $\mathbb{R}$ stands for time. Let us start with writing out the the Hamiltonian operator which can be expressed as:
\begin{equation}
\label{eq:hamiltonian}
\hat{H} = \frac{1}{2} Tr \left ( \hat{P}^2_i + \mu^2 \hat{X}_i^2- \frac{[L_a,\hat{X}_i]^2}{R^2} - \frac{\lambda}{2} \sum_{i,j} [\hat{X}_i,\hat{X}_j]^2 \right ).
\end{equation}
Here, $\hat{X}_i$ (with $i,j=1,2$) are $N\times N$ Hermitian matrices, $\hat{P}_i = \partial_t \hat{X}_i$ represent the conjugate momentum operators, which are also $N\times N$ Hermitian matrices: $\hat{X}_i^\dagger = \hat{X}_i$, $\hat{P}_i^\dagger = \hat{P}_i$, and they satisfy the canonical commutation relations in the standard quantum mechanical sense, as will be written explicitly below in an appropriate basis. $L_a$, $(a = 1,2,3)$ are the $N\times N$ Hermitian generators of $SU(2)$ in the spin $\frac{N-1}{2}$ irreducible representation satisfying the $SU(2)$ algebra commutation relations: $[L_a,L_b] = i\epsilon_{abc}L_c$. In above, $\lbrack L_a, \cdot \rbrack$ are the derivations on $S_F^2$ as usual. We may also note that $R$ is the radius of the fuzzy sphere, $\mu >0$ is the mass of the scalar fields, $\hat{X}_i$, and $\lambda>0$ stands for the coupling constant. The matrix fields $\hat{X}_i$ and their conjugate momenta $\hat{P}_i$ are functions of the time, i.e. their matrix entries are time-dependent. 

We see that this Hamiltonian has a global $SU(2)$ symmetry which is implemented by the adjoint action of $U \in SU(2)$: $\hat{X}_i \rightarrow U^\dagger \hat{X}_i  U$, $\hat{P}_i \rightarrow U^\dagger \hat{P}_i  U$, $\hat{L}_a \rightarrow U^\dagger \hat{L}_a  U$. It also has a rigid $O(2)$ symmetry rotating the matrices $\hat{X}_i$ among themselves: ${\hat X}_i \rightarrow R_{ij} {\hat X}_j$, where $R \in O(2)$. In what follow, we will also consider the model with only the planar interaction potential $\frac{\lambda}{2} Tr ({\hat X}_1  {\hat X}_1 {\hat X}_2 {\hat X}_2)$ which breaks the $O(2)$ symmetry to the dihedral group $D_4$.

It is convenient and also proves to be more suitable for the ensuing numerical analysis of the equations of motion to work with the real polarization operator, which we denote as $Z_{lm}$, with $l = 0, 1, ..., N-1$, $-l \leq m \leq l $, which are $N \times N$ Hermitian matrices, $Z_{lm}^\dagger = Z_{lm}$, carrying the spin-$l$ UIRR of $SU(2)$. They are complex linear combinations of the more familiar complex polarization operators $T_{lm}$ satisfying $T_{lm}^\dagger = (-1)^m T_{l -m}$ and like the latter form a basis for the $N \times N$ matrices. Definition and some useful properties of real polarization operator relevant for our purposes are given in the Appendix~\ref{sec:polarization-operator-basis}. We can expand the fields $\hat{X}_i$ and their conjugate momenta $\hat{P}_i$ in this basis as :
\be
\label{eq:2:1}
{\hat X}_i = \sum_{lm} {\hat X}_i^{lm} Z_{lm}, \quad {\hat P}_i =  \sum_{lm} {\hat P}_i^{lm} Z_{lm} \,.
\ee
Taking the trace, we are able to write the Hamiltonian in \eqref{eq:hamiltonian} as
\begin{align}
 \hat{H} &= \frac{1}{2} \left(\hat{P}_i^{lm} \hat{P}_i^{lm} + \left(\mu^2 + \frac{l(l+1)}{R^2} \right)  \left(\hat{X}_i^{lm} \hat{X}_i^{lm} \right) + \lambda \mathcal{H}^{l_1 l_2 l_3 l_4}_{m_1 m_2 m_3 m_4}  {\hat X}_1^{l_1 m_1}  {\hat X}_1^{l_2 m_2}  {\hat X}_2^{l_3 m_3}  {\hat X}_2^{l_4 m_4} \right) \,,
\end{align}
where we have used the fact that $Tr\left (Z_{lm} Z_{l'm'} \right) = \delta_{ll'} \delta_{mm'}$ (see Appendix~\ref{sec:polarization-operator-basis}) and defined
\begin{align}
\label{eq:H-symbol-definition}
\mathcal{H}^{l_1 l_2 l_3 l_4}_{m_1 m_2 m_3 m_4} &= \mathcal{K}^{l_1 l_2 l_3 l_4}_{m_1 m_2 m_3 m_4} - \mathcal{K}^{l_1 l_3 l_2 l_4}_{m_1 m_3 m_2 m_4},
\end{align}
with $\mathcal{K}^{l_1 l_2 l_3 l_4}_{m_1 m_2 m_3 m_4} := Tr \left(Z_{l_1 m_1} Z_{l_2 m_2} Z_{l_3 m_3} Z_{l_4 m_4} \right)$. Sum over the repeated indices $l_a$ and $m_a$ will be implicitly assumed from now on. Explicit expressions for computing the traces $\mathcal{K}^{l_1 l_2 l_3 l_4}_{m_1 m_2 m_3 m_4} $ in terms of the Wigner $6j$-symbols and the Clebsch-Gordan coefficients are given in the Appendix ~\ref{sec:polarization-operator-basis}.

In what follows, we are going to be essentially working in the 't Hooft limit, which is readily obtained by taking $N \rightarrow \infty$ and $\lambda \rightarrow 0$ while keeping $\lambda_{'t Hooft} = \lambda N $ fixed. Letting $\hat{X}_i^{lm} \rightarrow \lambda_t^{-1/6} \hat{X}_i^{lm}$, $\hat{P}_i^{lm} \rightarrow \lambda_t^{1/6} \hat{X}_i^{lm}$, $t \rightarrow \lambda_t^{-1/3} t$, $R \rightarrow \lambda_t^{-1/3} R $, $\mu \rightarrow \lambda_t^{1/3} \mu$, we can scale $\lambda_{'t Hooft}$ to $1$. Thus, practically 't Hooft limit is attained by simply letting $\lambda \rightarrow \frac{1}{N}$. In the analytical developments we will keep the coupling $\lambda$ for generality and set it to $\frac{1}{N}$ in numerical calculations.\footnote{Note that, since our model has only $SU(2)$ rather than an $SU(N)$ symmetry (as is usually the case in flat models like the BFSS), non-planar contributions will not be suppressed by $\frac{1}{N^2}$ as we will concretely see later. Nevertheless, it proves convenient to use the coupling strength as noted above to probe the chaotic dynamics.\label{fn:largeN}}

\subsection{Heisenberg Equations of Motion}\label{ssec:gaussian-state-approximation-model}

In the basis \eqref{eq:2:1} the canonical commutation relations between the fields and their conjugate momenta take the form
\begin{align}
[ {\hat X}_i^{l_1 m_1},  {\hat P}_j^{l_2 m_2}] = i\delta^{l_1 l_2} \delta^{m_1 m_2} \delta_{ij}.
\label{eq:cancom}
\end{align}
Thus, we can view our system to be composed of $2N^2$ canonical modes labeled with $l=0,1,\cdots\, N-1$, $|m| \leq l$, $i=1,2$. In the non-interacting limit $\lambda \rightarrow 0$ there are $2N^2$ harmonic oscillator with $l$-dependent frequencies $\omega_l^{(0)}= \sqrt{\mu^2 +\frac{l (l+1)}{R^2}}$, $l^{th}$ mode for each field being $2l+1$-fold degenerate. For future reference, let us also note that the zero mode is of the form ${\hat H}^{00}:= \frac{1}{2} (\hat{P}_i^{00})^2 + \frac{\mu^2}{2} (\hat{X}_i^{00})^2$, since $Z_{00} \propto \mathbb{I}_N$ commutes with all $Z_{lm}$ and hence decouples from the rest and will not have any effect on the dynamics of the $O(2)$ symmetric model. This does not happen for the model with the $D_4$-symmetric interaction potential $\frac{\lambda}{2} Tr ({\hat X}_1  {\hat X}_1 {\hat X}_2 {\hat X}_2)$ and therefore we keep the $l=0$ term in what follows.

Heisenberg equations of motion for $ {\hat X}_i^{l m}$ and $ {\hat P}_i^{l m}$ are
\begin{align}
\partial_t {\hat X}_i^{lm} = i [H,  {\hat X}_i^{lm}] \,, \quad 	 \partial_t{\hat P}_i^{lm} = i [ {\hat H},  {\hat P}_i^{lm}] \,.
\end{align}
Using the cyclic property of the trace and after some algebra, it can be shown that they take the form
\begin{align}
\label{eq:opeqs1}
	\partial_t {\hat X}_{1}^{lm} &=  {\hat P}_{1}^{l m},  \nn \\
	\partial_t {\hat X}_{2}^{lm} &= {\hat P} _{2}^{l m},   \\
	\partial_t {\hat P}_1^{lm} &= -\left[\left(\mu^2 + \frac{l(l+1)}{R^2}\right)  {\hat X}_1^{l m} + \frac{\lambda}{2} \mathcal{N}^{l l_1 l_2 l_3}_{m m_1 m_2 m_3} {\hat X}_1^{l_1 m_1} {\hat X}_2^{l_2 m_2}{\hat X}_2^{l_3 m_3}\right], \nn \\
	\partial_t {\hat P}_2^{lm} &= -\left[\left(\mu^2 + \frac{l(l+1)}{R^2}\right) {\hat X}_2^{l m} + \frac{\lambda}{2} \mathcal{N}^{l l_1 l_2 l_3}_{m m_1 m_2 m_3} {\hat X}_2^{l_1 m_1} {\hat X}_1^{l_2 m_2} {\hat X}_1^{l_3 m_3}\right]. \nn
\end{align}
where we have defined 
\begin{align}
	\mathcal{N}^{l_1 l_2 l_3 l_4}_{m_1 m_2 m_3 m_4} := \mathcal{K}^{l_1 l_2 l_3 l_4}_{m_1 m_2 m_3 m_4} + \mathcal{K}^{l_2 l_1 l_3 l_4}_{m_2 m_1 m_3 m_4} - \mathcal{K}^{l_1 l_3 l_2 l_4}_{m_1 m_3 m_2 m_4} - \mathcal{K}^{l_2 l_3 l_1 l_4}_{m_2 m_3 m_1 m_4},
\end{align}
to cast the equations in a relatively compact form. We have in total $2\times 2 N^2 = 4 N^2$ coupled first-order non-linear differential equations for $\hat{X}_i^{lm}$, $\hat{P}_i^{lm}$, which can not be solved in general in any practical sense. Thus, we consider an approximation which goes beyond the classical regime and captures features of the quantum dynamics as we describe next.

\subsection{Gaussian State Approximation}\label{ssec:gaussian-state-approximation}

In this paper, we use the Gaussian state approximation method to investigate the real time evolution of the system described by the Hamiltonian \eqref{eq:hamiltonian} and the operator equations of motion \eqref{eq:opeqs1}. The essence of this method relies on introducing the most general Gaussian state, which is fully characterized by the expectation values of  $\hat{X}_i^{lm}$, $\hat{P}_i^{lm}$ and their variances in this state, i.e. in terms of the $1$-point and $2$-point functions \cite{Mukunda:1988cb, Simon:1994hgu, Buividovich:2018scl}. Evaluating the expectation values of the Heisenberg equations motion in the most general Gaussian state, Wick's theorem ensures that a truncated set of coupled first order differential equations emerge for the $1$-point and $2$-point functions. 

To be more concrete, let us introduce a compact notation by writing the column vector with operator entities as $\hat{\xi}:= (\hat{X}_1^{lm}, \hat{X}_2^{lm}, \hat{P}_1^{lm}, \hat{P}_2^{lm})^T$, where $T$ stands for the matrix transpose. Correspondingly $\xi := (X_1^{lm}, X_2^{lm}, P_1^{lm}, P_2^{lm})^T$ is the column vector filled with the eigenvalues of  $\hat{X}_i^{lm}$, $\hat{P}_i^{lm}$, spanning the $4 N^2$-dimensional canonical phase space, with the usual symplectic form $\Omega$, whose explicit form will be given as a block matrix in what follows. In general, given a state described by a density matrix $\hat{\rho}$, the expectation value of operators $\hat{\mathcal O}$ are given in their usual form $\langle \hat{\mathcal O} \rangle := tr (\hat{\rho} \, \hat{\mathcal O})$. Using the Wigner function\footnote{In a notation adapted to our present setting and suppressing all the indices we may give the definition as \cite{Bertlmann:2023wol}
\begin{equation}
\mathcal{W}_{\hat{\rho}}(X, P) = \frac{1}{(2\pi)^{2N^2}} \int d Y e^{-i P^T Y}  \Bigg \langle X + \frac{Y}{2} \Bigg | {\hat{\rho}} \Bigg | X- \frac{Y}{2} \Bigg \rangle \,. \nn
\end{equation}
}
$\mathcal{W}_{\hat{\rho}}(\xi)$ associated to the density matrix $\hat{\rho}$, the expectation values can be expressed as  $\langle \hat{\mathcal O} \rangle := \int dV_\xi \mathcal{W}_{\hat{\rho}}(\xi) \mathcal{W}_{\hat{\mathcal O}}(\xi)$, where $dV_\xi$ stands for the volume element for the $4 N^2$-dimensional phase space.

The Wigner function $\mathcal{W}_{\hat{\rho}}(\xi)$ of the most general Gaussian density matrix $\hat{\rho}$ is \cite{Mukunda:1988cb, Simon:1994hgu, Bertlmann:2023wol}
\begin{align}
\mathcal{W}_{\hat{\rho}}(\xi) = \mathcal{N} e^{- \frac{1}{2} (\xi - \bar{\xi})^T \Sigma^{-1} (\xi - \bar{\xi}) }\,.
\end{align}
Here $\bar{\xi}_a := \braket{\hat{\xi}_a}$ are the $1$-point functions which are constituting the center of the Gaussian and 
\begin{align}
\Sigma_{ab} := \braket{\braket{\hat{\xi}_a \hat{\xi}_b}} = \frac{1}{2} \braket{\hat{\xi}_a \hat{\xi}_b + \hat{\xi}_b \hat{\xi}_a} - \braket{\hat{\xi}_a} \braket{\hat{\xi}_b} \,, 
\end{align}
is the covariance matrix, i.e. the $2$-point functions, which is arranged into a block matrix where $a, b$ are  the collective indices over $i,j$ and $l,m$ as appropriate. This notation is quite standard in the literature \cite{Mukunda:1988cb, Simon:1994hgu, Berges:2017hne,  Bertlmann:2023wol} and adapted to the index structure of the present problem.  We also note that $\mathcal{N} \sim \sqrt{\det(\Sigma^{-1})}$ is the normalization of the Gaussian Wigner function $\mathcal{W}_{\hat{\rho}}(\xi)$, which ensures that $\int dV_\xi \rho (\xi) =1$ and is equivalent to the condition\footnote{We may note that here the $tr$ operation is standing for the trace in the canonical Hilbert space over which $\hat{\xi}$ act, and should be distinguished from $Tr$, which denotes the trace over the $N \times N$-matrices.} $tr \hat{\rho} =1$.

We can concisely give the Wick's theorem in the form:
\begin{align}
&\langle (\hat{\xi}_{a_1} - \bar{\xi}_{a_1}) (\hat{\xi}_{a_2} - \bar{\xi}_{a_2}) \cdots (\hat{\xi}_{a_{2n}} - \bar{\xi}_{a_{2n}}) \rangle_s = \sum_{Pairings} \prod_{i,j} \langle (\hat{\xi}_{a_i} - \bar{\xi}_{a_i}) (\hat{\xi}_{a_j} - \bar{\xi}_{a_j})\rangle_s  \,, \nn \\
&\langle (\hat{\xi}_{a_1} - \bar{\xi}_{a_1}) (\hat{\xi}_{a_2} - \bar{\xi}_{a_2}) \cdots (\hat{\xi}_{a_{2n+1}} - \bar{\xi}_{a_{2n+1}}) \rangle_s = 0 \,.
\label{eq:Wick1}
\end{align}
where the subscript $s$ stands for symmetrization, which is relevant whenever $\hat{\xi}_{a_l}$ and $\hat{\xi}_{a_k}$ do not commute. Evidently, this occurs if one of them is $\hat{X}_i^{lm}$ and the other is $\hat{P}_i^{lm}$. In the Gaussian state the Hamiltonian reads
\begin{equation}
	\label{eq:averaged-hamiltonian_short}
	\braket{\hat{H}} = \frac{1}{2} \left (\braket{{\hat P}_i^{lm} {\hat P}_i^{lm}} + \left (\mu^2 + \frac{l(l+1)}{R^2} \right) \braket{{\hat X}_i^{lm} {\hat X} _i^{lm}} + \lambda \mathcal{H}^{l_1 l_2 l_3 l_4}_{m_1 m_2 m_3 m_4} \braket{{\hat X}_1^{l_1 m_1} {\hat X}_1^{l_2 m_2} {\hat X}_2^{l_3 m_3} {\hat X}_2^{l_4 m_4}} \right),
\end{equation}
whose form in terms of the $1$- and $2$-point functions after applying Wick's theorem \eqref{eq:Wick1} is relegated to the Appendix~\ref{sec:eoms}

Equations of motion for the $1$-point functions are obtained as 
\begin{align}
	\label{eq:eom-one-point}
	\partial_t\braket{{\hat X}_1^{l_1 m_1}}&=\braket{{\hat P}_1^{l_1 m_1}}, \nn \\
	\partial_t\braket{{\hat X}_2^{l_1 m_1}}&=\braket{{\hat P}_2^{l_1 m_1}}\,, 
\end{align}
\begin{align}
\label{eq:eom-one-pointp1}
\partial_t\braket{{\hat P}_1^{l_1 m_1}} & =-\Bigg\{\left(\mu^2+\frac{l_1\left(l_1+1\right)}{R^2}\right) \braket{\hat{X}_1^{l_1 m_1}} +\frac{\lambda}{2}\mathcal{N}^{l_1 l_3 l_4 l_5}_{m_1 m_3 m_4 m_5} 
\Big(\braket{{\hat X}_1^{l_3 m_3}}\braket{{\hat X}_2^{l_4 m_4}}\braket{{\hat X}_2^{l_5 m_5}} \nn \\ 
& +\braket{{\hat X}_2^{l_5 m_5}}\braket{\braket{{\hat X}_1^{l_3 m_3} {\hat X}_2^{l_4 m_4}}} +\braket{{\hat X}_1^{l_3 m_3}}\braket{\braket{{\hat X}_2^{l_4 m_4} {\hat X}_2^{l_5 m_5}}}+\braket{{\hat X}_2^{l_4 m_4}}\braket{\braket{{\hat X}_1^{l_3 m_3} {\hat X}_2^{l_5 m_5}}} \Big)\Bigg\}, 
\end{align}
\begin{align}	
		\label{eq:p2}
\partial_t\braket{{\hat P}_2^{l_1 m_1}}&=-\Bigg\{\left(\mu^2+\frac{l_1\left(l_1+1\right)}{R^2}\right) \braket{{\hat X}_2^{l_1 m_1}} +\frac{\lambda}{2}\mathcal{N}^{l_1 l_3 l_4 l_5}_{m_1 m_3 m_4 m_5}  
	\Big(\braket{{\hat X}_2^{l_3 m_3}}\braket{{\hat X}_1^{l_4 m_4}}\braket{{\hat X}_1^{l_5 m_5}} \nn \\
	& +\braket{{\hat X}_1^{l_5 m_5}}\braket{\braket{{\hat X}_2^{l_3 m_3} {\hat X}_1^{l_4 m_4}}} +\braket{{\hat X}_2^{l_3 m_3}}\braket{\braket{{\hat X}_1^{l_4 m_4} {\hat X}_1^{l_5 m_5}}}+\braket{{\hat X}_1^{l_4 m_4}}\braket{\braket{{\hat X}_2^{l_3 m_3} {\hat X}_1^{l_5 m_5}}} \Big)\Bigg\} \,.
\end{align}
As for the variances $\braket{\braket{{\hat X}_i^{l_1 m_1} {\hat X}_j^{l_2 m_2}}}$, $\braket{\braket{{\hat X}_i^{l_1 m_1} {\hat P}_j^{l_2 m_2}}}$ and $\braket{\braket{{\hat P}_i^{l_1 m_1} {\hat P}_j^{l_2 m_2}}}$, their equations  of motion are calculated in the same manner using the Wick's theorem. Since they are somewhat lengthy, their explicit form is presented in the Appendix~\ref{sec:eoms} The main point here is that  \eqref{eq:eom-one-point}-\eqref{eq:p2}, \eqref{eq:eom-two-point-xx}-\eqref{eq:eom-two-point-pp-22} together form a truncated set of first order coupled non-linear differential equations and govern the quantum dynamics of the system within the Gaussian state approximation. 

\subsection{Thermal Equilibrium and the Equation of State}

Although, it is possible to look for the solutions of the truncated equations of motion with an arbitrary set of initial conditions, proceeding in such a manner will not allow us in general to connect the resulting dynamics to the thermal properties of the system, since ultimately we would like to extract how the largest Lyapunov exponent changes with temperature. Thus, it is important  for us to determine the thermal equilibrium conditions and consider thermal or in fact nearly thermal states\footnote{In the next subsection, we will see that the thermal states are static solutions to the equations of motion, and hence nearly thermal initial states are sought after and identified to probe non-trivial dynamics.}as possible initial data for the equations of motion. We may first recall that the standard considerations in statistical mechanics imply that  a system becomes thermalized once the entropy is maximized while keeping the energy fixed. Thus, we may determine the thermal equilibrium configuration of our system using this basic principle. It is a well-known fact that for a Gaussian density matrix  $\hat{\rho}$, the von Neumann entropy $S = -Tr (\hat{\rho} \ln \hat{\rho})$ can be expressed in terms of the symplectic eigenvalues of the covariance matrix $\Sigma$ \cite{Saravani:2013nwa,Sorkin:2012sn,Bombelli:1986rw,Berges:2017hne, Jones:2026vkw}. If the symplectic form of the canonical coordinates $\xi$ is denoted by $\Omega$, Williamson's theorem \cite{Williamson, Bertlmann:2023wol} ensures that the eigenvalues of the product matrix $\Sigma \Omega$ come in complex pairs $\pm i f_k$, where the real positive numbers, $f_k$, are called the symplectic eigenvalues of $\Sigma$. The von Neumann entropy can be expressed in terms of the symplectic eigenvalues as
\begin{align}
S = \sum_k \left(f_k + \frac{1}{2}\right)\ln\left(f_k + \frac{1}{2}\right) - \left(f_k - \frac{1}{2}\right)\ln\left(f_k - \frac{1}{2}\right) \,.
\end{align}
Thus, $\Sigma$ describes a pure Gaussian state if and only if $f_k= 1/2$ $\forall\,k$, in which case the von Neumann entropy vanishes as expected. In the present model, we  may express the covariance matrix $\Sigma$ and the symplectic form as block matrices
\begin{align}
	\Sigma_{i j}^{[l m][l' m']} = \begin{pmatrix}
	\langle\langle {\hat X}_{i}^{l m} {\hat X}_{j}^{l' m'} \rangle\rangle & \langle\langle {\hat X}_{i}^{l m} {\hat P}_{j}^{l' m'} \rangle\rangle \\[6pt]
	\langle\langle {\hat P}_{i}^{l m} {\hat X}_{j}^{l' m'} \rangle\rangle & \langle\langle {\hat P}_{i}^{l m} {\hat P}_{j}^{l' m'} \rangle\rangle
	\end{pmatrix}\,, \quad 
	\Omega_{i j}^{[l m][l' m']} =
	\begin{pmatrix}
	0 & \delta^{l l'}  \delta^{m m'} \delta_{i j}\\[6pt]
	-\delta^{l l'}  \delta^{m m'} \delta_{i j} & 0
	\end{pmatrix}.
\end{align}
Both of these matrices are $2N_{\text{Tot}} \times 2N_{\text{Tot}} = 4N^{2} \times 4N^{2}$ dimensional with each block of size $2N^2 \times 2 N^2$. 

\begin{figure}[htbp]
	\centering
	\includegraphics[width=\linewidth]{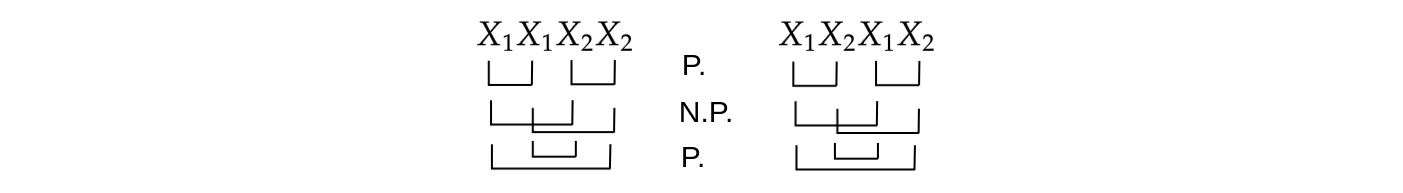}
	\caption{Planar and Non-Planar Contractions}\label{fig:E_T_1_Eq}
	\label{fig:contractions}
\end{figure}

The thermal equilibrium state is expected to share the symmetries of the Hamiltonian unless the symmetries are spontaneously broken.  For our model \eqref{eq:hamiltonian} (with $\mu^2 > 0$) each term of the potential $\hat{V} := \frac{1}{2}Tr \left (\mu^2 \hat{X}_i^2 - \frac{\lambda}{2} \sum_{i,j} [\hat{X}_i,\hat{X}_j]^2 \right)$ is separately positive definite (since $\hat{X}_i$ are Hermitian) and therefore its minimum value is zero, implying that each term has to vanish separately at the minimum, which is only possible with vanishing vacuum expectation value (VEV) for the fields ${\hat X}_{i}^{l m}$. Thus, both the $SU(2)$ and the $O(2)$ symmetries separately indicate that 
\begin{align}
\label{eq:initial-cond-one-point}
\braket{{\hat X}_{i}^{l m}} = 0, \quad \braket{{\hat P}_{i}^{l m}} = 0 \,,
\end{align}
while the latter is also readily implied by the time reversal symmetry.\footnote{For tachyonic mass $\mu^2 <0$, vanishing VEVs for ${\hat X}_{i}^{l m}$ is still an extremum of the potential but it is a local maximum rather than a minimum as the second derivative of the potential is negative definite at this value of the VEVs. In fact, for configurations such as the diagonal matrices or vanishing of one of the matrices $V \rightarrow  \frac{1}{2} \mu^2 \sum (\lambda_i^a)^2 < 0$ ($\lambda_i^a$ standing for the eigenvalues of $\hat{X}_i$) indicating that $V \rightarrow -\infty$ as $\lambda_a \rightarrow \pm \infty$ and hence that the potential is not bounded from below for $\mu^2 <0$. 
	
Let us also note that for a model with three matrices, in addition to the quadratic and quartic terms a cubic term is also possible and this can lead to non-trivial fuzzy sphere type vacua for $\mu^2 >0$. Indeed, we may note that the maximally supersymmetric massive deformation of the BFSS, namely the Berenstein-Maldacena-Nastase (BMN) model with $SU(N)$ gauge and $SO(3) \times SO(6)$ global symmetry is known to have fuzzy sphere vacua which spontaneously breaks the $SU(N)$ gauge symmetry either completely or to a product of $U(n_k)$'s depending on whether the vacua is a single fuzzy sphere at level $N$ or a direct sum of $k$ fuzzy spheres at levels $n_k$, respectively \cite{Berenstein:2002jq, Dasgupta:2002hx,Dasgupta:2002ru}. Quantum dynamics of the bosonic part of this model is currently under investigation using the Gaussian state approximation \cite{Soon}.} Due to the $SU(2)$ symmetry the $2$-point functions have the general  structure 
\begin{align}
\label{eq:initial-cond-two-point}
\braket{\braket{{\hat X}^{l_1 m_1}_i {\hat X}^{l_2 m_2}_j}} = \sigma_{xx}(l_1)  \delta_{ij}  \delta^{l_1 l_2} \delta^{m_1 m_2} \,, \quad
\braket{\braket{{\hat P}^{l_1 m_1}_i {\hat P}^{l_2 m_2}_j}} = \sigma_{pp}(l_1)  \delta_{ij}  \delta^{l_1 l_2} \delta^{m_1 m_2} \,, \quad
\end{align}
and $\braket{\braket{{\hat X}^{l_1 m_1}_i {\hat P}^{l_2 m_2}_j}} = 0$ as a consequence of the time reversal symmetry. Thus the covariance matrix has the structure
\begin{align}
	\Sigma_{i j}^{[l m][l' m']} = \begin{pmatrix}
		\sigma_{xx}(l) \delta^{l l'} \delta^{m m'} \delta_{ij} & 0 \\[6pt]
		0 & \sigma_{pp}(l) \delta^{l l'} \delta^{m m'} \delta_{ij}
	\end{pmatrix}.
\end{align}
It can be easily seen that the $\Sigma \Omega$ matrix has $4N^2$ eigenvalues $\pm i f_k$ with $k$ ($k = 1, ..., 2N^2$) being the collective index running over $i$, $l$ and $m$. Due to the $SU(2)$ and $O(2)$ symmetries, the eigenvalues depend only on $l$, they are $2l+1$-fold degenerate at each value of $l$ and can be written as
\begin{align}
f_l = \sqrt{\sigma_{xx}(l)\sigma_{pp}(l)},
\label{eq:fl}
\end{align}
In view of these considerations performing the sums over the indices $i$ and $m$ bring an overall factor of two and a factor of $2l+1$ under the summation symbol, respectively. 
The von Neumann entropy takes the form 
\be
S = 2 \sum_{l=0}^{N-1} (2l+1) \left[(f_l + \frac{1}{2}) \ln(f_l + \frac{1}{2}) - (f_l - \frac{1}{2}) \ln(f_l - \frac{1}{2})\right] \,.
\label{eq:entropy1}
\ee
Using \eqref{eq:initial-cond-one-point} and \eqref{eq:initial-cond-two-point}, we may write $\braket{\hat H}$ in \eqref{eq:averaged-hamiltonian_short} as 
\begin{align}
\label{eq:averaged-hamiltonian-ic}
\braket{\hat H} &= \sum_l (2l+1) \left ( \sigma_{pp}(l) + \left(\mu^2 + \frac{l(l+1)}{R^2}\right) \sigma_{xx}(l)  \right ) \nonumber \\ 
                &\;\;\;\;\;\;\;\;\;\;\;\;\;\;\;\;\;\;\;\;\;\;\;\;\;\;\;\;\;\;\;\;\;\;\;\;\;\;\;\;\;\;\;\;\;\;\;\;\;\;\;\;\;\;\;\;\;\;+ \frac{\lambda}{2} \mathcal{H}^{l_1 l_2 l_3 l_4}_{m_1 m_2 m_3 m_4}  \braket{{\hat X}_1^{l_1 m_1} {\hat X}_1^{l_2 m_2} {\hat X}_2^{l_3 m_3} {\hat X}_2^{l_4 m_4}} \,, \nn \\
&= \sum_l (2l+1) \left ( \sigma_{pp}(l) + \left(\mu^2 + \frac{l(l+1)}{R^2}\right) \sigma_{xx}(l)  \right ) + \frac{\lambda}{2} \mathcal{H}^{l_1 l_1 l_2 l_2}_{m_1 m_1 m_2 m_2} \sigma_{xx}(l_1) \sigma_{xx}(l_2) \,.
\end{align}
It is worthwhile to inspect the interaction term in more detail. From the first line of  \eqref{eq:averaged-hamiltonian-ic}, we have 
\begin{align}
  \mathcal{H}^{l_1 l_1 l_3 l_4}_{m_1 m_2 m_3 m_4}  \braket{{\hat X}_1^{l_1 m_1} {\hat X}_1^{l_2 m_2} {\hat X}_2^{l_3 m_3} {\hat X}_2^{l_4 m_4}} = &\mathcal{K}^{l_1 l_2 l_3 l_4}_{m_1 m_2 m_3 m_4} \braket{{\hat X}_1^{l_1 m_1} {\hat X}_1^{l_2 m_2} {\hat X}_2^{l_3 m_3} {\hat X}_2^{l_4 m_4}}\nonumber \\ 
                                                                                                                                                    &-  \mathcal{K}^{l_1 l_3 l_2 l_4}_{m_1 m_3 m_2 m_4} \braket{{\hat X}_1^{l_1 m_1} {\hat X}_1^{l_2 m_2} {\hat X}_2^{l_3 m_3} {\hat X}_2^{l_4 m_4}} \,.
\label{intpot1}
\end{align}
The last expression stems from $\frac{\lambda}{2} \left ( Tr ({\hat X}_1  {\hat X}_1 {\hat X}_2 {\hat X}_2)-  Tr( {\hat X}_1  {\hat X}_2 {\hat X}_1 {\hat X}_2) \right )$ after performing the traces and evaluating it in the Gaussian state. The possible Wick contractions are indicated in the diagram in Fig. \ref{fig:contractions}, where the planar and non-planar contributions are denoted by the abbreviations $P.$ and $N.P.$, respectively. Using $\braket{\braket{{\hat X}^{l_1 m_1}_i {\hat X}^{l_2 m_2}_j}} = 0$ for $i \neq j$, we see that the non-vanishing planar and non-planar contributions come only from the first and the second term of this expression (or equivalently of \eqref{intpot1}), respectively.

\vskip 1em

{\it Planar Interaction:}

\vskip 1em

We first consider the model with the planar interaction potential $\frac{\lambda}{2} Tr ({\hat X}_1  {\hat X}_1 {\hat X}_2 {\hat X}_2)$ as it is analytically more transparent and paves the way for the treatment of the model with the non-planar term included as we will see later on. We observe that this model breaks the $O(2)$ symmetry. Nevertheless,  it is broken only to a discrete dihedral subgroup $D_4$ generated by $({\hat X}_1, {\hat X}_2) \rightarrow (\pm {\hat X}_1,  \pm (\mp) {\hat X}_2), \; ({\hat X}_1, {\hat X}_2) \rightarrow (\pm {\hat X}_2, \pm (\mp) {\hat X}_1)$, which also continues to imply \eqref{eq:initial-cond-one-point}. We have
\begin{align}
& \mathcal{K}^{l_1 l_2 l_3 l_4}_{m_1 m_2 m_3 m_4} \braket{{\hat X}_1^{l_1 m_1} {\hat X}_1^{l_2 m_2} {\hat X}_2^{l_3 m_3} {\hat X}_2^{l_4 m_4}} = \mathcal{K}^{l_1 l_1 l_2 l_2}_{m_1 m_1 m_2 m_2} \sigma_{xx}(l_1) \sigma_{xx}(l_2)\nonumber \\
				&=\sum_{l_1 m_1}\sum_{l_2 m_2} \sigma_{xx}(l_1)\sigma_{xx}(l_2)\, Tr \left( Z_{l_1 m_1}  Z_{l_1 m_1} Z_{l_2 m_2} Z_{l_2 m_2} \right) \nonumber \\
				&=\sum_{l_1}\sum_{l_2}(2l_1 + 1)(2l_2 + 1)\sigma_{xx}(l_1) \, \sigma_{xx}(l_2) \, Tr \left(\frac{\mathbb{I}_N}{N^2}\right) \nonumber \\
				&=\frac{1}{N} \sum_{l_1}\sum_{l_2}(2l_1 + 1)(2l_2 + 1)\sigma_{xx}(l_1)\sigma_{xx}(l_2) \nonumber \\ 
				&=\frac{1}{N} \left(\sum_{l=0}^{N-1}(2l + 1)\sigma_{xx}(l)\right)^2 \,,
				\label{eq:planarint}
\end{align} 
where we have used $\sum_{m} Z_{lm} Z_{lm} = \frac{(2l+1)}{N} \mathbb{I}_N $ (see Appendix~\ref{sec:polarization-operator-basis} for a quick proof.)

Defining 
\be
C:= \sum_{l = 0}^{N-1} (2l+1)\sigma_{xx}(l) \,,
\label{eq:C}
\ee
we may write the energy as 
\begin{equation}
\label{eq:el-equation}
E: = \braket{H} = \sum_{l}(2l+1)\left[\sigma_{pp}(l) + \left(\mu^2 + \frac{l(l+1)}{R^2}\right) \sigma_{xx}(l) \right]+ \frac{\lambda C^2}{2N} .
\end{equation}
We would like to note that the quantity $C$ defined in \eqref{eq:C} is clearly independent of $l$ but depends on $N$. It will play a crucial role in our analysis  as it will become abundantly clear from the ensuing discussions. In particular, we may define the total coordinate dispersion as
\be
\frac{1}{N} \braket{ Tr \, {\hat X}_i \, {\hat X}_i} := \frac{1}{N} \left ( \braket{\braket {{\hat X}_i^{lm} {\hat X}_i^{lm}}} + \braket{{\hat X}_i^{lm}} \braket{{\hat X}_i^{lm}} \right ) \,. 
\label{eq:totalcoorddis}
\ee
We easily find that $\frac{1}{N} \braket{ Tr \, {\hat X}_i {\hat X}_i} = \frac{2}{N} C $ once \eqref{eq:initial-cond-one-point} and \eqref{eq:initial-cond-two-point} are used. We will revisit this point after we determine the thermal equilibrium configuration, which we take up next.

\subsubsection{Thermal Equilibrium:}
\label{ssec:thermal-equilibrium}

Our goal is to maximize the entropy of the system at fixed energy. By introducing the Lagrange multiplier $\beta$, let us write
\begin{align}
\mathcal{L} := S - \beta(E-E_{eq}) \,,
\label{eq:L}
\end{align}
where $E_{eq}$ denotes the total energy of the system at thermal equilibrium. We already know that the entropy of the system is given in the form \eqref{eq:entropy1} and from that expression it is readily seen that it is a monotonically increasing function of the symplectic eigenvalues $f_l$. Thus, the entropy will be maximized if the conditions
\begin{align}
\frac{\partial \mathcal{L}}{\partial \sigma_{xx}(l)} = \frac{\partial S}{\partial \sigma_{xx}(l)} -  \beta \frac{\partial E}{\partial \sigma_{xx}(l)} = 0 \,, \quad
\frac{\partial \mathcal{L}}{\partial \sigma_{pp}(l)} = \frac{\partial S}{\partial \sigma_{pp}(l)} - \beta \frac{\partial E}{\partial \sigma_{pp}(l)} = 0 \,,
\end{align}
are satisfied. 
These equations yield
\begin{align}
\frac{\partial S}{\partial f_l} \frac{\sigma_{pp}(l)}{2 f_l}  &= \beta (2 l +1) \left( \mu^2 + \frac{l (l+1)}{R^2} + \frac{\lambda C}{N}\right) \,, \nn \\
\frac{\partial S}{\partial f_l} \frac{\sigma_{xx}(l)}{2 f_l}  &= \beta(2l+1)  \,,
\label{eq:dsdfl}
\end{align}
from which we infer the thermal equilibrium condition to be:
\begin{align}
\frac{\sigma_{pp}(l)}{\sigma_{xx}(l)} = \left (\mu^2 + \frac{l(l+1)}{R^2} + \frac{\lambda C}{N}\right ) \,.
\label{eq:wl2}
\end{align}
In fact, it can be readily verified that together with this condition the configuration specified via \eqref{eq:initial-cond-one-point} and \eqref{eq:initial-cond-two-point} constitute a static solution of the set of equations \eqref{eq:eom-one-point}-\eqref{eq:p2}, \eqref{eq:eom-two-point-xx}-\eqref{eq:eom-two-point-pp-22}. This means that the system is indeed in mechanical equilibrium in the thermalized state.

It is very useful to define the characteristic frequency of the system at thermal equilibrium as $\omega_l := \sqrt{ \left (\mu^2 + \frac{l(l+1)}{R^2} + \frac{\lambda C}{N}\right)}$. As $\lambda \rightarrow 0$ this becomes $\omega_l ^{(0)} := \sqrt{\mu^2 + \frac{l(l+1)}{R^2}}$, which are nothing but the harmonic frequencies of the non-interacting system in the $\lambda \rightarrow 0$ limit. 
 
Using \eqref{eq:dsdfl} and \eqref{eq:wl2}, we have
\be
\label{eq:sigma-xx-k}
\sigma_{xx}(l)  =  \frac{f_l}{\omega_l} \,, \quad  \sigma_{pp}(l) = \omega_l f_l \,,
\ee
while from \eqref{eq:dsdfl} and \eqref{eq:sigma-xx-k} we get
\begin{align}
\beta \omega_l = \ln \left( \frac{f_l+\frac{1}{2}}{f_l-\frac{1}{2}} \right) \,,
\end{align}
which can be inverted to give the symplectic eigenvalues in the final form
\begin{align}
f_l = \frac{1}{2} \coth (\frac{\beta \omega_l}{2}) \,.
\label{eq:flT}
\end{align}
Let us also note that in the limit $\lambda \rightarrow 0$, $f_l \to \frac{1}{2} \coth (\frac{\beta \omega_l^{(0)}}{2})$, which are nothing but the symplectic eigenvalues of $2N^2$ harmonic oscillators with frequencies $ \omega_l^{(0)}$. 

We may now compute the temperature of the system in a standard manner by differentiating the entropy with respect to the energy. We have 
\be
\frac{1}{T} = \frac{\partial S}{\partial E} = \frac{d S}{d \beta} \left ( \frac{d E}{d \beta} \right )^{-1}  = \beta \,,   
\ee
where we have used\footnote{It is worthwhile to note that subsequent to the maximization of the entropy $S$ at fixed energy, the latter can be expressed as 
\be
E: = \sum_{l} 2 (2l+1) \omega_l f_l - \frac{\lambda C^2}{2N} \,.
\ee
} 
\begin{align}
\frac{d S}{d \beta} &= \sum_{l=0}^{N-1} \frac{\partial S}{\partial f_l} \frac{\partial f_l}{\partial \beta} =  - \frac{1}{2} \beta \sum_{l=0}^{N-1} (2l+1) \omega_l^2 \csch^2 \left ( \frac{\beta \omega_l}{2} \right )\,, \nn \\ 
\frac{d E}{d \beta} &= - \frac{1}{2} \sum_{l=0}^{N-1} (2l+1) \omega_l^2 \csch^2  \left ( \frac{\beta \omega_l}{2}  \right)\,.
\end{align}

\begin{figure}
  \centering
  \begin{subfigure}[b]{0.48\textwidth}
    \centering
    \includegraphics[width=\linewidth]{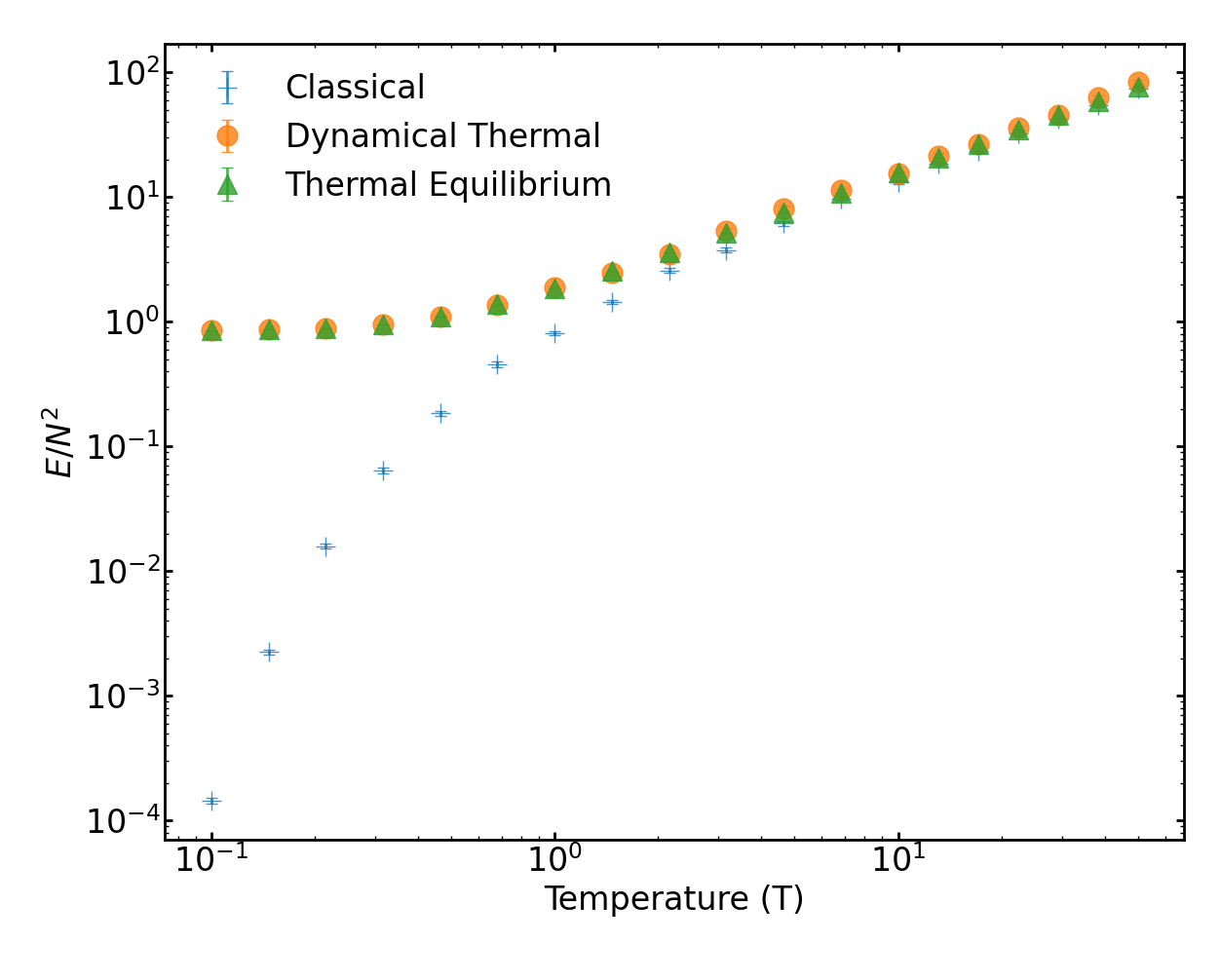}
    \caption{}
    \label{fig:E_T_Planar}
  \end{subfigure}
  \hfill
  \begin{subfigure}[b]{0.48\textwidth}
  	\centering
  	\includegraphics[width=\linewidth]{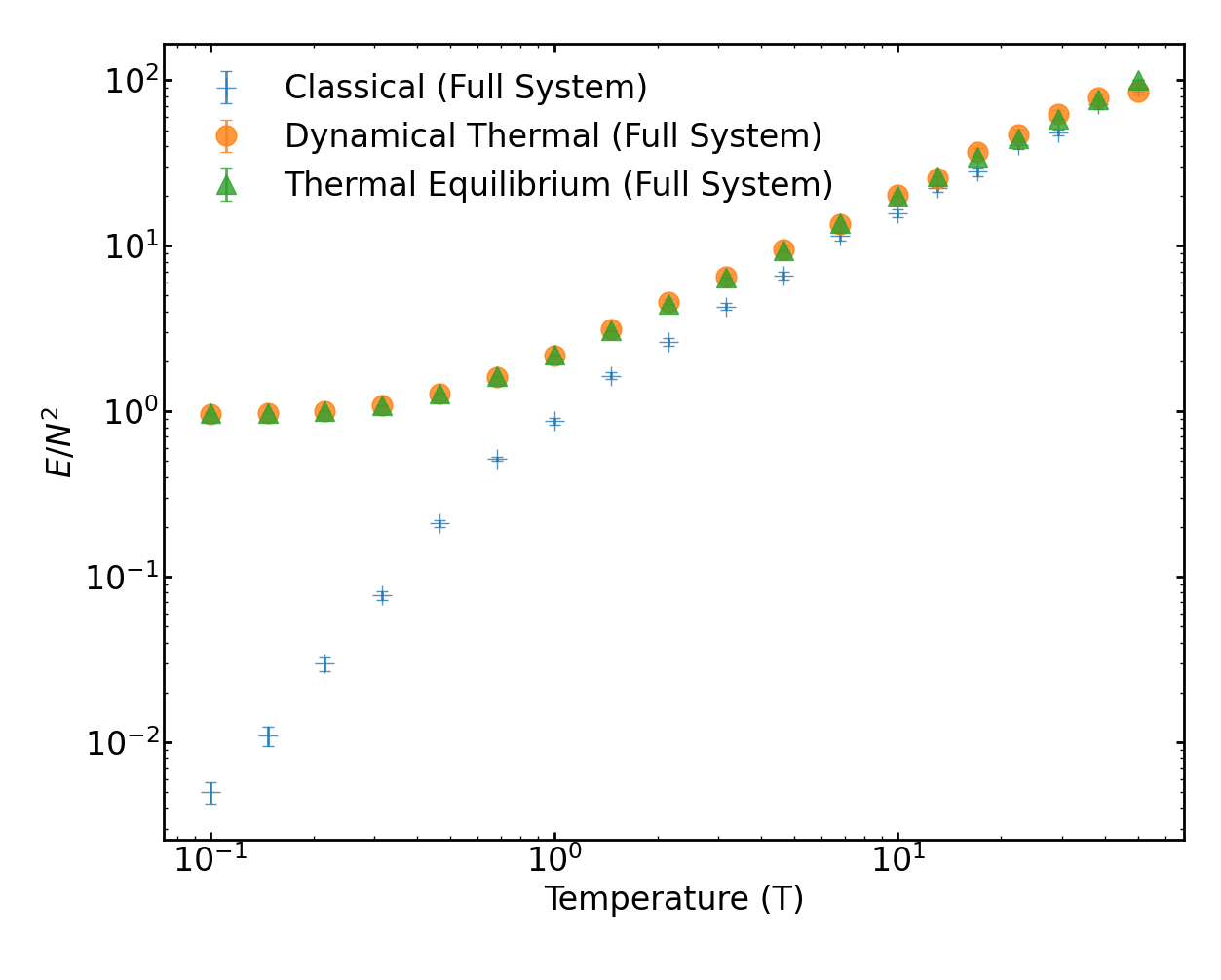}
  	\caption{}
 \label{fig:E_T_Full}
  \end{subfigure} 
  \begin{subfigure}[b]{0.48\textwidth}
  \centering
    \includegraphics[width=\linewidth]{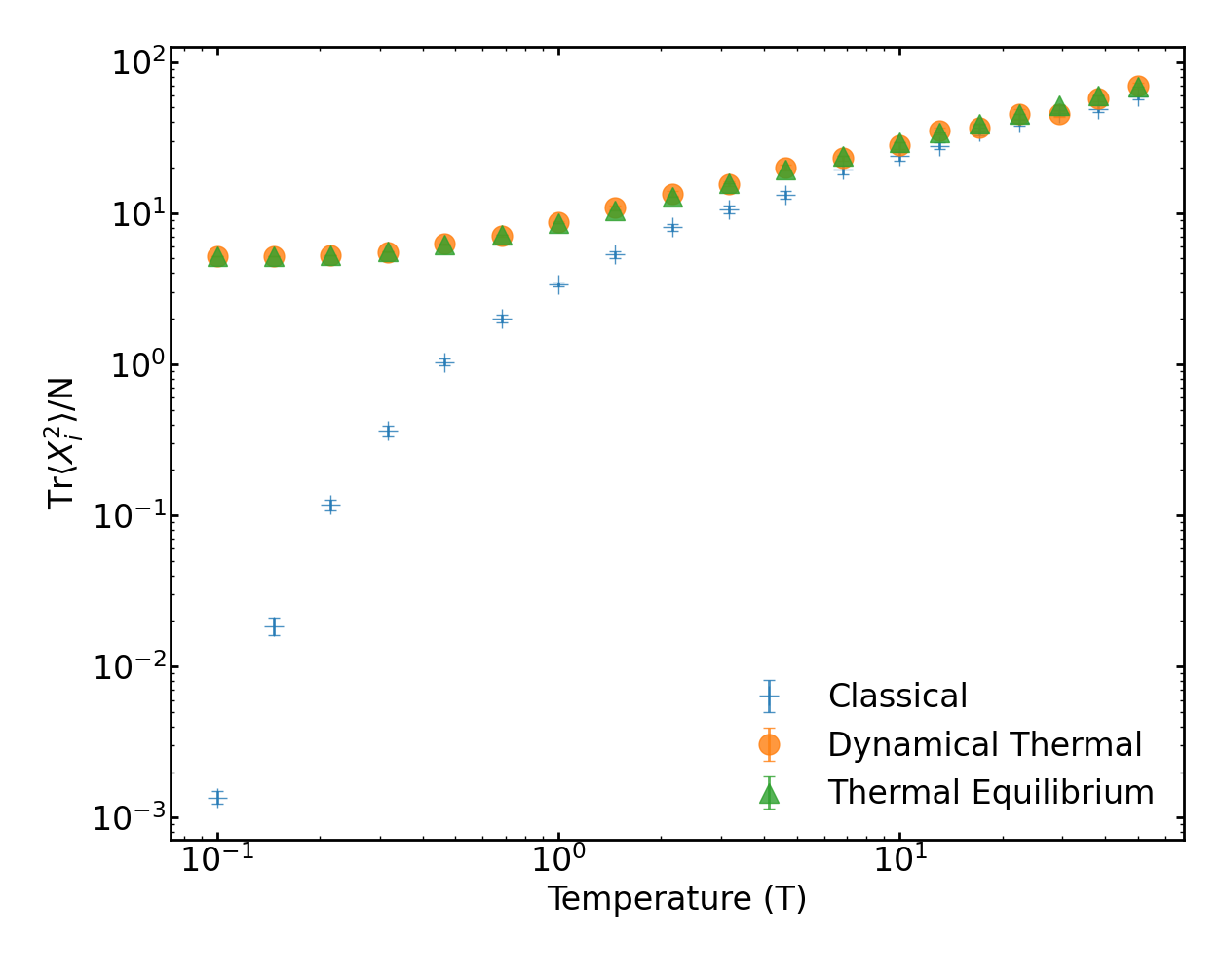}
   \caption{}
 \label{fig:CDvTPlanar}
  \end{subfigure}
  \hfill
  \begin{subfigure}[b]{0.48\textwidth}
	\centering
	\includegraphics[width=\linewidth]{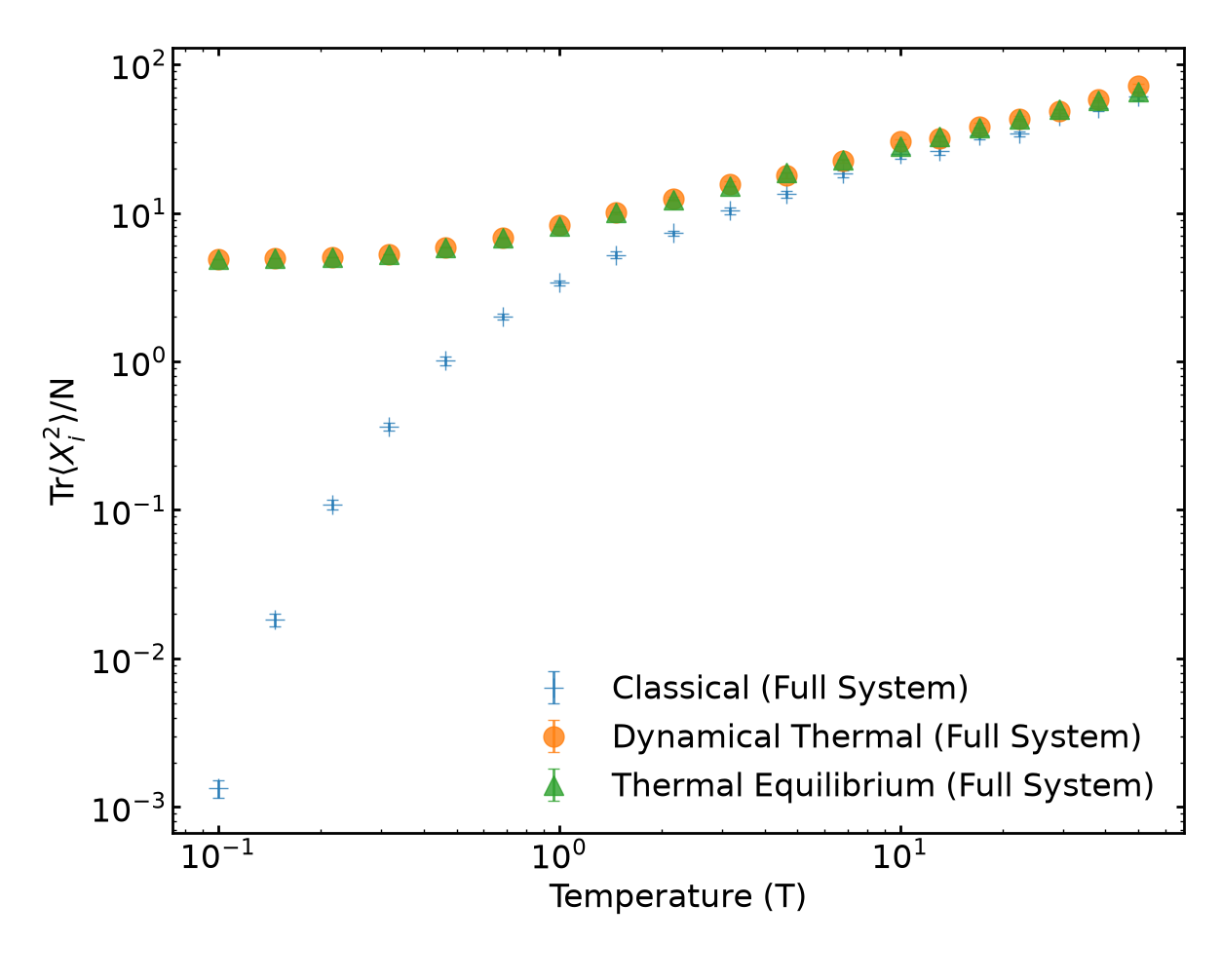}
	\caption{}
	 \label{fig:CDvTFull}
\end{subfigure}	
\caption{$E/N^2$ and Total Coordinate dispersion,$\frac{1}{N} \braket{ Tr \, {\hat X}_i {\hat X}_i}$, vs. $T$ ((a) \& (c) for planar interaction, (b) \& (d) for the full system) for $N=5$, $\mu = \frac{1}{10}$, $R=5$, $\lambda = \frac{1}{5}$.}
\label{fig:E_CDvT}
\end{figure}

Inserting for $f_l$ from \eqref{eq:flT} and multiply both sides of the first equation in \eqref{eq:sigma-xx-k} by $2l+1$ and summing over all $l$, we have 
\begin{align}
\label{eq:consistency-equation}
C = \frac{1}{2} \sum_{l=0}^{N-1} (2l+1)\frac{\coth \left(\frac{\beta}{2} \sqrt{\mu^2 + \frac{l(l+1)}{R^2} + \frac{\lambda C}{N}} \right)}{ \sqrt{\mu^2 + \frac{l(l+1)}{R^2} + \frac{\lambda C}{N}}}.
\end{align}
where we have used the definition of $C$ given in \eqref{eq:C}. Clearly, the same equation follows from \eqref{eq:C} by using the equation for $\sigma_{xx}(l)$  in \eqref{eq:sigma-xx-k}. The expression in \eqref{eq:consistency-equation} can be called a self-consistency equation for $C$ at thermal equilibrium. It can indeed be solved numerically to determine the value of $C$ at a given value of temperature $T$ and the other parameters $\mu$, $R$ and the size $N$ of the system, which can, in turn, be used to completely determine $\sigma_{xx}(l)$ and $\sigma_{pp}(l)$ as well as all the other physically relevant quantities, such as the energy $E$ given in \eqref{eq:el-equation} and the symplectic eigenvalues $f_l$ at a given temperature $T$. Thus, \eqref{eq:consistency-equation}  is the final and the most critical ingredient which allows us to unambiguously determine the equation of state.  In particular, in the zero temperature limit, $\lim_{\beta \to \infty} f_l = \frac{1}{2} \,\forall l$, \eqref{eq:consistency-equation} takes the form
\begin{align}
C \Big |_{T=0} = : C_0 = \frac{1}{2} \sum_{l=0}^{N-1} \frac{2l+1}{ \sqrt{\mu^2 + \frac{l(l+1)}{R^2} + \frac{\lambda C_0}{N}}} \,.
\label{eq:consistency-equationT0}
\end{align}
The ground state energy, i.e. the energy at $T=0$, is then given as 
\be
E_{0} = \sum_{l=0}^{N-1} (2l+1) \omega_{l}  - \frac{\lambda C_0^2}{2N} \,.
\label{eq:E0}
\ee
In Fig. \ref{fig:E_CDvT} the energy and the total coordinate dispersion versus the temperature plots are provided to demonstrate the thermal equilibrium profile of the system. 
\vskip 1em
{\it Non-Planar Term:}
\vskip 1em
We now return to the treatment of the non-planar term.  From \eqref{intpot1}, we have 
\begin{align}
	&- \mathcal{K}^{l_1 l_3 l_2 l_4}_{m_1 m_3 m_2 m_4} \braket{{\hat X}_1^{l_1 m_1} {\hat X}_1^{l_2 m_2} {\hat X}_2^{l_3 m_3} {\hat X}_2^{l_4 m_4}} = - \mathcal{K}^{l_1 l_2 l_1 l_2}_{m_1 m_2 m_1 m_2} \sigma_{xx}(l_1) \sigma_{xx}(l_2)\nonumber \\
	&= -\sum_{l_1 m_1}\sum_{l_2 m_2} \sigma_{xx}(l_1)\sigma_{xx}(l_2)\, Tr \left( Z_{l_1 m_1}  Z_{l_2 m_2} Z_{l_1 m_1} Z_{l_2 m_2} \right) \nonumber \\
	&= - \sum_{l_1 l_2} (-1)^{l_1+l_2+N-1} (2l_1 + 1)(2l_2 + 1)\sigma_{xx}(l_1)\sigma_{xx}(l_2) \sixj{l_1}{\frac{N-1}{2}}{\frac{N-1}{2}}{l_2}{\frac{N-1}{2}}{\frac{N-1}{2}} \,, \nn \\
	&=: - \sum_{l_1 l_2} (2l_1 + 1)(2l_2 + 1)\sigma_{xx}(l_1)\sigma_{xx}(l_2) G (l_1, l_2) \,.
	\label{eq:nonplanarint}
\end{align} 
Details of the calculation leading to this result are provided in the Appendix~\ref{sec:polarization-operator-basis}. As we anticipated and remarked earlier in the footnote \ref{fn:largeN} , we observe that \eqref{eq:nonplanarint} is not suppressed by $1/N^2$ compared to the planar term, since the contribution of the $6$-j symbol within the sum varies with $l_1$, $l_2$ and does not amount to a $1/N^2$ suppression.

Defining $D(l) := \sum_{l'=0}^{N-1}(2l'+1)\sigma_{xx}(l')G(l, l')$ we may write the energy as
\begin{multline}
	\label{eq:total-energy}
	E = \braket{\hat{H}} = \sum_{l=0}^{N-1}(2l+1)\left[\sigma_{pp}(l) + \left(\mu^2 + \frac{l(l+1)}{R^2}\right)\sigma_{xx}(l)\right] + \frac{\lambda}{2N}C^2 \\
-\frac{\lambda}{2}\sum_l (2l+1)\sigma_{xx}(l)\,D(l) \,.
\end{multline}
Maximization of the entropy works along the same lines as before and yields 
\be
\label{eq:thermal-ratio}
\frac{\sigma_{pp}(l)}{\sigma_{xx}(l)} = \mu^2 + \frac{l(l+1)}{R^2} + \frac{\lambda C}{N} - \lambda D(l) \,,
\ee
while the effective frequency becomes 
\begin{equation}
	\label{eq:omega-def}
	\omega_l := \sqrt{\mu^2 + \frac{l(l+1)}{R^2} + \frac{\lambda C}{N} - \lambda D(l)}\,,
\end{equation}
which are thus both modified by the $l$-dependent term: $-\lambda D(l)$. Expressions in \eqref{eq:sigma-xx-k} and \eqref{eq:flT} remain in the same form with the understanding that $\omega_l$ is now given by 	\eqref{eq:omega-def}. Also $\beta = T^{-1}$ follows identically from $\frac{1}{T} = \frac{\partial S}{\partial E}$ as before.

Substituting $\sigma_{xx}(l) = \frac{1}{2 \omega_l}\coth\left(\frac{\beta\omega_l}{2}\right)$ into $C = \sum_l(2l+1)\sigma_{xx}(l)$ and $D(l) = \sum_{l'}(2l'+1)\sigma_{xx}(l')G(l,l')$, we obtain a system of $N$ coupled nonlinear equations for the $N$ unknowns $\vec{\sigma} = (\sigma_{xx}(0),\sigma_{xx}(1), \ldots, \sigma_{xx}(N-1))$, which can be generically expressed as 
\begin{equation}
	\label{eq:consistency-system}
	\sigma_{xx}(l) = \frac{1}{2 \,\omega_l[\vec{\sigma}]}\coth\left(\frac{\beta\,\omega_l[\vec{\sigma}]}{2}\right)\,, \quad l = 0, 1, \ldots, N-1,
\end{equation}
where now
\begin{equation}
	\label{eq:omega-def2}
	\omega_l[\vec{\sigma}] = \sqrt{\mu^2 + \frac{l(l+1)}{R^2} + \frac{\lambda}{N}\sum_{l'}(2l'+1)\sigma_{xx}(l') - \lambda\sum_{l'}(2l'+1)\sigma_{xx}(l')\,G(l,l')}\,.
\end{equation}
Thus, the consistency equation for $C$ given in \eqref{eq:consistency-equation} generalizes to the system of coupled equations given in \eqref{eq:consistency-system}. Ignoring the $G(l,l')$-dependent term reduces to the planar case as expected, since then $\omega_l$ depends on $\vec{\sigma}$ only through $C = \sum_{l'}(2l'+1)\sigma_{xx}(l')$, and the system \eqref{eq:consistency-system} collapses to the equation for $C$. In particular, at $T = 0$ ($\beta \to \infty$), $\coth(\beta\omega_l/2) \to 1$ and $f_l \to 1/2$ and $\sigma_{xx}^{(T=0)}(l) = \frac{1}{2 \,\omega_l[\vec{\sigma}^{(T=0)}]}$ which still remains a system of $N$ coupled algebraic equations. We have solved \eqref{eq:consistency-system}  numerically and obtained the profiles of $E/N^2$ as well as $\frac{1}{N} \braket{ Tr \, {\hat X}_i {\hat X}_i}$ for the full interaction term. It is readily observed from Fig. \ref{fig:E_T_Full}, \ref{fig:CDvTFull} , that there is no substantial change in the thermal characteristics after the inclusion of the non-planar interaction term. 

At this point it is also worthwhile to comment on the connections between the present approach and the earlier work in the literature exploring the quantum field theory on the fuzzy sphere \cite{Vaidya:2001bt, Chu:2001xi, Dolan:2001gn, Vaidya:2003ew, Balachandran:2005ew}. These works considered the theory of a single real scalar field on the fuzzy sphere with a quartic interaction and obtained the $1$-loop effective action, whose results revealed that the continuum limit of the renormalized theory at $1$-loop differs from that on the continuum sphere $S^2 $ by a finite correction to the kinetic energy which arises from the non-planar loop integration. This is a mild $l$-dependent quantum contribution to the dispersion relation indicative of ultraviolet/infrared (UV/IR) mixing, i.e.  the emergence of an IR singularity from integrating out the UV degrees of freedom, encountered in the quantum effective actions of field theories on noncommutative spaces such as the noncommutative planes ${\mathbb R}_\theta^n$ \cite{Minwalla:1999px}. Thus, this mild correction to the kinetic energy can be considered as the regularization of the UV/IR mixing on the noncommutative plane ${\mathbb R}_\theta^2$, which indeed appears back in the scaling limit that takes $S_F^2$ to ${\mathbb R}_\theta^2$.  In comparison, the GSA approximation we are using in this work evaluates the 2-loop diagrams with a single vertex (i.e. the figure-8 diagram and its non-planar counterpart), where the full propagator appear in the loops (as opposed to the bare propagators used in 1PI effective action calculations) and it is obtained by solving the self-consistency equations \eqref{eq:consistency-system}. This is essentially equivalent to the 2PI self-consistent approach \cite{Berges:2004yj} mentioned in the introduction applied at the one vertex level. We can indeed make the connection between these two approaches somewhat more concrete by nothing that our result for the effective squared frequency $\omega_l^2$ given via \eqref{eq:omega-def} or \eqref{eq:omega-def2} together with the equations \eqref{eq:consistency-system} have precisely the renormalization contributions $\frac{\lambda C}{N}$ and $- \lambda D(l)$ to the bare squared frequencies $(\omega_l^{(0)})^2$ for the planar and non-planar interactions, respectively. These terms have a structure similar to the planar and non-planar 1-loop two-point functions \cite{Vaidya:2001bt, Chu:2001xi, Dolan:2001gn, Vaidya:2003ew, Balachandran:2005ew} with the full propagator $\sigma_{xx}(l)$ in place of the bare propagator $(\frac{l(l+1)}{R^2} + \mu^2)^{-1}$.  The $l$-dependent contributions to the frequencies via $D(l)$ parallels the mild $l$-dependent contribution to the dispersion relation of the 1-loop approach outlined above.

\begin{figure}[htbp]
	\centering
	\begin{subfigure}[b]{0.48\textwidth}
		\centering
		\includegraphics[width=\linewidth]{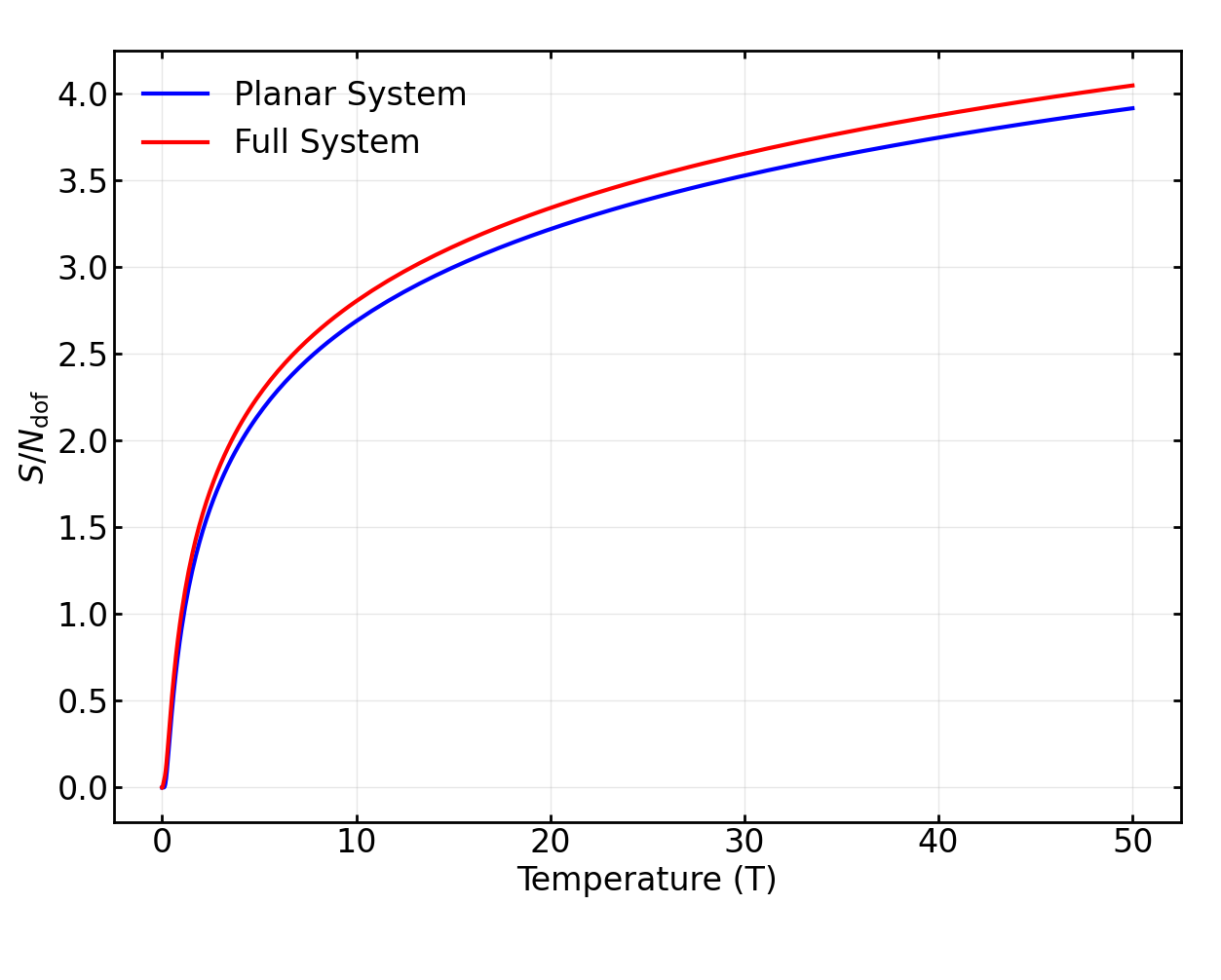}
		\caption{}
		\label{fig:S_TPP }
	\end{subfigure}
	\hfill
		\begin{subfigure}[b]{0.48\textwidth}
		\centering
		\includegraphics[width=\linewidth]{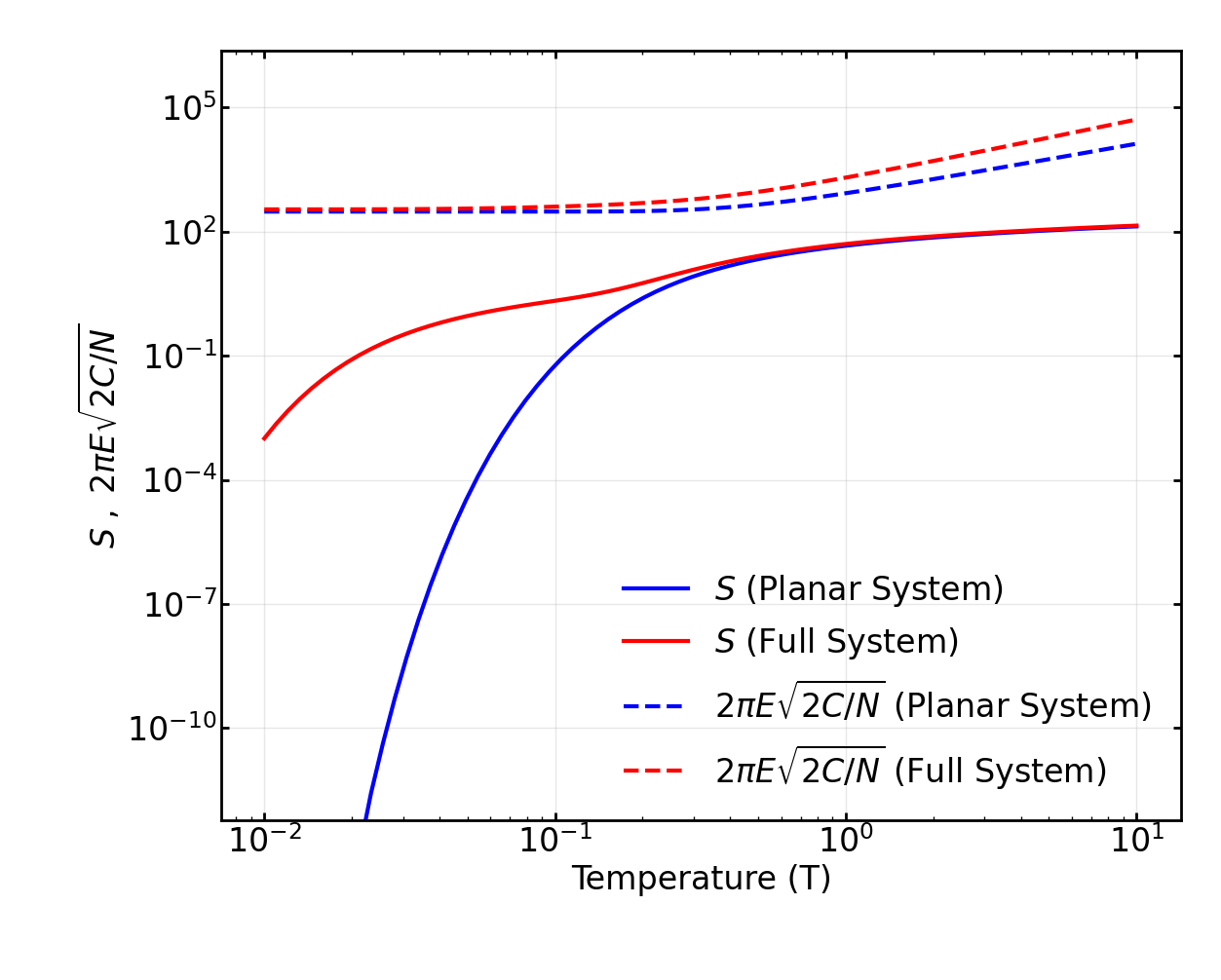}
		\caption{}
		\label{fig:S_T_Bounds}
	\end{subfigure}
\caption{(a) $S/N_{dof}$ vs $T$, (b) $S$, $2 \pi E K$ vs $T$ at $N=5$, $\mu = \frac{1}{10}$, $R=5$, $\lambda = \frac{1}{5}$ and $N_{dof} = 2 N^2 =50$.}
\label{fig:S_T}
\end{figure}

Variation of entropy with temperature is plotted in Fig. \ref{fig:S_T}, where we also demonstrate that the Bekenstein bound \cite{Bekenstein:1980jp, Page:2018yby} $S \leq 2 \pi E K$, with $K:= \sqrt{\frac{1}{N} \braket{ Tr \, {\hat X}_i {\hat X}_i}}$ is satisfied. Indeed, we find that $S \ll 2 \pi E \sqrt{2 C/ N}$. We may remark that the small dent in the entropy of the full configuration as seen in the red curve in Fig. \ref{fig:S_T_Bounds} around $T \approx 10^{-1}$ is caused by the contribution of the decoupled $l=0$ mode to the total entropy, which intersects the sum of the rest around this temperature. Total entropy is still monotonically increasing, as it should be, albeit with a slightly lower rate around this temperature. 

\subsection{Choosing the initial conditions}\label{ssec:choosing-the-initial-conditions}

We have already noted that the thermal equilibrium conditions yield a static solution for the equations of motion \eqref{eq:eom-one-point}-\eqref{eq:p2}, \eqref{eq:eom-two-point-xx}-\eqref{eq:eom-two-point-pp-22}. Linearization of the latter around this solution leads to small fluctuations as one can recognize both analytically and numerically. Thus, as may be naturally expected, by inspecting the small fluctuations about this thermal state we can not probe the chaotic dynamics of the system under study. Nevertheless, we can follow the approach given in \cite{Buividovich:2018scl} to choose initial data, which is close enough to the thermal state but yet dynamical so that the time evolution generated by the equations of motion allows us to study the emerging chaotic behavior. To serve this purpose, let us imagine a mixed state at temperature $T$ whose dispersions $\sigma_{xx}(l)$ and $\sigma_{pp}(l)$ are separated into quantum contributions due to a pure state at $T=0$ corresponding to $f_l = \frac{1}{2} \forall l$ and classical contributions describing the thermal fluctuations and write this as\footnote{As noted in \cite{Buividovich:2018scl}, it is possible to choose a different separation such that $\sigma_{xx}^0(l)$, $\sigma_{pp}^0(l)$ are not at zero temperature and the displacements are randomly picked from Gaussians with different variances. As this is not necessary in practice, we will not pursue it any further.} 
\begin{align}
\sigma_{xx}(l) = \sigma_{xx}^0(l) + \sigma_{xx}^c(l) \,, \quad \sigma_{pp}(l) = \sigma_{pp}^0(l) + \sigma_{pp}^c(l) \,.
\end{align}
This picture suggests that, we may construct quantum states that are classical mixtures of pure Gaussian states, which approximate the Gaussian thermal state at temperature $T$ quite well. To be more concrete, consider a pure state with non-vanishing $1$-point functions $\langle \hat {X}_i^{lm} \rangle = \bar{X}_i^{lm}$ and $ \langle \hat{P}_i^{lm} \rangle = \bar{P}_i^{lm}$ and denote it as $| \bar{X}_i^{lm}, \bar{P}_i^{lm} \rangle \langle \bar{X}_i^{lm}, \bar{P}_i^{lm} |$. This is not a thermal state, but can be obtained from the zero-temperature Gaussian pure state centered at the origin of the phase space, $|0, 0 \rangle \langle 0, 0 |$, by shifting the coordinates and momenta, via $| \bar{X}_i^{lm}, \bar{P}_i^{lm} \rangle = \exp(i  \bar{X}_i^{lm}  \hat{P}_i^{lm}  + i \bar{P}_i^{lm}  \hat{X}_i^{lm} ) |0,0 \rangle$.\footnote{The notation here simply means that the expectation values of $\hat{X}_i^{lm}$ and $\hat{P}_i^{lm}$ in this  state are $\bar{X}_i^{lm}$, $\bar{P}_i^{lm}$, respectively. Obviously, these are not the eigenstates of $\hat{X}_i^{lm}$ or $\hat{P}_i^{lm}$.} Subsequently, we can think of forming a classical mixture of such states, say with the density matrix $\hat {\rho}_c : = \left \langle | \bar{X}_i^{lm}, \bar{P}_i^{lm} \rangle \langle \bar{X}_i^{lm}, \bar{P}_i^{lm} | \right \rangle_c$, with $\langle \cdot \rangle_c$ representing the averaging over the classical probabilities and where the coordinate and momentum displacements are selected randomly from Gaussian distributions with zero mean and the variances $\braket{X_i^{lm} X_j^{l'm'}}_c = \sigma_{xx}^c(l) \delta^{l l'} \delta^{m m'} \delta_{ij}$ and $\braket{P_i^{lm} P_j^{l'm'}}_c = \sigma_{pp}^c(l)  \delta^{l l'} \delta^{m m'} \delta_{ij} $, respectively. The state obtained in this manner is dynamical as each of the shifted pure states are time-dependent and their time evolution is obtained by solving the equations \eqref{eq:eom-one-point}-\eqref{eq:p2}, \eqref{eq:eom-two-point-xx}-\eqref{eq:eom-two-point-pp-22} with the initial conditions 
\begin{align}
\braket{\braket{{\hat X}_i^{lm} {\hat X}_j^{l'm'}}} = \sigma_{xx}^0(l) \delta^{l l'} \delta^{m m'} \delta_{ij} \,, \quad \braket{\braket{{\hat P}_i^{lm} {\hat P}_j^{l'm'}}} = \sigma_{pp}^0(l) \delta^{l l'} \delta^{m m'} \delta_{ij} \,,
\end{align} 
and the non-vanishing $1$-point functions $\langle X_i^{lm} \rangle = \bar{X}_i^{lm}$ and $ \langle P_i^{lm} \rangle = \bar{P}_i^{lm}$ . 

\begin{figure}[htbp]
	\centering
	\begin{subfigure}[b]{0.48\textwidth}
		\centering
		\includegraphics[width=\textwidth]{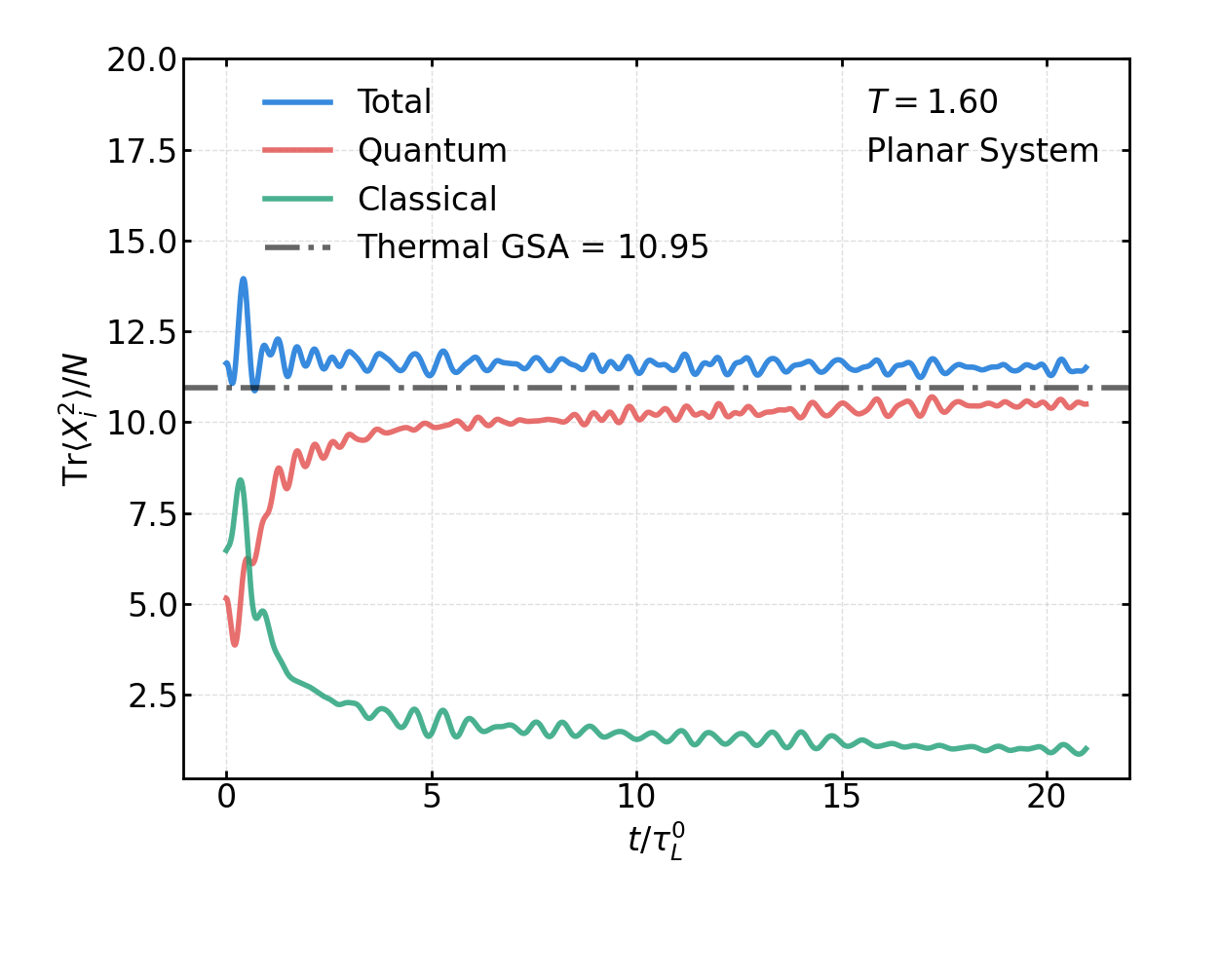}
		\caption{}\label{fig:Traces_T_16_Planar}
	\end{subfigure}
	\hfill
	\begin{subfigure}[b]{0.48\textwidth}
		\centering
		\includegraphics[width=\textwidth]{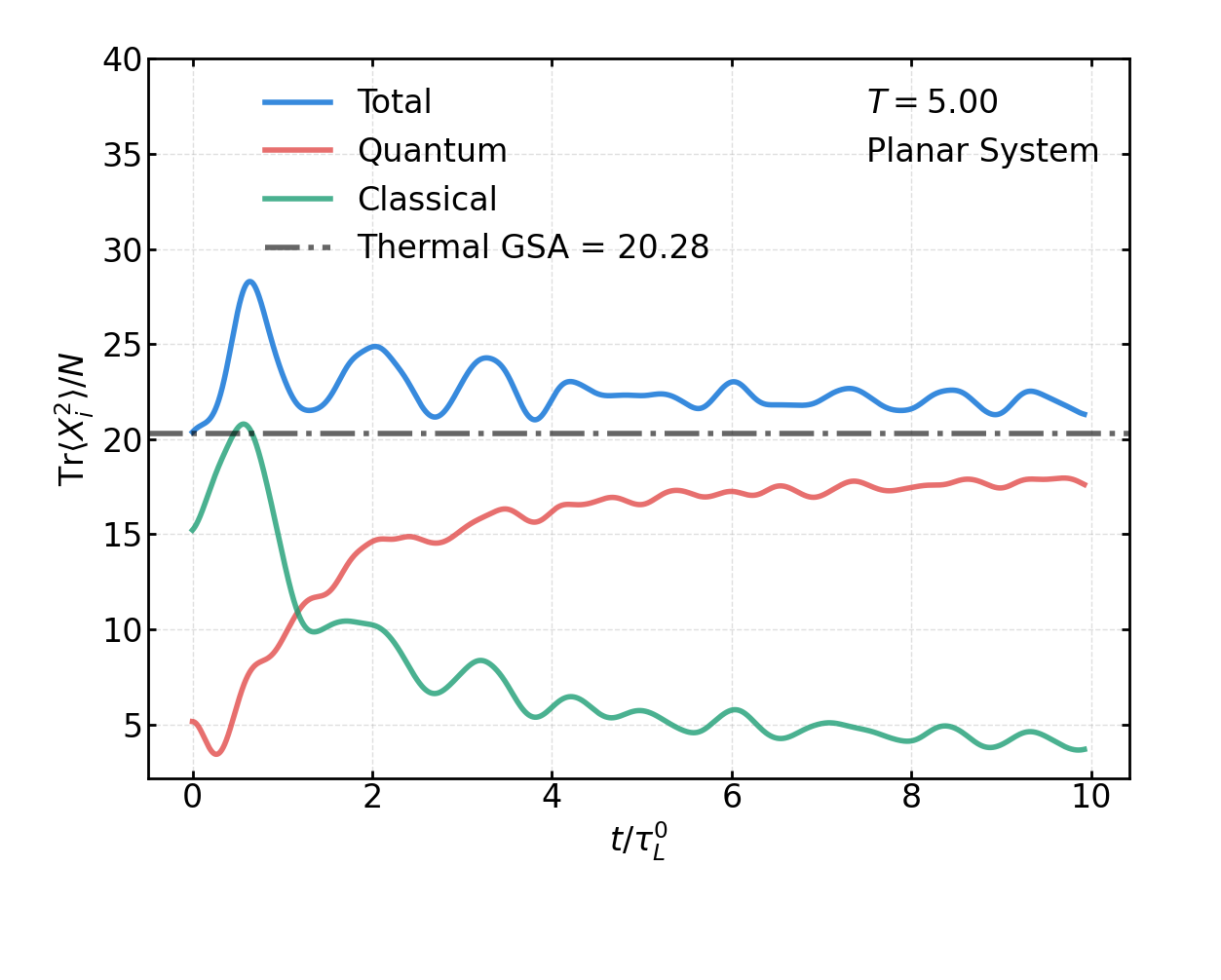}
		\caption{}\label{fig:Traces_T_5_Planar}
	\end{subfigure}
	\label{fig:TESV_two_point}
	\vspace{1em} 
	\centering
	\begin{subfigure}[b]{0.48\textwidth}
		\centering
		\includegraphics[width=\textwidth]{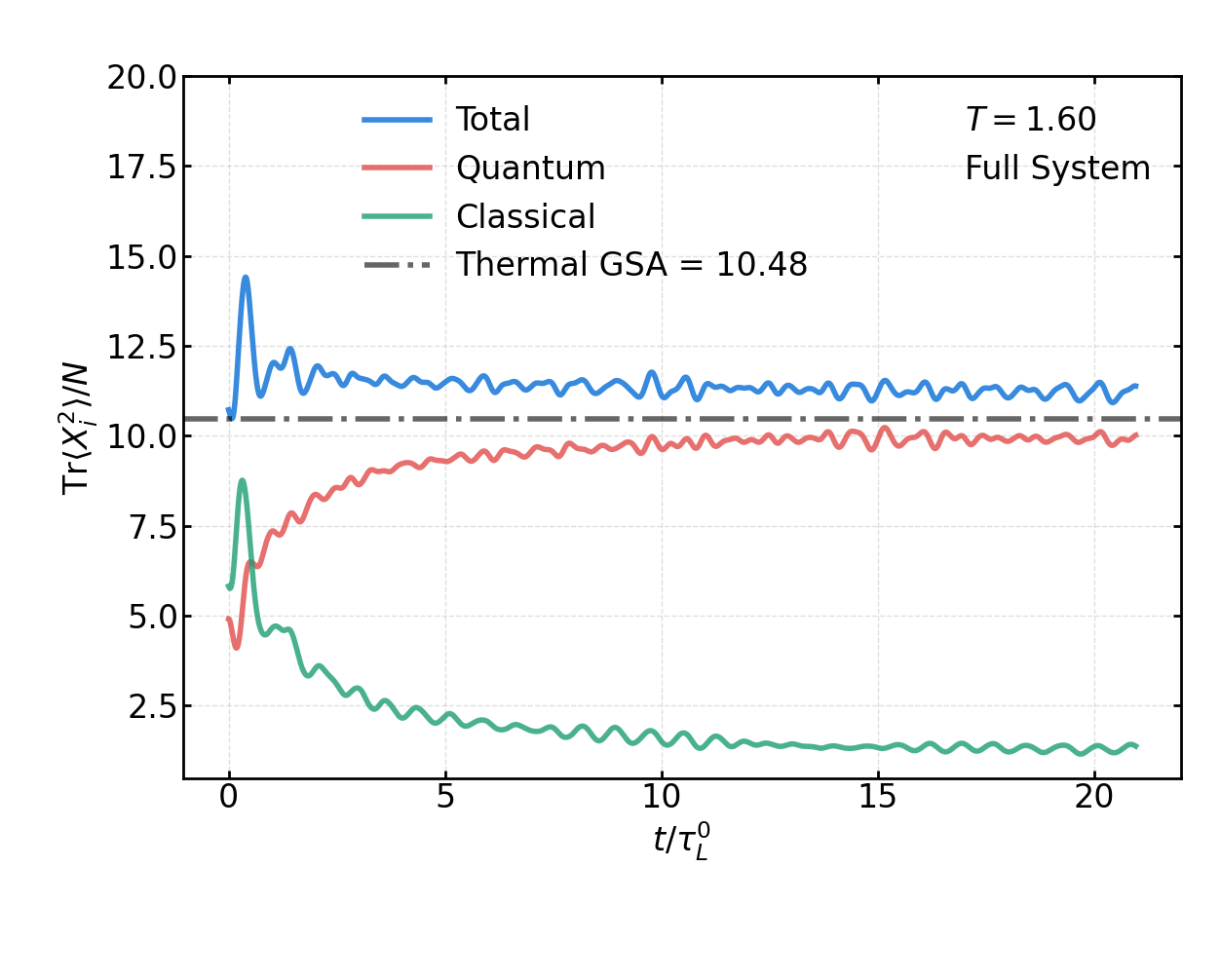}
		\caption{}\label{fig:Traces_T_16_Full}
	\end{subfigure}
	\hfill
	\begin{subfigure}[b]{0.48\textwidth}
		\centering
		\includegraphics[width=\textwidth]{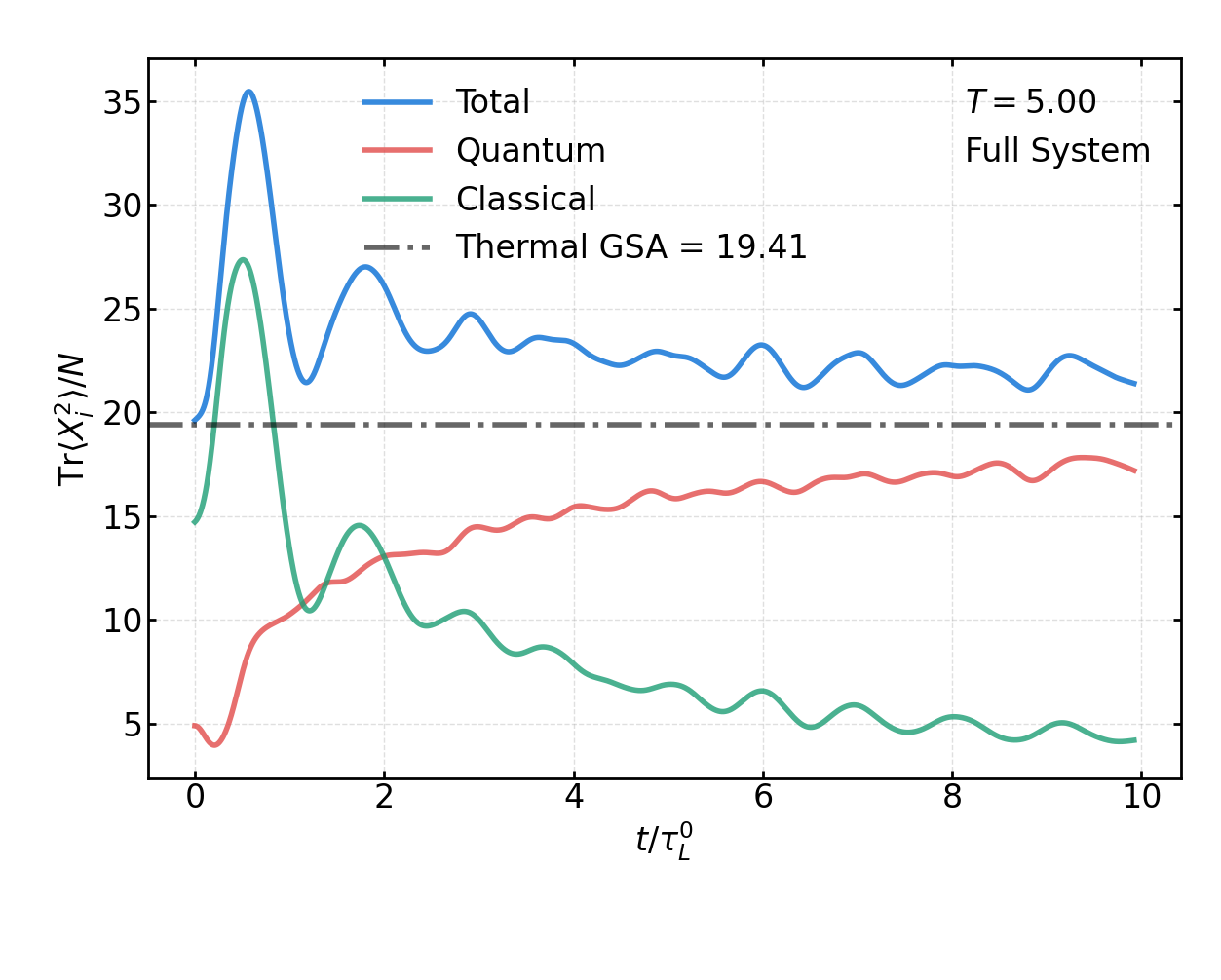}
		\caption{}\label{fig:Traces_T_5_Full}
	\end{subfigure} \\
\caption{Variation of $\braket{Tr(\hat{X}_i\hat{X}_i)}/N$ with time (in units of classical Lyapunov time $\tau_L = \lambda_L^{-1}$) plotted at $N=5$, $R=5$, $\mu=1/10$, $\lambda=1/5$.} 
\label{fig:TTC1}
\end{figure}

An initial state prepared in this manner, fits very well with the thermal equation of state with small statistical errors. Firstly, we observe from Fig.\ref{fig:E_CDvT} both the energy and total coordinate dispersion $\frac{1}{N} \braket{ Tr \, {\hat X}_i {\hat X}_i}$ calculated by averaging over the values obtained on the set of randomly shifted pure states (we use seven to nine randomly shifted initial conditions at each temperature) matches in a nice way with that of the thermal equilibrium state. Next, we also look at the time evolution of the $\frac{1}{N} \braket{ Tr \, {\hat X}_i {\hat X}_i}$. Unlike the energy, this is not a constant of the motion.  Nevertheless, as we see from Fig. \ref{fig:TTC1} it fluctuates with rather small amplitudes somewhat above its thermal equilibrium value at a given temperature. This behavior of the total coordinate dispersion is also encountered in the BFSS model and its bosonic part as discussed in \cite{Buividovich:2018scl}. Thus, in contrast to the thermal state, which is only a static solution of the equations \eqref{eq:eom-one-point}-\eqref{eq:p2}, \eqref{eq:eom-two-point-xx}-\eqref{eq:eom-two-point-pp-22}, the classical mixture of time-dependent pure states as we have constructed above can be interpreted to constitute a {\it dynamical thermal state}. Therefore, we expect that their non-trivial solutions encodes the chaotic motion, which will be discussed next.  

\section{Dynamics and Chaos}
\label{sec:DandC}

We have written a C++ code that solves the equations of motion \eqref{eq:eom-one-point}-\eqref{eq:p2}, \eqref{eq:eom-two-point-xx}-\eqref{eq:eom-two-point-pp-22}  where the dynamical thermal states are taken as initial conditions. In particular, we have obtained the solutions of these equations at several distinct values of the temperature at the matrix level $N=5$ and $R=5$, $\mu =\frac{1}{10}$ and $\lambda=\frac{1}{5}$, the latter corresponding to the 't Hooft limit as already noted earlier. At each temperature, for the given parameter values we average over seven to nine coordinate and momentum displaced states randomly picked from the Gaussian distributions with the variances $\sigma_{xx}^c$ and $\sigma_{pp}^c$ as explained in the previous section.

\begin{figure}[htbp]
		\begin{subfigure}[b]{0.48\textwidth}
			\centering
				\includegraphics[width=\linewidth]{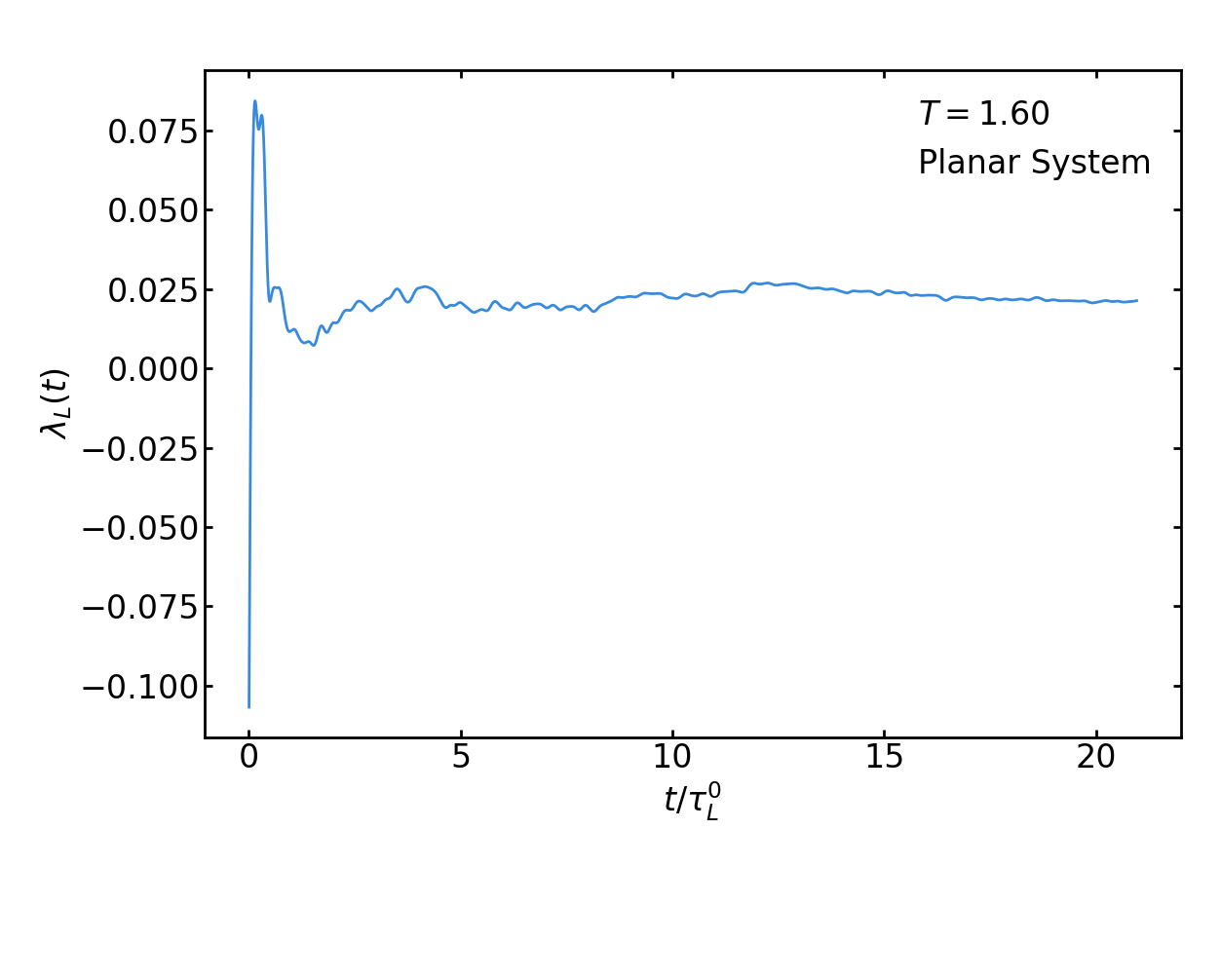}
			\caption{}\label{fig:Lyap_Time_Planar_T1.6}
		\end{subfigure}
		\hfill
		\begin{subfigure}[b]{0.48\textwidth}
			\centering
				\includegraphics[width=\linewidth]{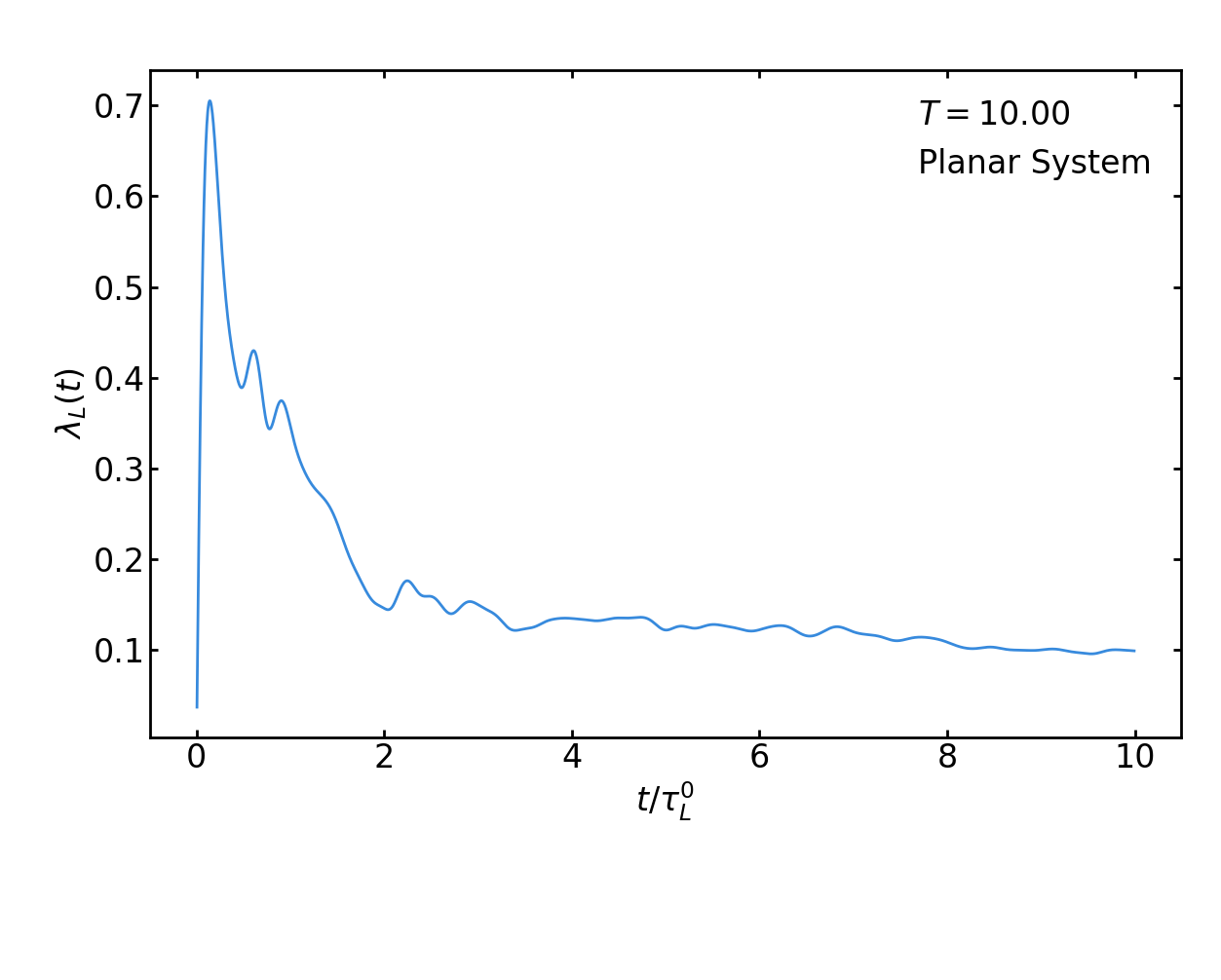}
			\caption{}\label{fig:Lyap_Time_Planar_T10}
		\end{subfigure}
		\begin{subfigure}[b]{0.45\textwidth}
			\centering
				\includegraphics[width=\linewidth]{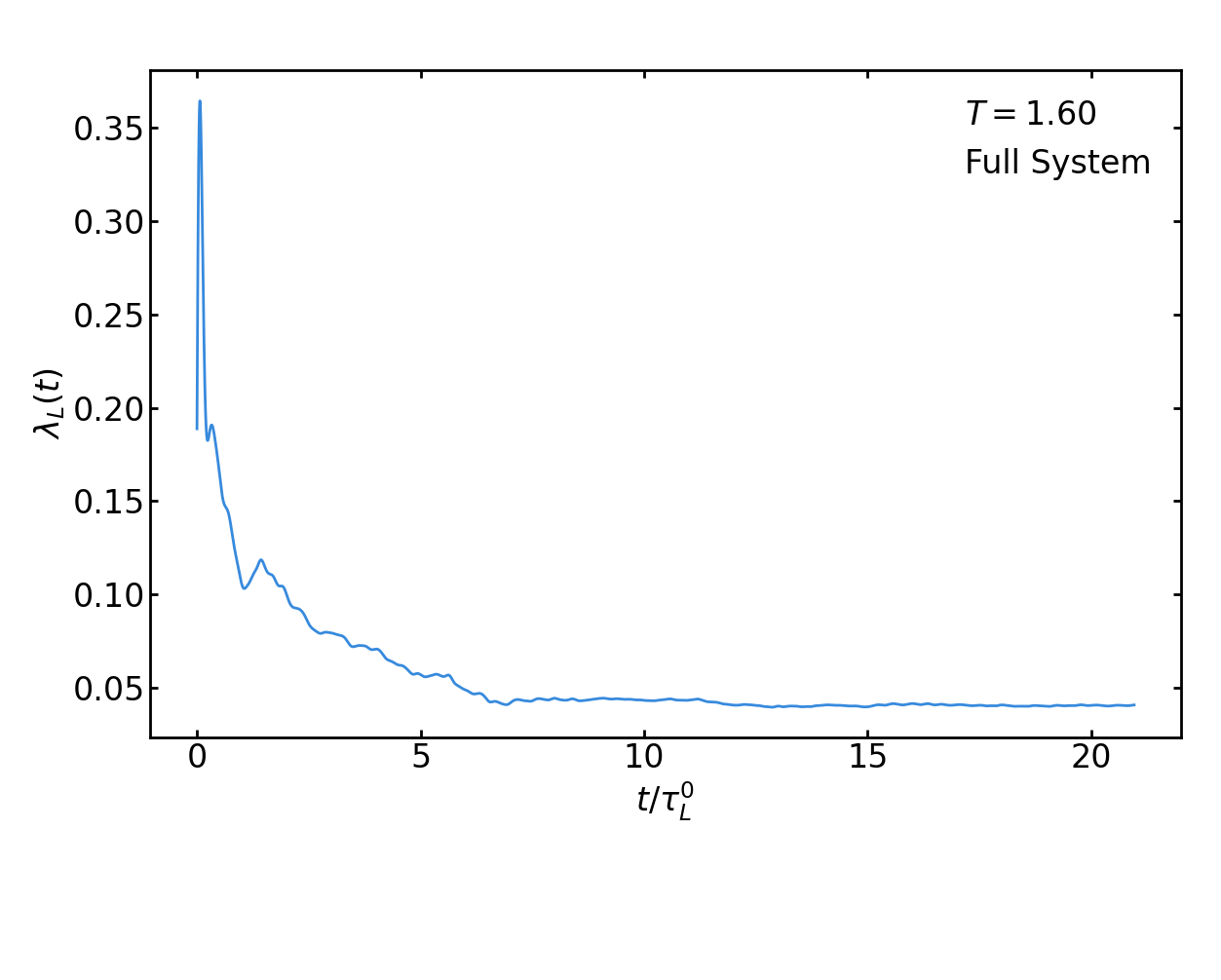}
			\caption{}\label{fig:Lyap_Time_NonPlanar_T1.6}
		\end{subfigure}
		\hfill
		\begin{subfigure}[b]{0.45\textwidth}
			\centering
				\includegraphics[width=\linewidth]{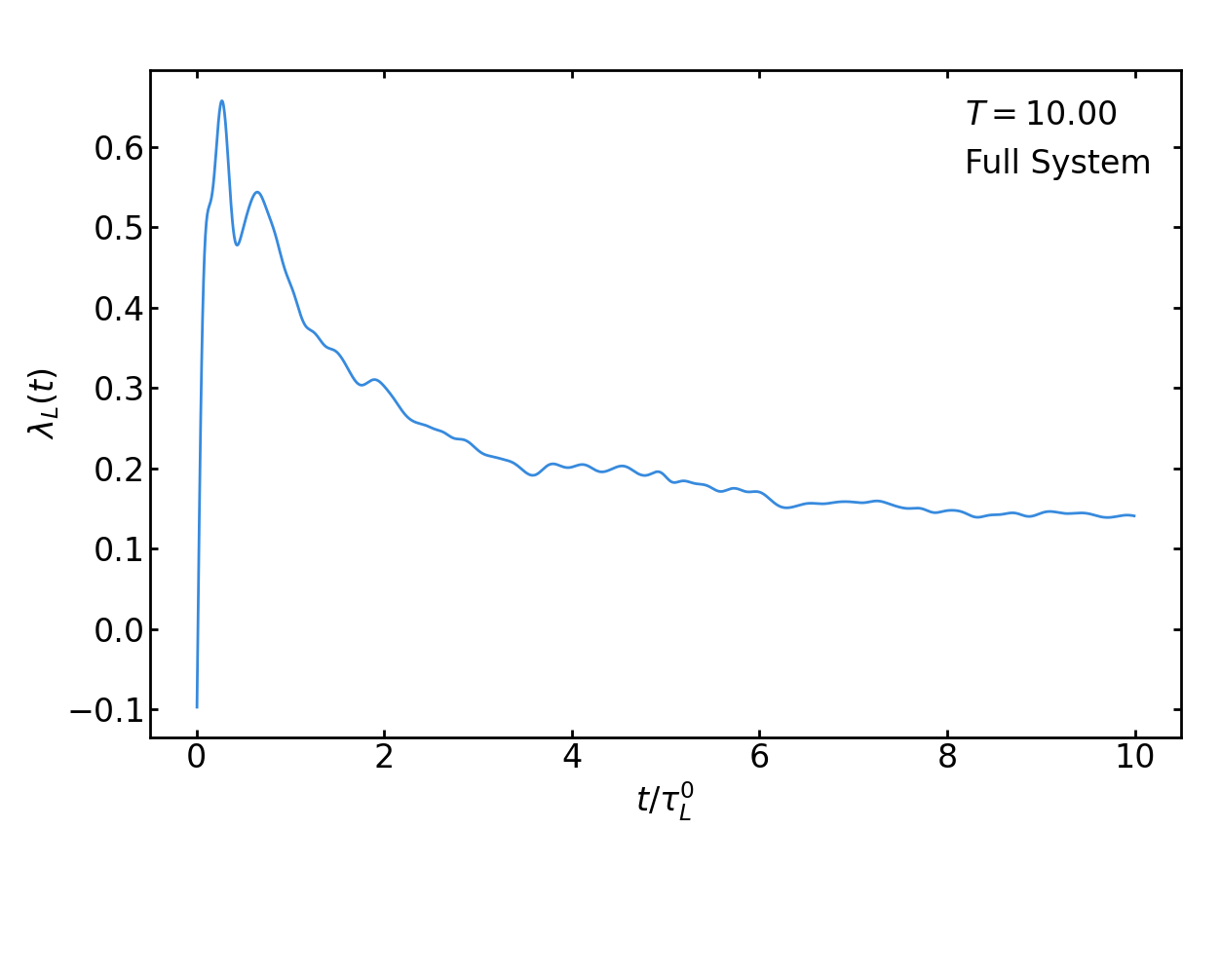}
			\caption{}\label{fig:Lyap_Time_NonPlanar_T10}
		\end{subfigure}
		\caption{$\lambda_L$ time series at $N=5$, $R=5$, $\mu=1/10$, $\lambda=1/5$. }
		\label{fig:Lyap_Time}
	\end{figure}

\begin{figure}[htbp]
			\centering
				\includegraphics[width=0.60\linewidth]{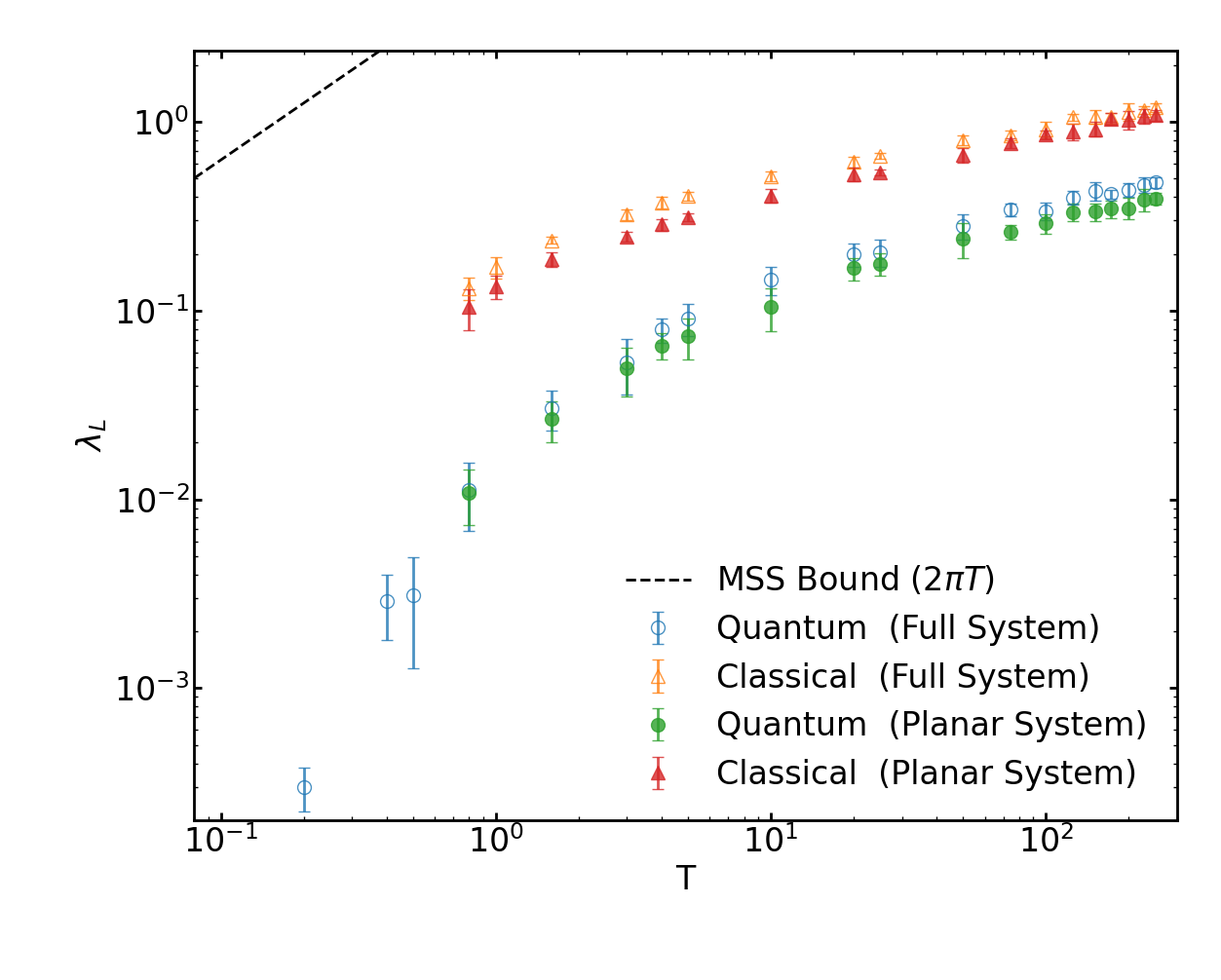}
		\caption{$\lambda_L$ as a function of temperature $T$ at $N=5$, $R=5$, $\mu=1/10$, $\lambda=1/5$.}
		\label{fig:Lyap_T_comparison}
	\end{figure}

In order to probe the chaotic dynamics, we have calculated the largest Lyapunov exponent, $\lambda_L$ by evaluating the exponential growth between two solutions whose initial ${\bar \xi}_a \equiv ({\bar X}_{lm}^{i}, {\bar P}_{lm}^{i})$ values differ by $\delta {\bar \xi}_a \equiv( \delta{\bar X}_{lm}^{i}, \delta{\bar P}_{lm}^{i})$, with the the norm ${\|\delta\xi(t)\|}\approx 10^{-5}$. Thus, for each of the seven to nine randomly selected initial condition at a fixed $T$, we performed a calculation of $\lambda_L$ as a function of time and subsequently averaged over them. In our simulations we have used time step-size $\Delta t \sim 10^{-3}$ at low temperatures $T \lesssim 0.2$ and $\Delta t \sim 10^{-4}\,, 10^{-5}$ for the rest. Some details of the steps followed in the evaluation of $\lambda_L$ are summarized in the Appendix \ref{sec:Lexp} The resulting time series for this process at $T=1.6$ and $T=10$ are presented in the Fig. \ref{fig:Lyap_Time}, where for convenience we use the scale of the Lyapunov time $\tau_L = \lambda_L^{-1}$ in the horizontal axis, which is evaluated from the classical Lyapunov exponent at the given temperature. We see that for both the planar and full interaction configurations there is a transient behavior within the first $2 \tau_L^{(0)}$ where $\lambda_L$ grows and decreases sharply and converges to a saturation value for $t \gtrsim 2 \tau_L^{(0)}$ in the former and $t \gtrsim 4 \tau_L^{(0)}$ in the latter case. 
One of our main results is given in Fig. \ref{fig:Lyap_T_comparison} where we show the variation of $\lambda_L$ with temperature.
\begin{figure}[htbp]
		\begin{subfigure}[b]{0.48\textwidth}
			\centering
				\includegraphics[width=\linewidth]{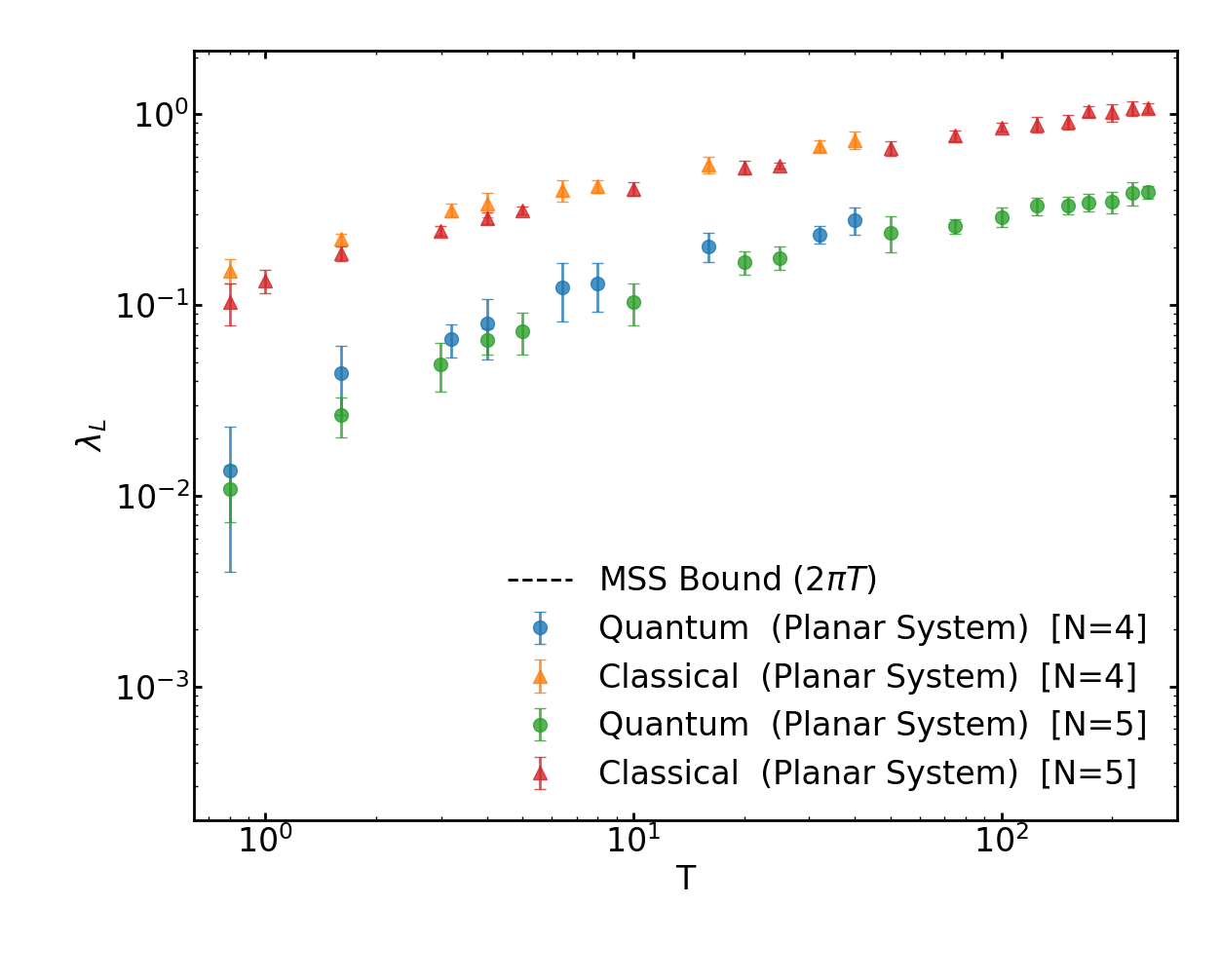}
			\caption{}\label{fig:Lyap_T_Planar_N45}
		\end{subfigure}
		\hfill
		\begin{subfigure}[b]{0.48\textwidth}
			\centering
				\includegraphics[width=\linewidth]{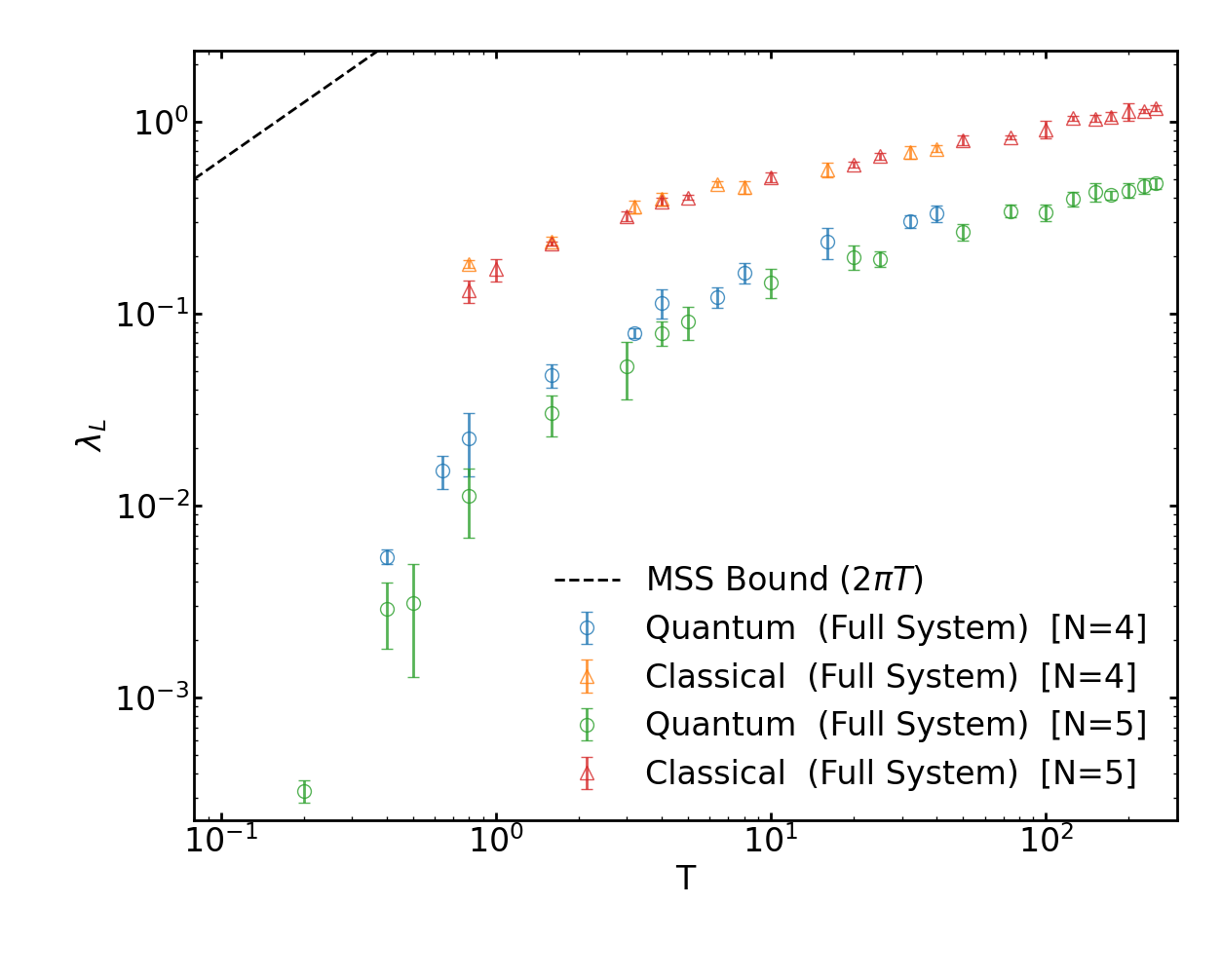}
			\caption{}\label{fig:Lyap_T_NonPlanar_N45}
		\end{subfigure}
		\caption{$\lambda_L$ as a function of temperature $T$ at $N=4$, $N=5$ comparison. $R=5$, $\mu=1/10$. (a) Planar Interaction and (b) Full Interaction.}
		\label{fig:Lyap_N4N5_Comparison}
	\end{figure}
From this figure, we see that the largest Lyapunov exponent $\lambda_L$ is vanishing around $T_c \lesssim 0.8$ for the planar and $T_c \lesssim 0.1$ for the full system. It is also observed that $\lambda_L$ values are increasing with temperature which means that quantum system is chaotic for $T\geq T_c$, and becomes more so with increasing temperature as naturally expected, while the quantum effects become dominant and starts to suppress chaos as the system gets colder. Thus the MSS bound \cite{Maldacena:2015waa} is respected across all temperatures  with $\lambda_L \rightarrow 0$ at a finite temperature. In Fig. \ref{fig:Lyap_N4N5_Comparison}, we present a comparison of the $N=4$ and $N=5$ results  from which we observe that  $\lambda_L$ scales quite alike in both cases for both the planar and full interaction configurations.

For comparison, we have also calculated the Lyapunov exponent for the corresponding classical systems (planar and full) using the solutions of the subsystem of equations \eqref{eq:eom-one-point}-\eqref{eq:p2} for the $1$-point functions with same set of initial conditions for temperatures $T \geq 1$, since as we see from Fig. \ref{fig:E_CDvT} that around\footnote{In the full interaction case separation of $\frac{1}{N} \braket{ Tr \, {\hat X}_i {\hat X}_i}$ for the classical configuration starts at a somewhat lower temperature $T \lesssim 0.8$.} $T \lesssim 1$ the classical configuration predicted by the displaced pure states starts to divert from the thermal equilibrium and hence are not reliable\footnote{We may still calculate the $\lambda_L$ at $T < 1$ for the classical case, bu the $\lambda_L$ values obtained may at most be lower bounds on their actual values.} for $T \lesssim 1$. These results demonstrate that the largest Lyapunov exponent for both the planar and full quantum system approach toward that of the respective classical system at higher temperatures $T \gtrsim 10^{2}$, while they remain safely below and do not show any tendency to exceed the latter. We also observe that the $\lambda_L$ values of the full system is a bit higher than that of the planar system across all temperatures.  

\begin{figure}[htbp]
\centering
\includegraphics[width=0.65\linewidth]{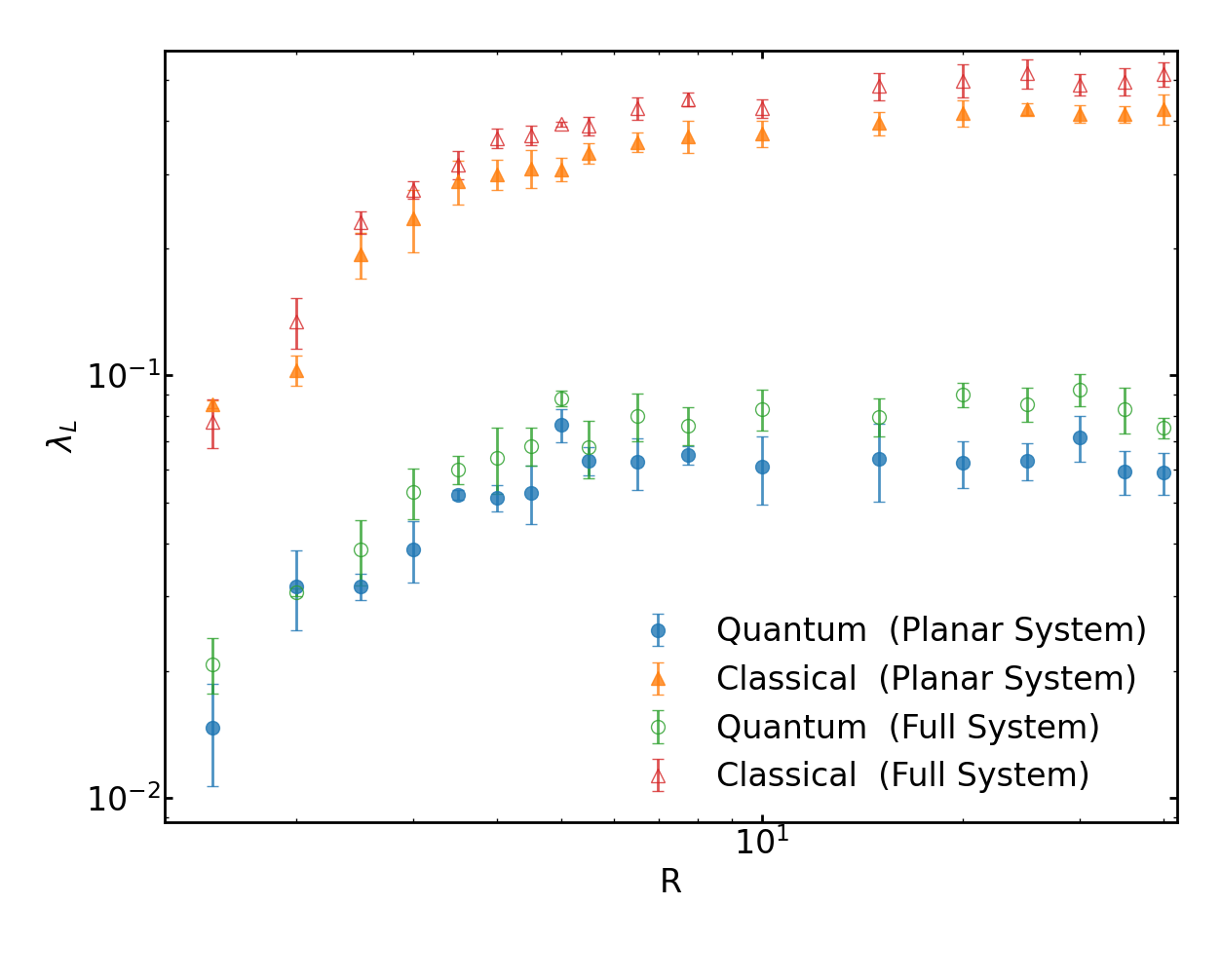}
\caption{$\lambda_L$ versus the radius $R$ of $S_F^2$ at $N=5$, $T=5$, $\mu=1/10$, $\lambda=1/5$.}
\label{fig:Lyap_R_comparison}
\end{figure}

Profile of $\lambda_L$ with respect to the radius of $R$ of $S_F^2$ is plotted  in Fig. \ref{fig:Lyap_R_comparison} at a fixed temperature $T=5$. We see that $\lambda_L$ remains essentially steady for $R \gtrsim 3$, while it starts to decrease slowly as $R$ goes below $3$. This results suggests that the chaotic dynamics is robust against the change in the radius of the fuzzy sphere within a large interval of the latter, while it is also plausible to observe some decrease in $\lambda_L$ as $R$ gets smaller, since $l(l+1)/R^2$ term in $\omega_l$ behaves as an additional effective mass-squared contribution due to curvature which increases with decreasing $R$, driving the system to be more harmonic and less chaotic. 

\section{Entanglement}
\label{sec:ent}

We now turn our attention to examine the dynamics of the entanglement within the present model, by suitably separating it into two subsystems in a sequential manner. In general, computing the entanglement entropy in real time is regarded as a difficult problem in the literature \cite{Buividovich:2018scl}. However, the point is that, although the Gaussian state approximation does not lead to unitary time evolution, it does ensure that pure states evolve into pure states in time, which is a fact already emphasized and utilized in \cite{Buividovich:2018scl} in the context of the BFSS model. For completeness, we give a brief proof of this fact in Appendix \ref{sec:Pure_pure} It is precisely this property of the GSA that makes the computation of entanglement entropy feasible for our model too. We may also note that there are real-time techniques applied to coupled harmonic oscillators, as well as free bosonic and fermionic models, that is models with only quadratic potentials in order to compute the entanglement entropy in \cite{Bombelli:1986rw, Srednicki:1993im, Casini:2009sr, Huerta:2011qi, Nishioka:2018khk}. The method we are going to outline and use below relies on the use and evaluation of the reduced covariance matrix and is therefore distinct and readily applies to the interacting theories, but at the same time it complements the aforementioned approaches in a sense since it has the fuzzy structure naturally playing the role of the lattice discretization while keeping the underlying symmetries ($SU(2)$ in the present case) intact.

As it is already well-known (See for instance \cite{Bertlmann:2023wol} and references therein as an introduction to the extensive literature) entanglement in a quantum system is a notion that has been developed to understand and quantify how the quantum information associated to the degrees of freedom become "scrambled" and most frequently characterized or measured by the entanglement entropy. To compute the latter, it is sufficient (but not necessary in general \cite{Witten:2021jzq}) that the Hilbert space $\mathcal{H}$ of the system can be decomposed as a direct product of two sub-Hilbert spaces $\mathcal{H}_A$ and $\mathcal{H}_B$ as $\mathcal{H} = \mathcal{H}_A \otimes \mathcal{H}_B$, i.e. the system has a bipartite composition \cite{Bertlmann:2023wol, Witten:2018zva}. It is understood that this separation is in general not unique and another decomposition will yield, in general a different entanglement entropy as may be naturally expected. In quantum systems exhibiting chaos, it is anticipated that the information gets rapidly "scrambled" between the states supported in $\mathcal{H}_A$ and $\mathcal{H}_B$ and quickly approaches toward a saturation value.  If $\mathcal{H}$ is finite-dimensional, the latter is called the "Haar-Scrambled" entanglement entropy in the literature \cite{Kunihiro:2008gv}.

Performing a partial trace over the full system's density matrix, $\hat{\rho}$, i.e. tracing out either of the $\mathcal{H}_A$ or $\mathcal{H}_B$ yields a reduced density matrix. For definiteness, we may consider tracing over the latter, which yields the reduced density matrix on $\mathcal{H}_A$ as $\hat{\rho}_A = \mathrm{tr}_B\, \hat{\rho}$. The entanglement entropy between the subsystems $A$ and $B$ is then nothing but the von Neumann entropy in $\mathcal{H}_A$, namely $S_A = -\mathrm{tr}\,(\rho_A \ln \rho_A)$. Reversing the argument yields $S_B = -\mathrm{tr}\,(\rho_B \ln \rho_B)$ and it can be shown that $S_A= S_B$ as expected (see, for instance, \cite{Bertlmann:2023wol,Witten:2018zva} for a general discussion).

If $\hat{\rho}$ stands for a density matrix of a Gaussian state, then the reduced density matrix $\hat{\rho}_A = \mathrm{tr}_B\,\hat{\rho}$ obtained after the partial trace over the subspace $\mathcal{H}_B$ is also a Gaussian density matrix and can be found by restricting the one- and two-point functions characterizing $\hat{\rho}$, or more precisely, $W_{\hat {\rho}}(\xi)$ to those in the $\mathcal{H}_A$ subsystem\footnote{Shortest way to see this is to use the Wigner characteristic function $\chi(\xi) = tr (\hat {\rho} e^{-i {\hat \xi}^T \Omega \xi})$, the Fourier transform of $W_{\xi}({\hat \rho})$, which for a Gaussian state is of the form $\chi(\xi) = e^{-\frac{1}{2} \xi^T \Omega^T \Sigma \Omega \xi + i {\bar \xi}^T \Omega \xi}$.  Using the notation $\xi \equiv (\xi^{A}, \xi^B)$, $\Omega \equiv \mbox{Diag}(\Omega^{A}, \Omega^{B})$, we have for $\hat{\rho}_A$ corresponding to $\chi_A(\xi_A) = tr_A (\rho_A e^{-i {\hat \xi}_A^T \Omega \xi_A})  = tr ({\hat \rho}  e^{-i {\hat \xi}_A^T \Omega \xi^{A}} \otimes I_B)$, which means that $\chi_A(\xi_A)= \chi (\xi_A, \xi_B=0)$ and this immediately yields $\chi(\xi_A) = e^{-\frac{1}{2} \xi_A^T \Omega_A^T \Sigma_A \Omega_A \xi + i  {\bar \xi}_A^T \Omega_A \xi_A}$, which is a Gaussian with the covariance matrix $\Sigma_A$ obtained from restricting $\Sigma$ to the subsystem $A$. This is equivalent to integrating $W_{\xi} ({\hat \rho})$ over the coordinates $\xi_B$, which is a lengthy calculation and therefore omitted here.} \cite{Weedbrook:2011wxo}. The important fact here is that, regardless $\hat{\rho}$ is pure or mixed, $\hat{\rho}_A$ or the corresponding $W^{A}_{\hat {\rho}}(\xi)$, generally describes a mixed Gaussian quantum state with non-vanishing von Neumann entropy, which is the entanglement entropy between the subsystems $A$ and $B$.
\begin{figure}[htbp]
	\centering
	\begin{subfigure}[b]{0.48\textwidth}
		\centering
		\includegraphics[width=\linewidth]{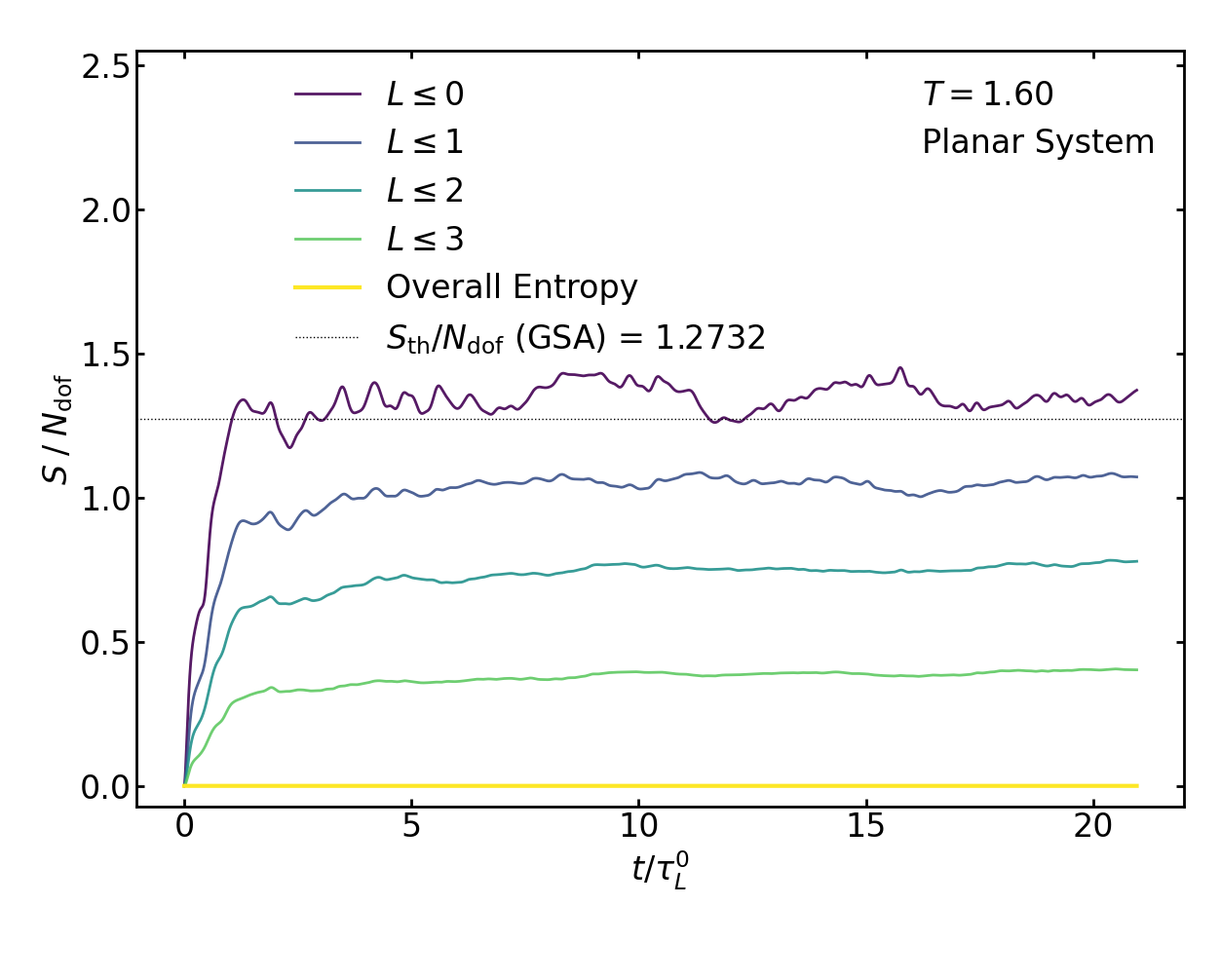}
			\caption{}\label{fig:SET16}
		\end{subfigure}
		\hfill
		\begin{subfigure}[b]{0.48\textwidth}
			\centering
			\includegraphics[width=\linewidth]{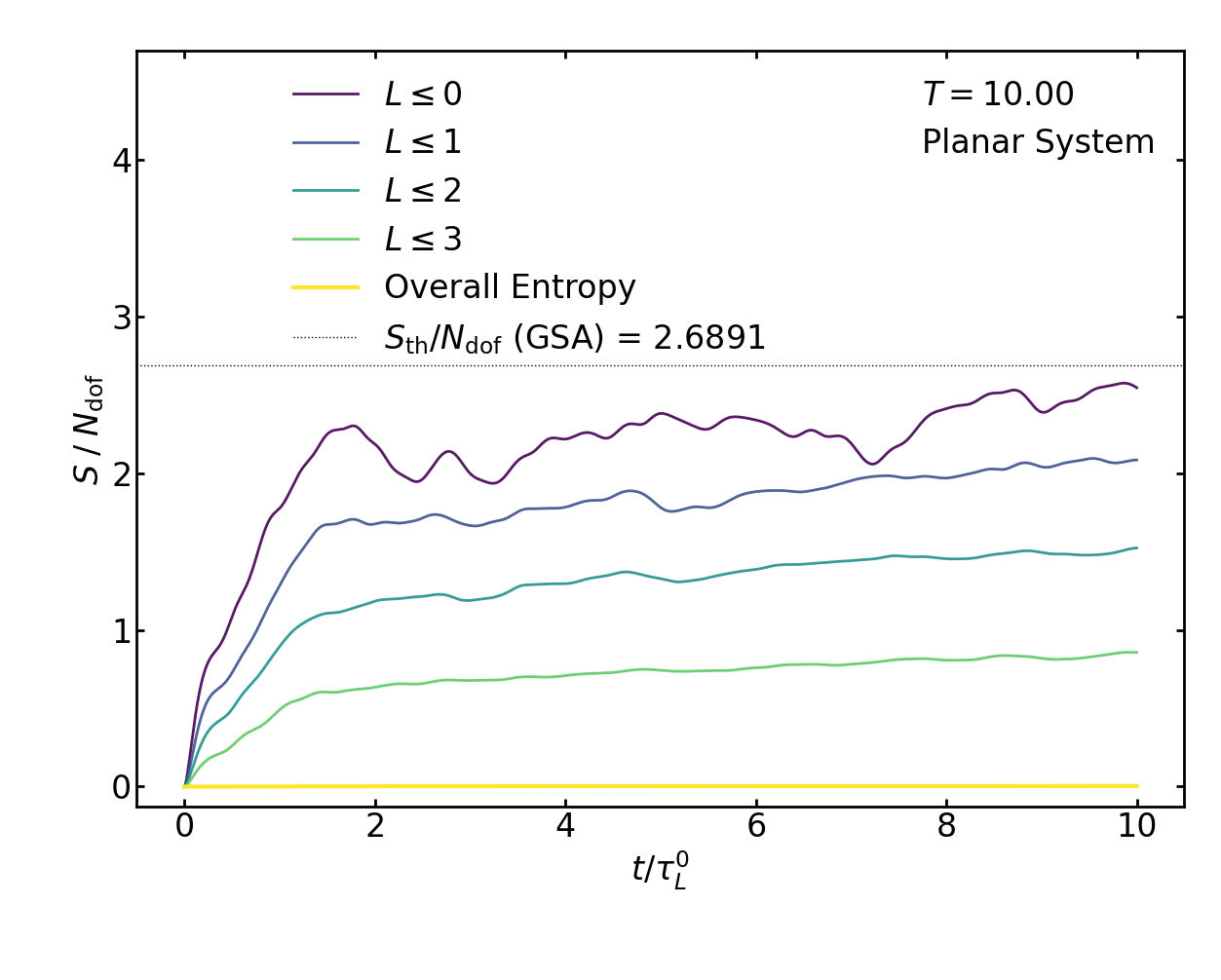}
			\caption{}\label{fig:SET10}
		\end{subfigure}
		\centering
		\begin{subfigure}[b]{0.48\textwidth}
			\centering
			\includegraphics[width=\linewidth]{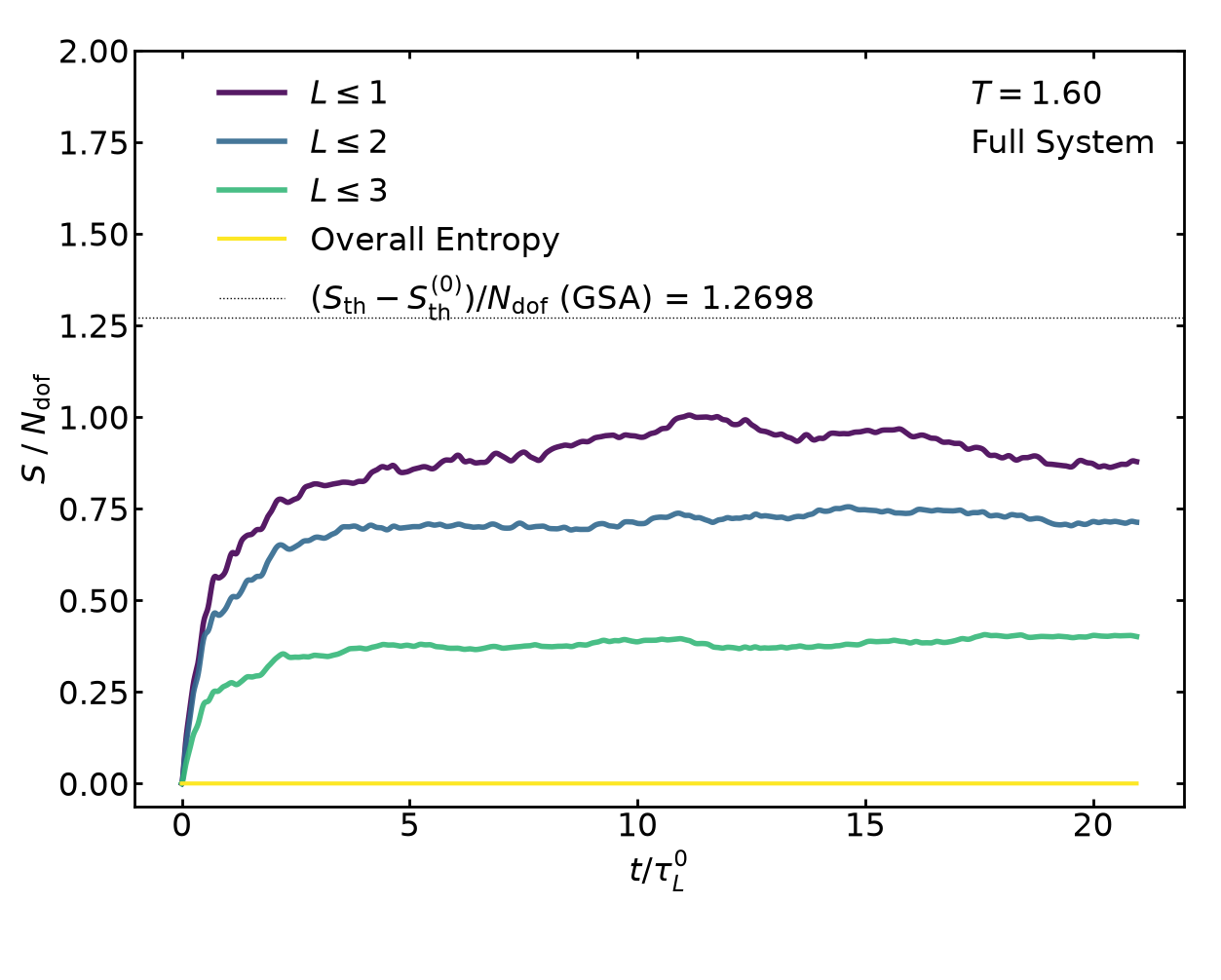}
			\caption{}\label{fig:SETNP16}
		\end{subfigure}
		\hfill
		\begin{subfigure}[b]{0.48\textwidth}
			\centering
			\includegraphics[width=\linewidth]{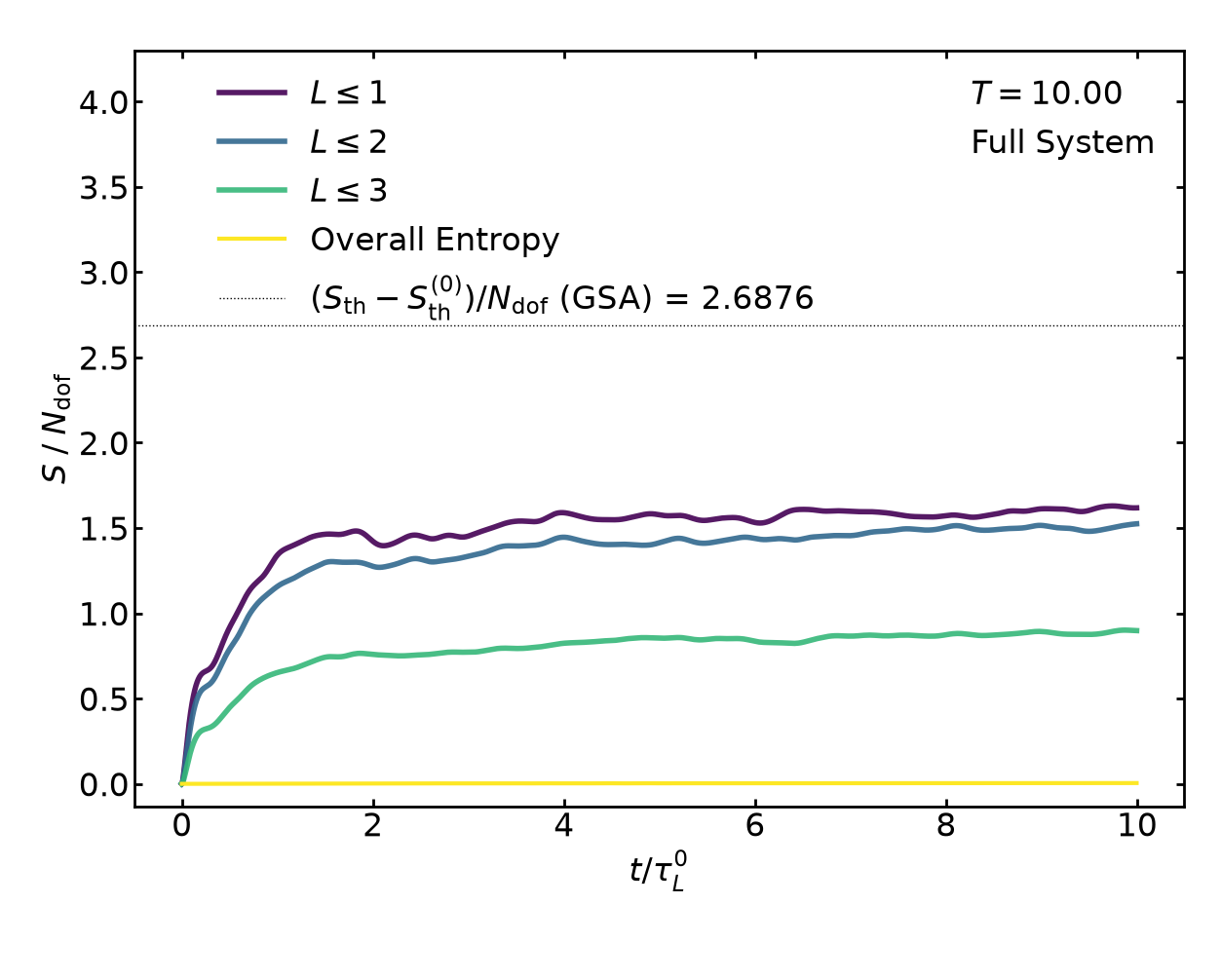}
			\caption{}\label{fig:SETNP10}
		\end{subfigure}
		\caption{$S/N_{dof}$ as a function of time.}
		\label{fig:Entanglement_time_series}
	\end{figure}
Since, the Hilbert space $\mathcal{H}$ our system is composed of $2N^2$ canonical modes generated by the canonically conjugate pairs $\hat{X}_i^{lm}$, $\hat {P}_i^{lm}$ labeled with $l=0,1,\cdots\, N-1$, $|m| \leq l$, $i=1,2$ with the $l^{th}$ mode of each field being $2l+1$-fold degenerate in its contribution to the entropy, we may consider a sequence of bipartite systems
\be
\mathcal{H} = \mathcal{H}_{\leq L} \otimes \mathcal{H}_{>L} \,,
\label{eq:bipartite}
\ee
where $\mathcal{H}_{\leq L}$ ($ \mathcal{H}_{>L} $) denote the Hilbert space composed of the $2(L+1)^2$ ($2(N^2- (L+1)^2)$) modes with $l=0,1,\cdots\, L$ ($l= L+1, \cdots N-1$). Note that, each subspace has the respective modes for both of the fields $X_1$ and $X_2$. We may as well form several other bipartite systems by randomly allocating some of the $N^2$ modes in the first subsystem and the rest in the second (keeping the same modes for both fields within a subsystem). Nevertheless, the separation in \eqref{eq:bipartite} is useful to capture and illustrate several features of the entanglement and to measure how the associated entropy responds to the partitions. To calculate the latter, we start from the Gaussian state which is of the form $| \bar{X}_i^{lm}, \bar{P}_i^{lm} \rangle \langle \bar{X}_i^{lm}, \bar{P}_i^{lm} |$. As emphasized earlier, this state is pure but not in thermal equilibrium and its time evolution within the GSA is governed by the equations \eqref{eq:eom-one-point}-\eqref{eq:p2}, \eqref{eq:eom-two-point-xx}-\eqref{eq:eom-two-point-pp-22} under which it remains pure with the von Neumann entropy vanishing at all times. For a given set of values of the parameters $T, R, \mu, N$, we compute the entanglement entropy $S_{\leq L}$ as a function of time by evaluating the von Neumann entropy from the symplectic eigenvalues of the reduced covariance matrix $\Sigma_{\leq L}$ obtained from $\Sigma$ by restricting it to the first $(L+1)^2$ modes. Practically, $\Sigma_{\leq L}$ is formed by omitting the rows and columns of the latter for $l > L$. Thus, $\Sigma_{\leq L}$ is a "bona fide" covariance matrix of a (mixed) Gaussian state and the formula \eqref{eq:entropy1} applies with $N-1$ replaced by $L$ and $f_l$ computed from the eigenvalues of $\Omega_{\leq L}\Sigma_{\leq L}(t)$ at time $t$. Our findings are illustrated in Fig. \ref{fig:Entanglement_time_series}, where we show the time series for entanglement entropy $S_{\leq L}/N_{dof}$ with $N_{dof} = 2(L+1)^2$ computed at $N=5$, $\lambda= \frac{1}{5}$, $R=5$, $\mu =\frac{1}{10}$ at the bipartite levels $L=0,1,2,3$ for the planar and for $L=1,2,3$ for the full case (recalling that the $l=0$ mode of the full system decouples from the rest completely) for $T=1.6$ and $T=10$. We observe that the entanglement entropy exhibits the expected scrambling features, namely a fast and roughly a linear growth at early times which tend toward a saturation at later times. Fast growth continues about $2 \tau_L^{0}$, before it lends to a saturation value with small variations in time. In the figure, we have also plotted the von Neumann entropy for the initial pure state, which remains zero at all times apart from very small fluctuations due to numerical discritization of time, confirming that pure states remain pure in time. As $L$ increases saturation value of $S_{\leq L}$ gets smaller which is expected since $ \mathcal{H}_{\leq L} \rightarrow \mathcal{H}$ and hence $S_{\leq L}$ tends to the von Neumann entropy as $L$ increases, or equally well from $S_{\leq L} = S_{>L}$ for the entanglement entropy. As a consistency check we also verified the latter fact numerically.

We see that for $(L+1)^2 \ll N^2$ the saturation value of $S_{\leq L}/2(L+1)^2$ for the planar model tends to the thermal entropy (i.e. the von Neumann entropy at thermal equilibrium) per degree of freedom, as can be seen from the plots in Fig. \ref{fig:SET16} and \ref{fig:SET10} with $S/2N^2$  plotted as a straight line, while we observe from \ref{fig:SETNP16} and \ref{fig:SETNP10} that in the full interaction configuration, saturation values are somewhat below the thermal entropy $(S-S_0)/2N^2$ (here since the zero mode decouples from the dynamics, its contribution to the total thermal entropy is subtracted). Thus, especially in the planar case our results support the real-time thermalization of pure states, which in a sense implies at the quantum level a rather non-trivial equivalence between pure states on one side indicating a microcanonical ensemble and a mixed state on the other side describing thermal equilibrium at $T$ i.e .canonical ensemble \cite{Sekino:2008he,Buividovich:2018scl}.

In connection to our earlier comments in section $2$ regarding the $l$-dependent non-planar contribution to dispersion indicative of UV/IR mixing, we may note that our results for the entanglement entropy for the model with the full interaction term takes this $l$-dependent non-planar contribution $-\lambda D(l)$ to the frequencies, $\omega_l$, completely into account via the solutions of the self-consistency equations \eqref{eq:consistency-system} and as noted above and observed from comparing \ref{fig:SET16} and \ref{fig:SET10} with \ref{fig:SETNP16} and \ref{fig:SETNP10}, respectively, its overall effect appears to be a slight decrease in the saturation values of the entanglement entropy in comparison to the purely planar case.
  
\begin{figure}[htbp]
	\centering
	\begin{subfigure}[b]{0.48\textwidth}
		\centering
		\includegraphics[width=\linewidth]{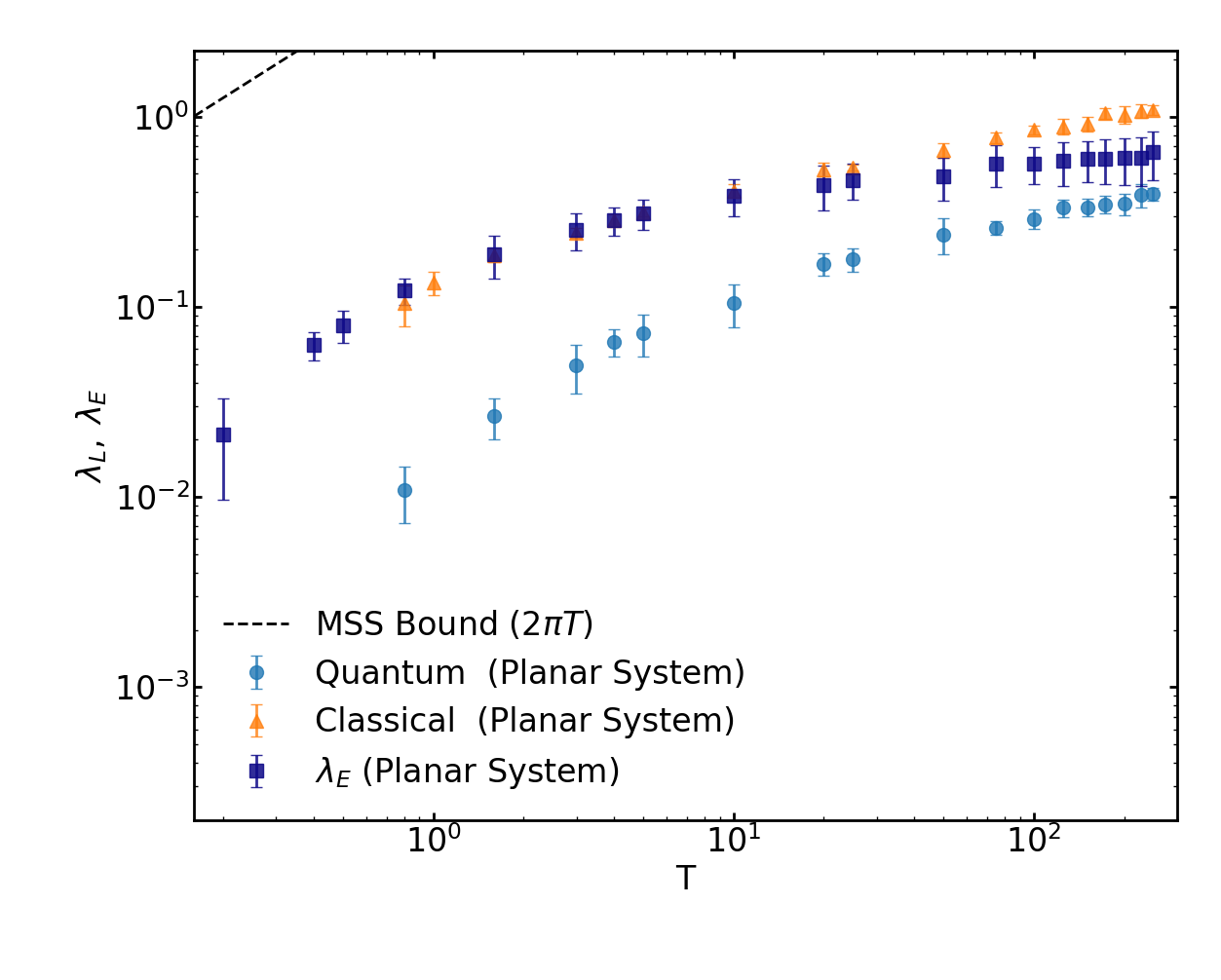}
		\caption{}\label{fig:LLP}
	\end{subfigure}
	\hfill
	\begin{subfigure}[b]{0.48\textwidth}
		\centering
		\includegraphics[width=\linewidth]{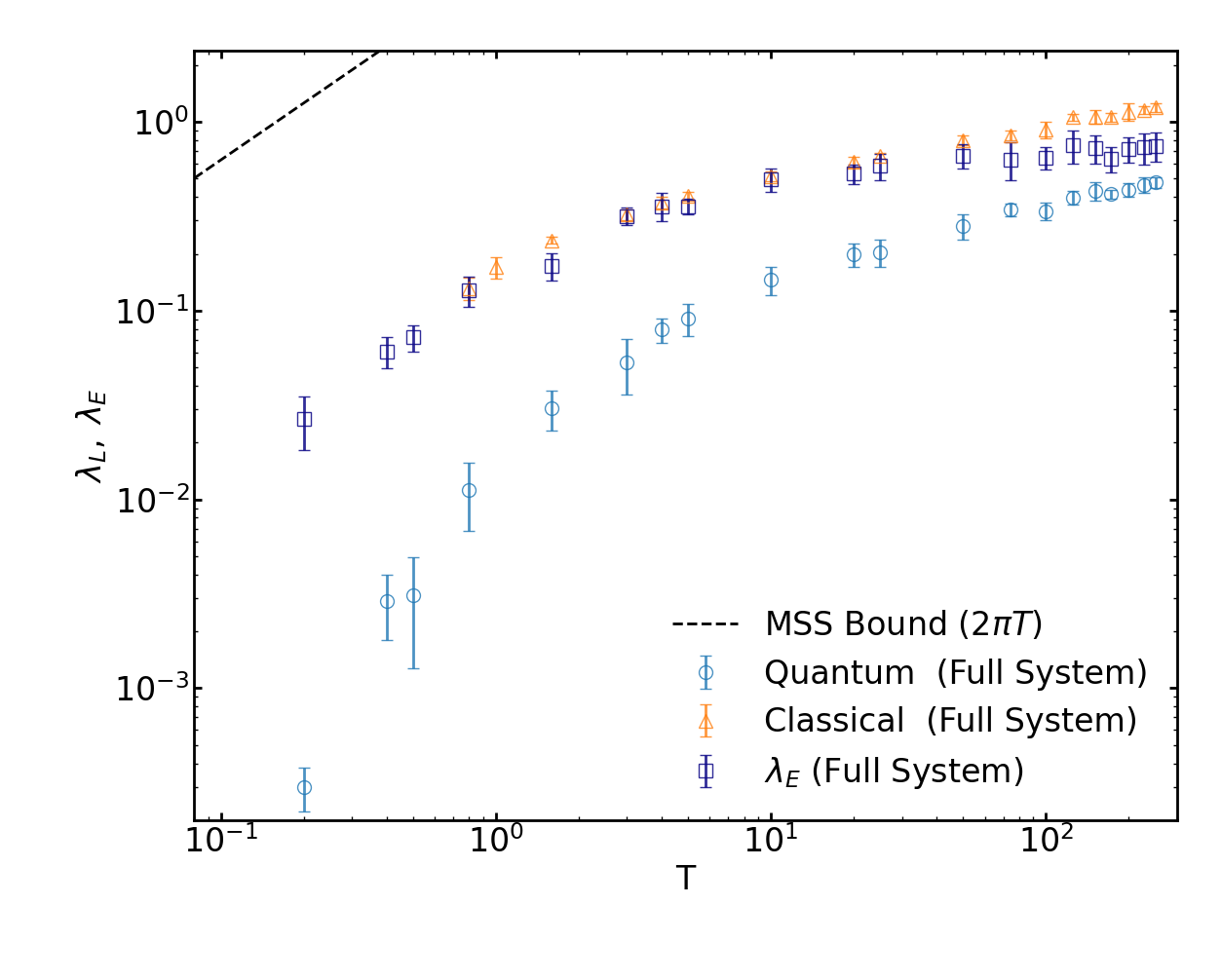}
		\caption{}\label{fig:LLNP}
	\end{subfigure}
\caption{$\lambda_L$ and $\lambda_E$ (Averaged over all available $\mathcal{H}_{\leq L}$ subsystems) as a function of temperature.}
\label{fig:LET}
\end{figure}

Following \cite{Buividovich:2018scl}, we have also defined an entanglement saturation time $\tau_E$ by fitting the profiles of entanglement entropy to a function of the form ${\tilde S} \tanh (t/\tau_E)$. Here ${\tilde S}$ adjust for fits in each sub-partition $\mathcal{H}_{\leq L}$ to capture the saturation value of the entanglement entropy and at the same time lets us to compare $\tau_E$ at different temperatures and $L$ values. We may introduce the inverse of the entanglement saturation time $\lambda_E := \tau_E^{-1}$, which can be compared with the Lyapunov exponent $\lambda_L$. We plot $\lambda_L$ and $\lambda_E$ as a function of $T$, where $\lambda_E$ at a given $T$ is obtained by averaging its values over all the available bipartions ($L=0,1,2,3$ for the planar and for $L=1,2,3$ the full configurations), since we already recognize that the saturation time for entanglement entropy does not vary much with temperature, this averaging is quite suitable. We see from Fig. \ref{fig:LET} that $\lambda_E$ is quite close to the classical value of $\lambda_L$ at all temperatures  for which the latter is evaluated (i.e. $T \gtrsim 1 $). We may note the feature that $\lambda_E > \lambda_L$ across all temperatures while they become close to each other at higher temperatures $T \approx 10^{2}$. In other words, the entanglement saturation time is always shorter than the Lyapunov time and the difference becomes more significant as the system gets colder. 

\section{Conclusions and Outlook}

In this work, we have applied the Gaussian state approximation (GSA) to investigate the real-time quantum dynamics of a bosonic two-matrix model on the fuzzy sphere. By truncating the Heisenberg equations of motion via Wick's theorem, we derived a closed system of nonlinear coupled ordinary differential equations governing the one- and two-point correlation functions in the most general Gaussian state. 

A central component of our analysis was the thermodynamic characterization of the system. By maximizing the von Neumann entropy of the most general Gaussian density matrix at fixed energy, we established thermal equilibrium conditions and derived the equation of state. In order to probe the chaotic dynamics of the model, we constructed dynamical initial configurations drawn from classical Gaussian ensembles with variances calibrated so that  the dynamical configurations are close to the thermal equation of state, which we confirmed by observing their agreement with the equilibrium energy and total coordinate dispersion $\frac{1}{N}\langle \mathrm{Tr}\, \hat{X}_i \hat{X}_i \rangle$ versus temperature curves. Averaging over several such initial configurations, we computed the largest Lyapunov exponent $\lambda_L$ across a wide range of temperatures in the 't Hooft limit at matrix level $N = 5$ and also at $N=4$.

Our results reveal a clear picture of the chaotic dynamics within this model. At higher temperatures, both the classical and quantum dynamics are found to be comparably and significantly chaotic, consistent with the expectation that the quantum effects become subdominant in this regime. As the temperature is lowered, $\lambda_L$ decreases monotonically and vanishes at a finite temperature, which we have estimated to be $T_c \approx 0.8$ and $T_c = 0.1$ for the planar and full interactions respectively at $N=5$, (an similar $T_c$ values at $N=4$) signaling the suppression of chaos by quantum fluctuations and the crossover to a non-chaotic regime as the system becomes colder. We may emphasize that, this behavior complies with the MSS bound $\lambda_L \leq 2\pi T $ at all temperatures and also implies that the GSA captures essential features of quantum effects on chaos. 

Finally, we investigated the entanglement dynamics of the system through a sequence of bipartitions $\mathcal{H} = \mathcal{H}_{\leq L} \otimes \mathcal{H}_{>L}$, where the subsystem $\mathcal{H}_{\leq L}$ collects the $2(L+1)^2$ canonical modes with angular momentum index $l \leq L$. The entanglement entropy $S_{\leq L}$ was computed from the symplectic eigenvalues of the appropriate reduced covariance matrix. For both the planar and the full interaction and across bipartite levels ($L = 0, 1, 2, 3$, $L = 1, 2, 3$, respectively) the entanglement entropy exhibits the hallmark behavior expected of a fast scrambler: a sharp initial growth followed by saturation toward a constant value. This behavior, common to strongly chaotic quantum systems, further corroborates the chaotic nature of the model and underscores the consistency of the GSA as a framework for capturing quantum information-theoretic features of matrix models.

It will also be interesting to explore connections of our results with the recently emerging new measures of chaos and entanglement such as the Krylov complexity \cite{Parker2019,Balasubramanian2022,Huh:2024ytz,Nandy2025}. Several other aspects of the application of the GSA to matrix models with fuzzy geometry are under investigation, including the dynamics and entanglement in the BMN model \cite{Berenstein:2002jq, Dasgupta:2002hx,Dasgupta:2002ru} and will be reported \cite{Soon} in the near future.

\vskip 1em

\section*{Acknowledgments}

The authors acknowledge the support of Türkiye Bilimsel ve Teknolojik Araştırma Kurumu (TÜBİTAK) under the 2515-COST Action project with the contract number 124F284 and 
the COST Action 21109 CaLISTA supported by COST (European Cooperation in Science and Technology). The numerical calculations reported in this paper were fully performed at TÜBİTAK ULAKBİM, High Performance and Grid Computing Center (TRUBA resources). S. K. has benefited from discussions with V. P. Nair, D. Karabali, B. Dolan on several occasions and he thanks them for their constructive comments and suggestions. S. K. has also benefited from the discussions with the late professor A.P Balachandran during the early stages of this work and gratefully acknowledges the mentorship and friendship he has shown for over quarter of a century. This work is dedicated to his memory with love and respect.

\appendices

\section{Fuzzy Sphere and  the Polarization Operators}
\label{sec:polarization-operator-basis}

\subsection{Fuzzy Sphere}

Definition of the fuzzy sphere \cite{Madore:1991bw} is quite well known in the literature. Here we give the basic definition and refer the reader to \cite{Balachandran:2005ew, Ydri:2016dmy} and an extensive list of references there in for its various aspects.

Let us denote the $SU(2)$ representations in the spin-$j=\frac{N-1}{2}$ irreducible representation by $L_a$ $(a=1,2,3)$. They satisfy the commutation relations  
\be
\lbrack L_a,, L_b \rbrack = i \epsilon_{abc} L_c \,.
\ee
The fuzzy two-sphere, $S_F^2$ of radius $R$ at the matrix level $N$ is defined in terms of the rescaled generators of $SU(2)$ 
\be
\chi_a := \frac{R}{\sqrt{j(j+1)}} L_a \,,
\ee
and their commutation relations are given as 
\be
\label{NnCmm}
\lbrack \chi_a , \chi_b \rbrack = i \frac{R}{\sqrt{j(j+1)}} \epsilon_{abc} \chi_c.
\ee
They satisfy $\chi_a^2 = R^2 1_N$. As $N$ goes to infinity, the standard commutative $S^2$ is recovered. 

An $N \times N$ matrix, say $M  \in Mat(N)$, is an element of the fuzzy sphere $S_F^2$ at the matrix level $N$.  Derivations on the fuzzy sphere are given by the adjoint action of $L_a$'s on $M$:
\be
ad L_a \, M : = \lbrack L_a \,, M \rbrack \,,
\ee
and they obey the Leibniz rule.  In particular, the Laplacian on the fuzzy sphere takes the double commutator form ${\cal L}^2 M :=  \frac{1}{R^2} \lbrack L_a \,, \lbrack L_a \,, M \rbrack \rbrack$.

Any $N \times N$ matrix $M$ can be expanded in a basis provided for $Mat(N)$. For instance, $M$ can be freely generated by taking as many products of $\chi_a$ as may be necessary:  $M = M_{i_1 i_2 \cdots i_k} \chi_{i_1} \chi_{i_2} \cdots  \chi_{i_k}$, where sum over the repeated indices is implied. A particularly useful basis is that of the polarization operators \cite{Varshalovich:1988ifq}. They can be thought of as the analogue of the spherical harmonics on the fuzzy sphere. They form a basis for $N \times N$-matrices, in a manner similar to spherical harmonics forming a basis of ${\mathbb C}^\infty$ functions on the sphere. Concretely, the polarization operators $T_{lm}$ are $ N \times N$ matrices  with $l = 0, 1, \ldots, N-1$ and $m = {-l,-(l-1) \cdots, l-1,l}$. They obey the orthogonality relation
\begin{align}
	Tr (T_{lm} T^\dagger_{l'm'}) = \delta_{ll'} \delta_{mm'} \,,
	\label{eq:o1}
\end{align}
where the Hermitian conjugate is given by
\begin{align}
	\label{eq:hermitian-conjugate}
    {T}_{lm}^{\dagger} = (-1)^m T_{l,-m}.
\end{align}
Completeness relation, or the resolution of identity, takes the form
\begin{align}
	\sum_{m} T_{lm} T^\dagger_{lm} = \frac{(2l+1)}{N} \mathbb{I}_N \,.
	\label{eq:resid1}
\end{align}
This easily follows by noting that, being summed over the index $m$, the l.h.s must be a rotational invariant and therefore must be proportional to the identity matrix $\mathbb{I}_N$, while the constant of proportionality is fixed by taking the trace of both sides and using \eqref{eq:o1}. 

We may therefore write $M = \sum_{l, m} M_{lm} T_{lm}$ for any $M \in Mat(N)$. 

Derivations and the Laplacian on $S_F^2$ in this basis take the form
\begin{align}
	\mathcal{L}_3 T_{lm} &= [L_3, T_{lm}] = m T_{lm}\,,\nn \\
	\mathcal{L}_\pm T_{lm} &= [L_\pm, T_{lm}] = \sqrt{(l\mp m)(l \pm m+1)} T_{l m \pm 1},\nn \\
	\mathcal{L}^2 T_{lm} &= [L_i, [L_i, T_{lm}]] = l(l+1) T_{lm}.
	\label{eq:dervSF2}
\end{align}

Product of two polarization operators expand as \cite{Varshalovich:1988ifq}
\begin{equation}
T_{l_1 m_1}T_{l_2 m_2} = \sum^{N-1}_{l} \sum_{m= -l}^l \sqrt{(2l_1+1)(2l_2+1)}\,(-1)^{N-1+l}\sixj{l_1}{l_2}{l}{\frac{N-1}{2}}{\frac{N-1}{2}}{\frac{N-1}{2}}C^{lm}_{l_1 m_1 l_2 m_2}\,T_{lm}\,.
\label{eq:Texp}
\end{equation}

\subsection{Real Polarization Operator Basis}
\label{ssec:real-polarization-operator-basis}

For computational practicality in computer codes, we have used the real polarization operator basis in our work. The real polarization operator basis is defined analogously to the real spherical harmonics in the following manner 
\begin{equation}
    Z_{lm} = 
    \begin{cases}
      \frac{1}{\sqrt{2}} (T_{lm} + T_{lm}^\dagger) & \text{if } m > 0, \\
        T_{l0} & \text{if } m = 0 , \\
        \frac{i}{\sqrt{2}} (T_{lm} - T_{lm}^\dagger) & \text{if } m < 0 \,, 
    \end{cases}
\end{equation}
It is readily seen that then  $Z_{lm}^\dagger  = Z_{lm}$.  Transformation between $Z_{lm} $ and $T_{lm}$ can also be compactly expressed as,  
\begin{align}
Z_{lm} = \sum_{k} \mathcal{U}^{(l)}_{mk} T_{lk}
\end{align}
where $\mathcal{U}^{(l)}_{mk}$ takes the obvious form
\begin{align}
  \mathcal{U}^{(l)}_{mk} =
  \begin{cases}
    \frac{1}{\sqrt{2}} (\delta_{mk} + (-1)^m \delta_{m,-k}) & \text{if } m > 0, \\
    \delta_{0k} & \text{if } m = 0, \\
    \frac{i}{\sqrt{2}} (\delta_{mk} - (-1)^m \delta_{m,-k}) & \text{if } m < 0.
  \end{cases}
\end{align}
 The transformation matrices satisfy
 \begin{equation}
 	\label{eq:UU-identity}
 	\sum_{m=-l}^{l} \mathcal{U}^{(l)}_{mk}\,\mathcal{U}^{(l)}_{mk'} = (-1)^k\,\delta_{k,-k'}\,.
 \end{equation}
We can easily see that 
\begin{align}
Tr\left (Z_{lm} Z_{l'm'} \right) = \delta_{ll'} \delta_{mm'} \,,
\label{eq:o2}
\end{align}
\begin{align}
	\sum_{m} {Z}_{lm} {Z}_{lm} = \frac{(2l+1)}{N}\mathbb{I}_N.
	\label{eq:Zcomp}
\end{align}
are the orthogonality and completeness relations in this basis. 

Likewise, we have the derivations analogous to \eqref{eq:dervSF2}
\begin{align}
	\mathcal{L}^2 Z_{lm} = [L_i, [L_i, Z_{lm}]] = l(l+1) Z_{lm}\,.
	\label{eq:dervSF2real}
\end{align}

\subsection{Traces}

In general, we have
\begin{align}
  \mathcal{K}^{l_1 l_2 l_3 l_4}_{m_1 m_2 m_3 m_4} &= Tr \left( Z_{l_1 m_1} Z_{l_2 m_2} Z_{l_3 m_3} Z_{l_4 m_4} \right) \nonumber \\ 
  &= \sum_{k_1 k_2 k_3 k_4} \mathcal{U}^{(l_1)}_{m_1, k_1} \mathcal{U}^{(l_2)}_{m_2, k_2} \mathcal{U}^{(l_3)}_{m_3, k_3} \mathcal{U}^{(l_4)}_{m_4, k_4} Tr \left( T_{l_1 k_1} T_{l_2 k_2} T_{l_3 k_3} T_{l_4 k_4} \right) \nonumber \\
 &= \sum_{k_1 k_2 k_3 k_4} \mathcal{U}^{(l_1)}_{m_1, k_1} \mathcal{U}^{(l_2)}_{m_2, k_2} \mathcal{U}^{(l_3)}_{m_3, k_3} \mathcal{U}^{(l_4)}_{m_4, k_4} K^{l_1 l_2 l_3 l_4}_{k_1 k_2 k_3 k_4},
\end{align}
where from \cite{Varshalovich:1988ifq} or using \eqref{eq:o1} and \eqref{eq:Texp}
\begin{align}
& K^{l_1 l_2 l_3 l_4}_{m_1 m_2 m_3 m_4} = Tr \left( T_{l_1 m_1} T_{l_2 m_2} T_{l_3 m_3} T_{l_4 m_4} \right) \nonumber \\
&= \prod_{i=1}^4 \sqrt{(2l_i + 1)} \sum_{l \, m} (-1)^{m} \sixj{l_1}{l_2}{l}{\frac{N-1}{2}}{\frac{N-1}{2}}{\frac{N-1}{2}} \sixj{l_3}{l_4}{l}{\frac{N-1}{2}}{\frac{N-1}{2}}{\frac{N-1}{2}} C^{l m}_{l_1 m_1 l_2 m_2} C^{l -m}_{l_3 m_3 l_4 m_4}.
\end{align}
For the trace involved in the planar interaction term \eqref{eq:planarint}, we find
\begin{align}
 \mathcal{K}^{l_1 l_1 l_2 l_2}_{m_1 m_1 m_2 m_2} &= \sum_{m_1 m_2}  Tr \left( Z_{l_1 m_1} Z_{l_1 m_1} Z_{l_2 m_2} Z_{l_2 m_2} \right)  \,,  \nonumber \\ 
 &= \sum_{m_1 m_2}  Tr \left( T_{l_1 m_1} T^\dagger_{l_1 m_1} T_{l_2 m_2} T^\dagger_{l_2 m_2} \right) \,,  \nonumber \\ 
 &= \frac{(2 l_1+1)(2 l_2+1)}{N} \,.
 \end{align}
 We may note that this result follows from either directly using the identity \eqref{eq:Zcomp} or noting first that the rotational invariance of $\mathcal{K}^{l_1 l_1 l_2 l_2}_{m_1 m_1 m_2 m_2}$ implies, or use of \eqref{eq:UU-identity} explicitly converts the trace in the $Z_{lm}$ basis to that in the $T_{lm}$ basis and subsequently using \eqref{eq:resid1} in the latter. 
 
 For the trace involved in the non-planar interaction term \eqref{eq:nonplanarint}, we have 
 \begin{align}
 	\label{eq:NP-expand}
 	&\sum_{m_1 m_2} Tr\,(Z_{l_1 m_1}Z_{l_2 m_2}Z_{l_1 m_1}Z_{l_2 m_2}) \,, \nn \\
 	&= \sum_{m_1 m_2}\sum_{k_1 k_2 k_1' k_2'}\mathcal{U}^{(l_1)}_{m_1 k_1}\mathcal{U}^{(l_2)}_{m_2 k_2}\mathcal{U}^{(l_1)}_{m_1 k_1'}\mathcal{U}^{(l_2)}_{m_2 k_2'}\,Tr\,(T_{l_1 k_1}T_{l_2 k_2}T_{l_1 k_1'}T_{l_2 k_2'})\,,  \nn \\
 	&= \sum_{k_1 k_2} Tr\,(T_{l_1 k_1} T_{l_2 k_2} T^\dagger_{l_1 k_1} T^\dagger_{l_2 k_2})  \,,  \nonumber \\ 
 	&= \sum_{k_1 k_2} (2l_1+1)(2l_2+1)(-1)^{k_1+k_2}\sum_{l,q}(-1)^{2(N-1+l)}(-1)^{q} \sixj{l_1}{l_2}{l}{\frac{N-1}{2}}{\frac{N-1}{2}}{\frac{N-1}{2}}^{\!2} C^{lq}_{l_1 k_1 l_2 k_2}\,C^{l-q}_{l_1 -k_1 l_2 -k_2} \,,  \nonumber \\ 
 	&=\sum_{k_1 k_2} (2l_1+1)(2l_2+1)(-1)^{k_1+k_2} \sum_{l,q}(-1)^{l_1+l_2-l}(-1)^q\sixj{l_1}{l_2}{l}{\frac{N-1}{2}}{\frac{N-1}{2}}{\frac{N-1}{2}}^{\!2}\left(C^{lq}_{l_1 k_1 l_2 k_2}\right)^2\,, \nn \\
 	&= (2l_1+1)(2l_2+1) \sum_{l=0}^{N-1}(2l+1)(-1)^{l_1+l_2-l}\sixj{l_1}{l_2}{l}{\frac{N-1}{2}}{\frac{N-1}{2}}{\frac{N-1}{2}}^{\!2}\,, \nn \\
 	&= (2l_1+1)(2l_2+1) (-1)^{l_1+l_2+N-1} \sixj{l_1}{\frac{N-1}{2}}{\frac{N-1}{2}}{l_2}{\frac{N-1}{2}}{\frac{N-1}{2}} \,, \nn \\
 	&=:  (2l_1 + 1)(2l_2 + 1) G(l_1, l_2) \,,
 	\end{align}
where the equivalence of the traces in the first and third lines also follow from rotational invariance, with the overall constant of proportionality matched, for instance, by using the fact that $Z_{l0}= T_{l0}$.  In arriving this result, we have used the symmetry and orthogonality of the Clebsch-Gordan coefficients:
\begin{equation}
C^{l-q}_{l_1 - k_1 l_2 -k_2} = (-1)^{l_1+l_2-l}\,C^{lq}_{l_1 k_1 l_2 k_2}\,, \quad \quad \sum_{k_1 k_2}\left(C^{lq}_{l_1 k_1 l_2 k_2}\right)^2 =1  \,,
\end{equation}
the selection rule $k_1 + k_2 = q$ and in the last line the formula \cite{Varshalovich:1988ifq}
\begin{align}
\sum_{l=0}^{N-1}(2l+1)(-1)^{N-1-l} \sixj{l_1}{l_2}{l}{\frac{N-1}{2}}{\frac{N-1}{2}}{\frac{N-1}{2}}  \sixj{l_1}{l_2}{l}{\frac{N-1}{2}}{\frac{N-1}{2}}{\frac{N-1}{2}} = \sixj{l_1}{\frac{N-1}{2}}{\frac{N-1}{2}}{l_2}{\frac{N-1}{2}}{\frac{N-1}{2}} \,.
\end{align}

\section{Equations of Motion}

\subsection{Equations of motion for $1$- and $2$-point functions}

\label{sec:eoms}

Starting from the expression \eqref{eq:averaged-hamiltonian_short} and applying Wick's theorem, we have the expectation value of the Hamiltonian in the Gaussian state expressed as:
\begin{align}
	\label{eq:averaged-hamiltonian}
	\braket{\hat{H}} &= \frac{1}{2} \Bigg[ \braket{{\hat P}_i^{lm}{\hat P}_i^{lm}} + \left( \mu^2 + \frac{l(l+1)}{R^2} \right) \braket{{\hat X}_i^{lm}{\hat X}_i^{lm}} + \lambda \mathcal{H}^{l_1 l_2 l_3 l_4}_{m_1 m_2 m_3 m_4} \braket{{\hat X}_1^{l_1 m_1} {\hat X}_1^{l_2 m_2} {\hat X}_2^{l_3 m_3} {\hat X}_2^{l_4 m_4}} \Bigg] \nonumber \\
	&= \frac{1}{2} \Bigg[ \braket{{\hat P}_i^{lm}{\hat P}_i^{lm}} + \left( \mu^2 + \frac{l(l+1)}{R^2} \right) \braket{{\hat X}_i^{lm}{\hat X}_i^{lm}} + \lambda \mathcal{H}^{l_1 l_2 l_3 l_4}_{m_1 m_2 m_3 m_4} \Big( \braket{{\hat X}_1^{l_1 m_1}}\braket{{\hat X}_1^{l_2 m_2}}\braket{{\hat X}_2^{l_3 m_3}}\braket{{\hat X}_2^{l_4 m_4}} \nonumber \\
	+&\braket{\braket{{\hat X}_1^{l_1 m_1} {\hat X}_1^{l_2 m_2}}} \left(\braket{\braket{{\hat X}_2^{l_3 m_3} {\hat X}_2^{l_4 m_4}}} + \braket{{\hat X}_2^{l_3 m_3}}\braket{{\hat X}_2^{l_4 m_4}}\right) + \braket{{\hat X}_1^{l_1 m_1}}\braket{{\hat X}_1^{l_2 m_2}}\braket{\braket{{\hat X}_2^{l_3 m_3}{\hat X}_2^{l_4 m_4}}} \nonumber \\
	+&\braket{\braket{{\hat X}_1^{l_1 m_1} {\hat X}_2^{l_3 m_3}}} \left(\braket{\braket{{\hat X}_1^{l_2 m_2} {\hat X}_2^{l_4 m_4}}} + \braket{{\hat X}_1^{l_2 m_2}}\braket{{\hat X}_2^{l_4 m_4}}\right) + \braket{{\hat X}_1^{l_1 m_1}}\braket{{\hat X}_2^{l_3 m_3}}\braket{\braket{{\hat X}_1^{l_2 m_2}{\hat X}_2^{l_4 m_4}}} \nonumber \\
	+&\braket{\braket{{\hat X}_1^{l_1 m_1} {\hat X}_2^{l_4 m_4}}} \left(\braket{\braket{{\hat X}_1^{l_2 m_2} {\hat X}_2^{l_3 m_3}}} + \braket{{\hat X}_1^{l_2 m_2}}\braket{{\hat X}_2^{l_3 m_3}}\right) + \braket{{\hat X}_1^{l_1 m_1}}\braket{{\hat X}_2^{l_4 m_4}}\braket{\braket{{\hat X}_1^{l_2 m_2}{\hat X}_2^{l_3 m_3}}} \Bigg]. 
\end{align}
In a similar manner, evaluating \eqref{eq:opeqs1} in the Gaussian state and applying Wick's theorem we obtain the equations in \eqref{eq:eom-one-point}-\eqref{eq:p2} as given in the text
and the equations for the correlation functions which take the form
\begin{align}
	\label{eq:eom-two-point-xx}
	\partial_t\braket{\braket{{\hat X}_1^{l_1 m_1} {\hat X}_1^{l_2 m_2}}} &= \braket{\braket{ {\hat X}_1^{l_2 m_2} {\hat P}_1^{l_1 m_1} }} + \braket{\braket{{\hat X}_1^{l_1 m_1} {\hat P}_1^{l_2 m_2}}}  \,, \nn \\
	\partial_t\braket{\braket{{\hat X}_1^{l_1 m_1} {\hat X}_2^{l_2 m_2}}} &= \braket{\braket{ {\hat X}_2^{l_2 m_2} {\hat P}_1^{l_1 m_1} }} + \braket{\braket{{\hat X}_1^{l_1 m_1} {\hat P}_2^{l_2 m_2}}}  \,, \\
	\partial_t\braket{\braket{{\hat X}_2^{l_1 m_1} {\hat X}_2^{l_2 m_2}}} &= \braket{\braket{ {\hat X}_2^{l_2 m_2} {\hat P}_2^{l_1 m_1} }} + \braket{\braket{{\hat X}_2^{l_1 m_1} {\hat P}_2^{l_2 m_2}}} \,, \nn
\end{align}
\vspace*{-2em}
\begin{align}
	\label{eq:eom-two-point-xp1}
	\partial_t\braket{\braket{{\hat X}_1^{l_1 m_1} {\hat P}_1^{l_2 m_2}}} &= \braket{\braket{{\hat P}_1^{l_1 m_1} {\hat P}_1^{l_2 m_2}}} -\Bigg\{\left(\mu^2+\frac{l_2\left(l_2+1\right)}{R^2}\right) \braket{\braket{{\hat X}_1^{l_2 m_2} {\hat X}_1^{l_1 m_1}}} +\frac{\lambda}{2} \mathcal{N}^{l_2 l_3 l_4 l_5}_{m_2 m_3 m_4 m_5} \nonumber \\
	&\times \Big(\braket{\braket{{\hat X}_1^{l_1 m_1} {\hat X}_1^{l_3 m_3}}}\braket{\braket{{\hat X}_2^{l_4 m_4} {\hat X}_2^{l_5 m_5}}}+\braket{\braket{{\hat X}_1^{l_1 m_1} {\hat X}_1^{l_3 m_3}}}\braket{{\hat X}_2^{l_4 m_4}}\braket{{\hat X}_2^{l_5 m_5}}\nonumber \\
	& \;\;\;+\braket{\braket{{\hat X}_1^{l_3 m_3} {\hat X}_2^{l_5 m_5}}}\braket{\braket{{\hat X}_1^{l_1 m_1} {\hat X}_2^{l_4 m_4}}}+\braket{\braket{{\hat X}_1^{l_1 m_1} {\hat X}_2^{l_4 m_4}}}\braket{{\hat X}_1^{l_3 m_3}}\braket{{\hat X}_2^{l_5 m_5}}\nonumber \\
	& \;\;\;+\braket{\braket{{\hat X}_1^{l_1 m_1} {\hat X}_2^{l_5 m_5}}}\braket{\braket{{\hat X}_1^{l_3 m_3} {\hat X}_2^{l_4 m_4}}}+\braket{\braket{{\hat X}_1^{l_1 m_1} {\hat X}_2^{l_5 m_5}}}\braket{{\hat X}_1^{l_3 m_3}}\braket{\hat{X}_2^{l_4 m_4}}\Big)\Bigg\} \,, 
\end{align}
\vspace*{-2em}
\begin{align}
	\label{eq:eom-two-point-xp2}
	\partial_t\braket{\braket{{\hat X}_2^{l_1 m_1} {\hat P}_1^{l_2 m_2}}} &= \braket{\braket{{\hat P}_2^{l_1 m_1} {\hat P}_1^{l_2 m_2}}} -\Bigg\{\left(\mu^2+\frac{l_2\left(l_2+1\right)}{R^2}\right) \braket{\braket{{\hat X}_1^{l_2 m_2} {\hat X}_2^{l_1 m_1}}} +\frac{\lambda}{2} \mathcal{N}^{l_2 l_3 l_4 l_5}_{m_2 m_3 m_4 m_5} \nonumber \\
	&\times \Big(\braket{\braket{{\hat X}_2^{l_1 m_1} {\hat X}_1^{l_3 m_3}}}\braket{\braket{{\hat X}_2^{l_4 m_4} {\hat X}_2^{l_5 m_5}}}+\braket{\braket{{\hat X}_2^{l_1 m_1} {\hat X}_1^{l_3 m_3}}}\braket{{\hat X}_2^{l_4 m_4}}\braket{{\hat X}_2^{l_5 m_5}}\nonumber \\
	& \;\;\;+\braket{\braket{{\hat X}_1^{l_3 m_3} {\hat X}_2^{l_5 m_5}}}\braket{\braket{{\hat X}_2^{l_1 m_1} {\hat X}_2^{l_4 m_4}}}+\braket{\braket{{\hat X}_2^{l_1 m_1} {\hat X}_2^{l_4 m_4}}}\braket{{\hat X}_1^{l_3 m_3}}\braket{{\hat X}_2^{l_5 m_5}}\nonumber \\
	& \;\;\;+\braket{\braket{{\hat X}_2^{l_1 m_1} {\hat X}_2^{l_5 m_5}}}\braket{\braket{{\hat X}_1^{l_3 m_3} {\hat X}_2^{l_4 m_4}}}+\braket{\braket{{\hat X}_2^{l_1 m_1} {\hat X}_2^{l_5 m_5}}}\braket{{\hat X}_1^{l_3 m_3}}\braket{{\hat X}_2^{l_4 m_4}}\Big)\Bigg\} \,, 
\end{align}
\vspace*{-2em}
\begin{align}
	\label{eq:eom-two-point-xp3}
	\partial_t\braket{\braket{{\hat X}_1^{l_1 m_1} {\hat P}_2^{l_2 m_2}}}&=\braket{\braket{{\hat P}_1^{l_1 m_1} {\hat P}_2^{l_2 m_2}}} -\Bigg\{\left(\mu^2+\frac{l_2\left(l_2+1\right)}{R^2}\right)\braket{\braket{{\hat X}_2^{l_2 m_2} {\hat X}_1^{l_1 m_1}}} +\frac{\lambda}{2} \mathcal{N}^{l_2 l_3 l_4 l_5}_{m_2 m_3 m_4 m_5} \nonumber \\
	&\times \Big(\braket{\braket{{\hat X}_1^{l_1 m_1} {\hat X}_2^{l_3 m_3}}}\braket{\braket{{\hat X}_1^{l_4 m_4} {\hat X}_1^{l_5 m_5}}}+\braket{\braket{{\hat X}_1^{l_1 m_1} {\hat X}_2^{l_3 m_3}}}\braket{{\hat X}_1^{l_4 m_4}}\braket{{\hat X}_1^{l_5 m_5}}\nonumber \\
	& \;\;\;+\braket{\braket{{\hat X}_2^{l_3 m_3} {\hat X}_1^{l_5 m_5}}}\braket{\braket{{\hat X}_1^{l_1 m_1} {\hat X}_1^{l_4 m_4}}}+\braket{\braket{{\hat X}_1^{l_1 m_1} {\hat X}_1^{l_4 m_4}}}\braket{{\hat X}_2^{l_3 m_3}}\braket{{\hat X}_1^{l_5 m_5}}\nonumber \\
	& \;\;\;+\braket{\braket{{\hat X}_1^{l_1 m_1} {\hat X}_1^{l_5 m_5}}}\braket{\braket{{\hat X}_2^{l_3 m_3} {\hat X}_1^{l_4 m_4}}}+\braket{\braket{{\hat X}_1^{l_1 m_1} {\hat X}_1^{l_5 m_5}}}\braket{{\hat X}_2^{l_3 m_3}}\braket{{\hat X}_1^{l_4 m_4}}\Big)\Bigg\} \,, 
\end{align}
\vspace*{-2em}
\begin{align}
	\label{eq:eom-two-point-xp4}
	\partial_t\braket{\braket{{\hat X}_2^{l_1 m_1} {\hat P}_2^{l_2 m_2}}}&=\braket{\braket{{\hat P}_2^{l_1 m_1} {\hat P}_2^{l_2 m_2}}}-\Bigg\{\left(\mu^2+\frac{l_2\left(l_2+1\right)}{R^2}\right) \braket{\braket{{\hat X}_2^{l_2 m_2} {\hat X}_2^{l_1 m_1}}} +\frac{\lambda}{2} \mathcal{N}^{l_2 l_3 l_4 l_5}_{m_2 m_3 m_4 m_5} \nonumber \\
	&\times \Big(\braket{\braket{{\hat X}_2^{l_1 m_1} {\hat X}_2^{l_3 m_3}}}\braket{\braket{{\hat X}_1^{l_4 m_4} {\hat X}_1^{l_5 m_5}}}+\braket{\braket{{\hat X}_2^{l_1 m_1} {\hat X}_2^{l_3 m_3}}}\braket{{\hat X}_1^{l_4 m_4}}\braket{{\hat X}_1^{l_5 m_5}}\nonumber \\
	& \;\;\;+\braket{\braket{{\hat X}_2^{l_3 m_3} {\hat X}_1^{l_5 m_5}}}\braket{\braket{{\hat X}_2^{l_1 m_1} {\hat X}_1^{l_4 m_4}}}+\braket{\braket{{\hat X}_2^{l_1 m_1} {\hat X}_1^{l_4 m_4}}}\braket{{\hat X}_2^{l_3 m_3}}\braket{{\hat X}_1^{l_5 m_5}}\nonumber \\
	& \;\;\;+\braket{\braket{{\hat X}_2^{l_1 m_1} {\hat X}_1^{l_5 m_5}}}\braket{\braket{{\hat X}_2^{l_3 m_3} {\hat X}_1^{l_4 m_4}}}+\braket{\braket{{\hat X}_2^{l_1 m_1} {\hat X}_1^{l_5 m_5}}}\braket{{\hat X}_2^{l_3 m_3}}\braket{{\hat X}_1^{l_4 m_4}}\Big)\Bigg\} \,, 
\end{align}
\vspace*{-3em}
\begin{align}
	\label{eq:eom-two-point-pp}
	\partial_t\braket{\braket{{\hat P}_1^{l_1 m_1} {\hat P}_1^{l_2 m_2}}}&=-\Bigg\{\left(\mu^2+\frac{l_1\left(l_1+1\right)}{R^2}\right) \braket{\braket{{\hat X}_1^{l_1 m_1} {\hat P}_1^{l_2 m_2}}} +\frac{\lambda}{2} \mathcal{N}^{l_1 l_3 l_4 l_5}_{m_1 m_3 m_4 m_5}\nonumber \\
	&\times \Big(\braket{\braket{{\hat X}_1^{l_3 m_3} {\hat P}_1^{l_2 m_2}}}\braket{\braket{{\hat X}_2^{l_4 m_4} {\hat X}_2^{l_5 m_5}}}+\braket{\braket{{\hat X}_1^{l_3 m_3} {\hat P}_1^{l_2 m_2}}}\braket{{\hat X}_2^{l_4 m_4}}\braket{{\hat X}_2^{l_5 m_5}}\nonumber \\
	&\;\;\;+\braket{\braket{{\hat X}_1^{l_3 m_3} {\hat X}_2^{l_5 m_5}}}\braket{\braket{{\hat X}_2^{l_4 m_4} {\hat P}_1^{l_2 m_2}}}+\braket{{\hat X}_1^{l_3 m_3}}\braket{{\hat X}_2^{l_5 m_5}}\braket{\braket{{\hat X}_2^{l_4 m_4} {\hat P}_1^{l_2 m_2}}}\nonumber \\
	&\;\;\;+\braket{\braket{{\hat X}_2^{l_5 m_5} {\hat P}_1^{l_2 m_2}}}\braket{\braket{{\hat X}_1^{l_3 m_3} {\hat X}_2^{l_4 m_4}}}+\braket{\braket{{\hat X}_2^{l_5 m_5} {\hat P}_1^{l_2 m_2}}}\braket{{\hat X}_1^{l_3 m_3}}\braket{{\hat X}_2^{l_4 m_4}}\Big)\nonumber \\
	&+\left(\mu^2+\frac{l_2\left(l_2+1\right)}{R^2}\right) \braket{\braket{{\hat X}_1^{l_2 m_2} {\hat P}_1^{l_1 m_1}}}+\frac{\lambda}{2}\mathcal{N}^{l_2 l_3 l_4 l_5}_{m_2 m_3 m_4 m_5} \nonumber \\
	&\times \Big(\braket{\braket{{\hat X}_1^{l_3 m_3} {\hat P}_1^{l_1 m_1}}}\braket{\braket{{\hat X}_2^{l_4 m_4} {\hat X}_2^{l_5 m_5}}}+\braket{\braket{{\hat X}_1^{l_3 m_3} {\hat P}_1^{l_1 m_1}}}\braket{{\hat X}_2^{l_4 m_4}}\braket{{\hat X}_2^{l_5 m_5}}\nonumber \\
	&\;\;\;+\braket{\braket{{\hat X}_1^{l_3 m_3} {\hat X}_2^{l_5 m_5}}}\braket{\braket{{\hat X}_2^{l_4 m_4} {\hat P}_1^{l_1 m_1}}}+\braket{{\hat X}_1^{l_3 m_3}}\braket{{\hat X}_2^{l_5 m_5}}\braket{\braket{{\hat X}_2^{l_4 m_4} {\hat P}_1^{l_1 m_1}}}\nonumber \\
	&\;\;\;+\braket{\braket{{\hat X}_2^{l_5 m_5} {\hat P}_1^{l_1 m_1}}}\braket{\braket{{\hat X}_1^{l_3 m_3} {\hat X}_2^{l_4 m_4}}}+\braket{\braket{{\hat X}_2^{l_5 m_5} {\hat P}_1^{l_1 m_1}}}\braket{{\hat X}_1^{l_3 m_3}}\braket{{\hat X}_2^{l_4 m_4}}\Big)\Bigg\} \,, 
\end{align}
\vspace*{-3em}
\begin{align}
	\label{eq:eom-two-point-pp-12}
	\partial_t\braket{\braket{{\hat P}_1^{l_1 m_1} {\hat P}_2^{l_2 m_2}}}&=-\Bigg\{\left(\mu^2+\frac{l_1\left(l_1+1\right)}{R^2}\right) \braket{\braket{{\hat X}_1^{l_1 m_1} {\hat P}_2^{l_2 m_2}}} + \frac{\lambda}{2} \mathcal{N}^{l_1 l_3 l_4 l_5}_{m_1 m_3 m_4 m_5}\nonumber \\
	&\times \Big(\braket{\braket{{\hat X}_1^{l_3 m_3} {\hat P}_2^{l_2 m_2}}}\braket{\braket{{\hat X}_2^{l_4 m_4} {\hat X}_2^{l_5 m_5}}}+\braket{\braket{{\hat X}_1^{l_3 m_3} {\hat P}_2^{l_2 m_2}}}\braket{{\hat X}_2^{l_4 m_4}}\braket{{\hat X}_2^{l_5 m_5}}\nonumber \\
	&\;\;\;+\braket{\braket{{\hat X}_1^{l_3 m_3} {\hat X}_2^{l_5 m_5}}}\braket{\braket{{\hat X}_2^{l_4 m_4} {\hat P}_2^{l_2 m_2}}}+\braket{{\hat X}_1^{l_3 m_3}}\braket{{\hat X}_2^{l_5 m_5}}\braket{\braket{{\hat X}_2^{l_4 m_4} {\hat P}_2^{l_2 m_2}}}\nonumber \\
	&\;\;\;+\braket{\braket{{\hat X}_2^{l_5 m_5} {\hat P}_2^{l_2 m_2}}}\braket{\braket{{\hat X}_1^{l_3 m_3} {\hat X}_2^{l_4 m_4}}}+\braket{\braket{{\hat X}_2^{l_5 m_5} {\hat P}_2^{l_2 m_2}}}\braket{{\hat X}_1^{l_3 m_3}}\braket{{\hat X}_2^{l_4 m_4}}\Big)\nonumber \\
	&+\left(\mu^2+\frac{l_2\left(l_2+1\right)}{R^2}\right) \braket{\braket{{\hat X}_2^{l_2 m_2} {\hat P}_1^{l_1 m_1}}}+\frac{\lambda}{2}\mathcal{N}^{l_2 l_3 l_4 l_5}_{m_2 m_3 m_4 m_5} \nonumber \\
	&\times \Big(\braket{\braket{{\hat X}_2^{l_3 m_3} {\hat P}_1^{l_1 m_1}}}\braket{\braket{{\hat X}_1^{l_4 m_4} {\hat X}_1^{l_5 m_5}}}+\braket{\braket{{\hat X}_2^{l_3 m_3} {\hat P}_1^{l_1 m_1}}}\braket{{\hat X}_1^{l_4 m_4}}\braket{{\hat X}_1^{l_5 m_5}}\nonumber \\
	&\;\;\;+\braket{\braket{{\hat X}_2^{l_3 m_3} {\hat X}_1^{l_5 m_5}}}\braket{\braket{{\hat X}_1^{l_4 m_4} {\hat P}_1^{l_1 m_1}}}+\braket{{\hat X}_2^{l_3 m_3}}\braket{{\hat X}_1^{l_5 m_5}}\braket{\braket{{\hat X}_1^{l_4 m_4} {\hat P}_1^{l_1 m_1}}}\nonumber \\
	&\;\;\;+\braket{\braket{{\hat X}_1^{l_5 m_5} {\hat P}_1^{l_1 m_1}}}\braket{\braket{{\hat X}_2^{l_3 m_3} {\hat X}_1^{l_4 m_4}}}+\braket{\braket{{\hat X}_1^{l_5 m_5} {\hat P}_1^{l_1 m_1}}}\braket{{\hat X}_2^{l_3 m_3}}\braket{{\hat X}_1^{l_4 m_4}}\Big)\Bigg\} \,, 
\end{align}
\vspace*{-3em}
\begin{align}
	\label{eq:eom-two-point-pp-22}
	\partial_t\braket{\braket{{\hat P}_2^{l_1 m_1} {\hat P}_2^{l_2 m_2}}}&=-\Bigg\{\left(\mu^2+\frac{l_1\left(l_1+1\right)}{R^2}\right) \braket{\braket{{\hat X}_2^{l_1 m_1} {\hat P}_2^{l_2 m_2}}} + \frac{\lambda}{2} \mathcal{N}^{l_1 l_3 l_4 l_5}_{m_1 m_3 m_4 m_5}\nonumber \\
	&\times \Big(\braket{\braket{{\hat X}_2^{l_3 m_3} {\hat P}_2^{l_2 m_2}}}\braket{\braket{{\hat X}_1^{l_4 m_4} {\hat X}_1^{l_5 m_5}}}+\braket{\braket{{\hat X}_2^{l_3 m_3} {\hat P}_2^{l_2 m_2}}}\braket{{\hat X}_1^{l_4 m_4}}\braket{{\hat X}_1^{l_5 m_5}}\nonumber \\
	&\;\;\;+\braket{\braket{{\hat X}_2^{l_3 m_3} {\hat X}_1^{l_5 m_5}}}\braket{\braket{{\hat X}_1^{l_4 m_4} {\hat P}_2^{l_2 m_2}}}+\braket{{\hat X}_2^{l_3 m_3}}\braket{{\hat X}_1^{l_5 m_5}}\braket{\braket{{\hat X}_1^{l_4 m_4} {\hat P}_2^{l_2 m_2}}}\nonumber \\
	&\;\;\;+\braket{\braket{{\hat X}_1^{l_5 m_5} {\hat P}_2^{l_2 m_2}}}\braket{\braket{{\hat X}_2^{l_3 m_3} {\hat X}_1^{l_4 m_4}}}+\braket{\braket{{\hat X}_1^{l_5 m_5} {\hat P}_2^{l_2 m_2}}}\braket{{\hat X}_2^{l_3 m_3}}\braket{{\hat X}_1^{l_4 m_4}}\Big)\nonumber \\
	&+\left(\mu^2+\frac{l_2\left(l_2+1\right)}{R^2}\right) \braket{\braket{{\hat X}_2^{l_2 m_2} {\hat P}_2^{l_1 m_1}}} + \frac{\lambda}{2}\mathcal{N}^{l_2 l_3 l_4 l_5}_{m_2 m_3 m_4 m_5} \nonumber \\
	&\times \Big(\braket{\braket{{\hat X}_2^{l_3 m_3} {\hat P}_2^{l_1 m_1}}}\braket{\braket{{\hat X}_1^{l_4 m_4} {\hat X}_1^{l_5 m_5}}}+\braket{\braket{{\hat X}_2^{l_3 m_3} {\hat P}_2^{l_1 m_1}}}\braket{{\hat X}_1^{l_4 m_4}}\braket{{\hat X}_1^{l_5 m_5}}\nonumber \\
	&\;\;\;+\braket{\braket{{\hat X}_2^{l_3 m_3} {\hat X}_1^{l_5 m_5}}}\braket{\braket{{\hat X}_1^{l_4 m_4} {\hat P}_2^{l_1 m_1}}}+\braket{{\hat X}_2^{l_3 m_3}}\braket{{\hat X}_1^{l_5 m_5}}\braket{\braket{{\hat X}_1^{l_4 m_4} {\hat P}_2^{l_1 m_1}}}\nonumber \\
	&\;\;\;+\braket{\braket{{\hat X}_1^{l_5 m_5} {\hat P}_2^{l_1 m_1}}}\braket{\braket{{\hat X}_2^{l_3 m_3} {\hat X}_1^{l_4 m_4}}}+\braket{\braket{{\hat X}_1^{l_5 m_5} {\hat P}_2^{l_1 m_1}}}\braket{{\hat X}_2^{l_3 m_3}}\braket{{\hat X}_1^{l_4 m_4}}\Big)\Bigg\} \,.
\end{align}

\subsection{Lax Structure of the Equations of motion}
\label{sec:Pure_pure}

We observe that the equations of motion can be expressed as 
\begin{eqnarray}
\partial_t {\bar \xi } &=& \Omega  \frac{\partial \braket{\hat{H}}}{\partial {\bar \xi}} \,, \nn \\
\partial_t \Sigma &=& \Omega \frac{\partial \braket{\hat{H}}}{\partial \Sigma} \Sigma - \Sigma \frac{\partial \braket{\hat{H}}}{\partial \Sigma}  \Omega \,,
\end{eqnarray}
where the components of $\frac{\partial \braket{\hat{H}}}{\partial \Sigma}$ can be obtained by computing the total variation of $\braket{\hat{H}}$. The equation for the one-point functions is therefore of the standard Hamiltonian form, while multiplying both sides of the second equation with $\Omega$ from the right and using $\Omega^2 = - \mathbb{I}$ twice we may cast it into
\begin{align}
\partial_t (\Sigma \Omega) &= \Omega \frac{\partial \braket{\hat{H}}}{\partial \Sigma} (\Sigma \Omega) - (\Sigma\Omega) \frac{\partial \braket{\hat{H}}}{\partial \Sigma}  \Omega \,, \nn \\
 &= \lbrack \Omega \frac{\partial \braket{\hat{H}}}{\partial \Sigma}, (\Sigma \Omega) \rbrack \,,
\end{align}
from which we infer that the time evolution of $\Sigma \Omega$ is governed by a Lax-type equation which preserves the eigenvalues of $(\Sigma \Omega)$, which are the symplectic eigenvalues of $\Sigma$. Thus, it follows that pure states evolve into pure states under these time evolution equations. 

\section{Computational Methods and Results} 
\label{sec:computational-methods-and-results}

In this section we outline the computational methods we have used in solving the equations of motion given explicitly in Appendix \ref{sec:eoms} and also those for obtaining the largest Lyapunov exponent $\lambda_L$.

\subsection{Integration Method}
\label{sec:int_method}

We have used the symplectic Euler method for our integrations, which, as the name suggests, is a symplectic integrator\cite{Buividovich:2018scl}. We can write our equations of motion as given in Appendix \ref{sec:eoms} in the implicit form as
\begin{align}
	\partial_t \braket{{\hat X}} &= F^{(1)}\left(\braket{{\hat P}}\right) \,, \nn \\
	\partial_t \braket{{\hat P}} &= F^{(2)}\left(\braket{{\hat X}}, \braket{\braket{{\hat X}{\hat X}}}\right)  \,, \nn \\
	\partial_t\braket{\braket{{\hat X}{\hat X}}} &= F^{(3)}\left(\braket{\braket{{\hat X} {\hat P}}}\right) \,, \nn \\
	\partial_t\braket{\braket{{\hat X}{\hat P}}} &= F^{(4)}\left(\braket{{\hat X}}, \braket{\braket{{\hat X}{\hat X}}}, \braket{\braket{{\hat P} {\hat P}}}\right) \,, \nn\\
	\partial_t\braket{\braket{{\hat P} {\hat P}}} &= F^{(5)}\left(\braket{{\hat X}}, \braket{\braket{{\hat X}{\hat X}}}, \braket{\braket{{\hat X} {\hat P}}}\right)  \,.
\end{align}
Then the symplectic Euler method can be written as
\begin{align}
	\braket{{\hat P}}_{n+1}  &= \braket{{\hat P}}_n + h F^{(2)}\left(\braket{{\hat X}}_n, \braket{\braket{{\hat X}{\hat X}}}_n\right)  \,, \nn \\
	\braket{\braket{{\hat X}{\hat P}}}_{n+1} &= \braket{\braket{{\hat X}{\hat P}}}_n + h F^{(4)}\left(\braket{{\hat X}}_n, \braket{\braket{{\hat X}{\hat X}}}_n, \braket{\braket{{\hat P}{\hat P}}}_n\right)  \,, \nn \\
	\braket{\braket{{\hat P}{\hat P}}}_{n+1} &= \braket{\braket{{\hat P}{\hat P}}}_n + h F^{(5)}\left(\braket{{\hat X}}_n, \braket{\braket{{\hat X}{\hat X}}}_n, \braket{\braket{{\hat X}{\hat P}}}_{n+1}\right)  \,, \nn \\
	\braket{\braket{{\hat X}{\hat X}}}_{n+1} &= \braket{\braket{{\hat X}{\hat X}}}_n + h F^{(3)}\left(\braket{\braket{{\hat X}{\hat P}}}_{n+1}\right)  \,, \nn \\
	\braket{{\hat X}}_{n+1}  &= \braket{{\hat X}}_n + h F^{(1)}\left(\braket{{\hat P}}_{n+1}\right)  \,.
\end{align}
This is our integration method and update order for the equations of motion.

\subsection{Lyapunov Exponent}
\label{sec:Lexp}

\begin{figure}[htbp]
	\begin{center}
\includegraphics[width=0.95\textwidth]{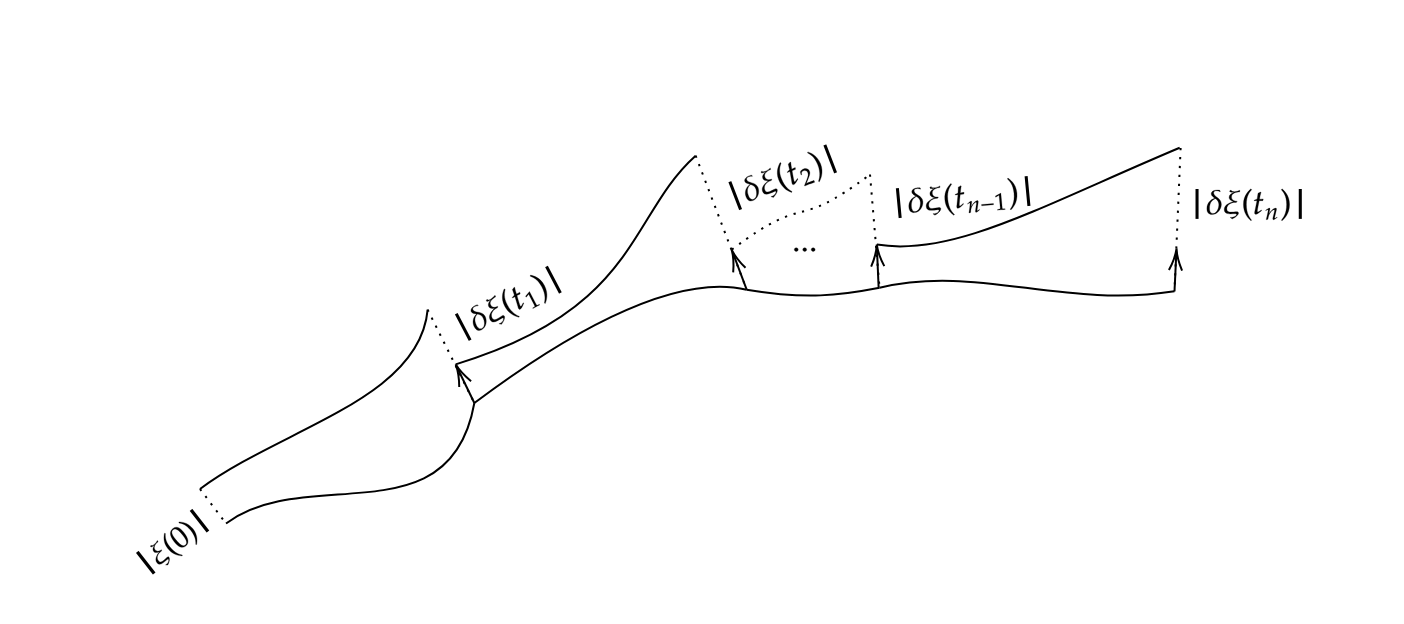}
	\end{center}
	\caption{Lyapunov Exponent Computation}\label{fig:lyapunov-exponent}
\end{figure}

Computing the largest Lyapunov exponent, although simple once the Hamilton equations of motion are solved numerically, has to be done with some care. We can define the largest Lyapunov exponent by
\begin{align}
	\lambda_L \;=\; 
	\lim_{t\to\infty} \frac{1}{t}\,
	\log\!\left(\frac{\|\delta\xi(t)\|}{\|\delta\xi(0)\|}\right)\,.
\end{align}
In practice, to control numerical errors we use the renormalize‐and‐sum procedure illustrated in Fig.~\ref{fig:lyapunov-exponent}. We evolve two nearby trajectories, $\xi_1(t)$ and $\xi_2(t)$, over equal time‐steps

\begin{align}
	t_i = i\,\Delta t\,,\qquad i=0,1,\dots,n,
\end{align}
and at each step we measure the separation
\begin{align}
	\|\delta\xi(t_i)\| = \bigl\lVert\xi_1(t_i)-\xi_2(t_i)\bigr\rVert,
\end{align}
then rescale it back to the initial length
\(
\|\delta\xi(0)\|
\),
setting
\begin{align}
	\alpha_i = \frac{\|\delta\xi(t_i)\|}{\|\delta\xi(0)\|}\,, 
	\quad
	S = \sum_{i=1}^n \ln(\alpha_i)\,.
\end{align}
The discrete estimator for \(\lambda_L\) is then
\begin{align}
	\lambda_L
	\;\approx\;
	\frac{S}{t_n}
	\;=\;
	\frac{1}{t_n}
	\sum_{i=1}^n 
	\ln\!\left(\frac{\|\delta\xi(t_i)\|}{\|\delta\xi(0)\|}\right)
	\;=\;
	\frac{1}{n\,\Delta t}
	\sum_{i=1}^n 
	\ln(\alpha_i)\,.
\end{align}
Optionally, introducing the “local” exponent
\(
\lambda_i = \tfrac{1}{\Delta t}\,\ln(\delta_i/\delta_{i-1}),
\)
one finds
\be
\lambda_L 
\;\approx\;
\frac{1}{n}\sum_{i=1}^n\lambda_i\,.
\ee

\end{document}